\documentclass[11pt]{article}
\usepackage{jheppub}
\usepackage{amsmath,amssymb,amsfonts,graphicx,slashed,amsthm,mathtools,upgreek, enumerate, tensor, subfig}
\usepackage[dvipsnames]{xcolor}
\usepackage{arydshln}

\usepackage{comment}
\usepackage{hyperref}
\usepackage[utf8]{inputenc}
\usepackage[titletoc]{appendix}

\usepackage{physics}
\usepackage{bm}

\usepackage{tikz}

\numberwithin{equation}{section} 
\allowdisplaybreaks

\allowdisplaybreaks

\newcommand{\be}{\begin{equation}}
\newcommand{\ee}{\end{equation}}
\newcommand{\f}{\frac}

\newcommand{\bea}{\begin{eqnarray}}
\newcommand{\eea}{\end{eqnarray}}
\newcommand{\ba}{\begin{align}}
\newcommand{\ea}{\end{align}}

\newcommand{\la}{\langle}
\newcommand{\ra}{\rangle}
\newcommand{\beq}{\begin{equation}}
\newcommand{\eeq}{\end{equation}}
\newcommand{\mn}{\mathcal{N}}
\newcommand{\mr}{\mathcal{R}}

\newcommand{\epr}{\text{EPR}}
\newcommand{\tfd}{\text{TFD}}
\newcommand{\ktbra}[2]{\ket{ #1 }_{ #2 } \! \bra{ #1 }}
\newcommand{\ktbrad}[3]{\ket{ #1 }_{ #3 } \! \bra{ #2 }}

\newcommand{\ketTFD}[1]{\ket{\text{TFD}}_{ #1 }}
\newcommand{\braTFD}[1]{ _{ #1 }\!\bra{\text{TFD}}}

\newcommand{\ipGrapsi}[4]{\braket{ \psi_{ #1 }^{ #2 }}{ \psi_{ #3 }^{#4 }}}

\title{The Petz (lite) recovery map for scrambling channel}

\author[a]{Yasuaki Nakayama}
\author[b]{,\! Akihiro Miyata}
\author[c]{,\! Tomonori Ugajin}

\affiliation[\,a]{Department of Physics, Kyoto University, Kyoto 606-8502, Japan}
\affiliation[\,b]{Kavli Institute for Theoretical Sciences, University of Chinese Academy of Sciences, Beijing 100190, China}
\affiliation[\,c]{Department of Physics, Rikkyo University, Toshima, Tokyo 171-8501, Japan }

\emailAdd{nakayama@gauge.scphys.kyoto-u.ac.jp}
\emailAdd{a.miyata@ucas.ac.cn}
\emailAdd{ugajin@rikkyo.ac.jp}

\preprint{KUNS-2985}

\abstract{
We study properties of the Petz recovery map in chaotic systems, such as the Hayden-Preskill setup for evaporating black holes and the SYK model. Since these systems exhibit the phenomenon called scrambling, we expect that the expression of the recovery channel $\mathcal{R}$ gets simplified, given by just the adjoint $\mathcal{N}^{\dagger}$ of the original channel $\mathcal{N}$ which defines the time evolution of the states in the code subspace embedded into the physical Hilbert space. We check this phenomenon in two examples. The first one is the Hayden-Preskill setup described by  Haar random unitaries. We compute the relative entropy $S(\mathcal{R}\left[\mathcal{N}[\rho]\right] ||\rho)$ and show that it vanishes when the decoupling is archived. We further show that the simplified recovery map is equivalent to the protocol proposed by  Yoshida and Kitaev. The second example is the SYK model where the two dimensional code subspace is defined by an insertion of a fermionic operator, and the system is evolved by the SYK Hamiltonian. We check the recovery phenomenon by relating some matrix elements of an output density matrix $\bra{T}\mathcal{R}[\mathcal{N}[\rho]]\ket{T'}$ to R\'enyi-two modular flowed correlators, and show that they coincide with the elements for the input density matrix with small error after twice the scrambling time.}


\makeatletter
\gdef\@fpheader{}
\makeatother

\begin{document}

\maketitle

\parskip=3pt

\section{Introduction}

Advances in our understanding of the relationship between quantum information theory and holographic principles have revealed the connection between the structure of spacetime and quantum entanglement. In particular, the island formula \cite{Penington:2019npb,Almheiri:2019psf,Almheiri:2019hni,Penington:2019kki,Almheiri:2019qdq} for the entropy of Hawking radiation implies the island region in the interior of an old black hole is reconstructed from the information of Hawking radiation. 

However, it still remains to be understood the precise way to  recover black hole interior region from Hawking radiation. It has been realized that for this purpose it is convenient to regard the black hole interior as a code subspace embedded in the Hilbert space of Hawking radiation as a quantum error correcting code \cite{ Hayden:2007cs,Verlinde:2012cy,Papadodimas:2012aq}. For instance, the decoupling theorem by Hayden and Preskill \cite{Hayden:2007cs} implies that the black hole interior region is  protected against the erasure of black hole degrees of freedom, which assures the recovery. Once we regard an evaporating black hole as a quantum error correcting (QEC) code, then the general argument of QEC \cite{Barnum2000ReversingQD} tells us that the recovery is achieved by applying  the Petz recovery map 
\cite{Petz:1986tvy,ohya2004quantum}.

In this paper, we study properties of the Petz recovery map in chaotic  systems, such as the Hayden-Preskill (HP) setup for evaporating black holes  and the SYK model. Since these systems exhibit the phenomenon  called scrambling, we expect that the recovery channel $\mathcal{R}$ gets simplified,  given by just the adjoint $\mathcal{N}^{\dagger}$ of the original channel $\mathcal{N}$ which defines the embedding of the black hole interior into the Hawking radiation. Therefore schematically  we have 
\be
\mathcal{R} \sim a \; \mathcal{N}^{\dagger},
\ee
where $a$ is some numerical factor depending on the dimensions of the Hilbert spaces of black hole and Hawking radiation.

We will see this phenomenon in two examples. The first one is  the Hayden-Preskill setup where the dynamics of an evaporating black hole and Hawking radiation is described by  Haar random unitaries. We do this by computing the relative entropy $S(\mathcal{R}\left[\mathcal{N}[\rho]\right] ||\rho)$ and show that it is vanishing when the decoupling is archived. We further show that the simplified recovery map is equivalent to the Yoshida-Kitaev protocol\footnote{This equivalence has not been directly shown, but such a equivalence is suggested by Yoshida in  \cite{Yoshida:2021xyb,Yoshida:2021haf}.}. 
The second example is one of the SYK model versions of the Hayden-Preskill setup, discussed in \cite{Chandrasekaran:2022qmq}\footnote{In \cite{Nakata:2023hwg}, the authors discuss another Hayden-Preskill setup in the SYK model, and the setup is different from our setup.}. In this setup, code information is expressed as excitations, and a system is evolved by the SYK Hamiltonian. We check the recovery phenomenon by relating some elements of an output density matrix $\bra{T}\mathcal{R}[\mathcal{N}[\rho]]\ket{T'}$ to R\'enyi-two modular flowed correlators, and show that they gives an input density matrix $\bra{T}\rho\ket{T'}$ with small error after twice the scrambling time. However, there are still remaining matrix element, which we need to check, but it is difficult to evaluate them directly. In upcoming paper \cite{WIP}, we will give their direct evaluations. In this paper, we do not evaluate them directly, but indirectly guess their expectations for them based on our obtained results.

Our paper is organized as follows. 
 In section \ref{sec:HaarHP}, we start with introducing a quantum channel induced by the Hayden-Preskill setup, and explain how we write down the simplified recovery map in the original Hayden-Preskill setup, which is applicable to the SYK case. We also explain a convenient notation to treat quantum channels induced by the Hayden-Preskill setup, and in the notation, one can imagine gravitational interpretation simply.
 In section \ref{sec:sufficiency}, by using the convenient notation, we compute some relative entropies to check the sufficiency that we can use the simplified recovery map as a recovery map. Also, we show that the Yoshida-Kitaev protocol can be written as the recovery map.
  In section \ref{sec:SYKHP}, we explain one of the Hayden-Preskill setup using the SYK model, and introduce a corresponding quantum channel. After that, we give the simplified recovery map, and show that some matrix elements of output results can be written as ``R\'enyi-two modular flowed correlator". By evaluating the ``R\'enyi-two modular flowed correlators" analytically, we show some matrix elements of output results by the simplified recovery map give desired results.
  In section \ref{sec:SYKconjecture}, from the previous section result we have computed, we estimate the remaining matrix elements of output results, which we are evaluating. The detail of the remaining ones will be reported  in upcoming paper \cite{WIP}. 
  In section \ref{sec:discussion}, we conclude this paper by the discussion of our results and future directions.
  In appendix \ref{app:Kraus}, we give another derivation of the simplified recovery map using a Kraus representation.
  In appendix \ref{app:opetransEPR}, we show the relation that holds for an EPR state, which is used in section \ref{sec:sufficiency}.
    In appendix \ref{app:conventionSYKHP}, conventions used in section \ref{sec:SYKHP} are listed. 
 In appendix \ref{app:derivationChanneltoCorrelator}, we show that, in the SYK version of the Hayden-Preskill setup, some recovery results can be written as ``R\'enyi-two modular flowed correlators" .

\section{Recovery map for the Hayden-Preskill channel} \label{sec:HaarHP}

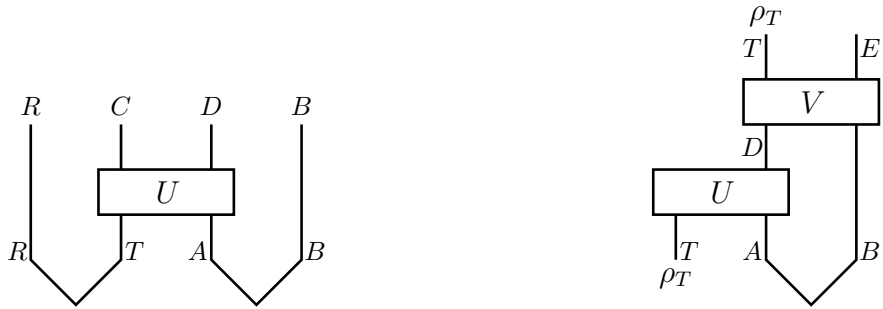
\begin{figure}
\vspace{-2cm}
    \begin{minipage}[b]{0.45\linewidth}
      \centering
        \begin{tikzpicture}[line width=1pt,x=0.6cm,y=0.6cm]
            \draw(0,4)--(0,1)--(1,0)--(2,1)--(2,2);
            \draw(2,3)--(2,4);
            \draw(6,4)--(6,1)--(5,0)--(4,1)--(4,2);
            \draw(4,3)--(4,4);
            \draw(1.5,2) rectangle (4.5,3);
            \draw(3.04,2.5) node {\large{$U$}};
            \draw(-0.3,1.2) node {\small{$R$}};
            \draw(0,4.4) node {\small{$R$}};
            \draw(2.3,1.2) node {\small{$T$}};
            \draw(2,4.4) node {\small{$C$}};
            \draw(3.7,1.2) node {\small{$A$}};
            \draw(4,4.4) node {\small{$D$}};
            \draw(6.3,1.2) node {\small{$B$}};
            \draw(6,4.4) node {\small{$B$}};
        \end{tikzpicture}
    \end{minipage}
    \hspace{2em}
    \begin{minipage}[b]{0.45\linewidth}
      \centering
        \begin{tikzpicture}[line width=1pt,x=0.6cm,y=0.6cm]
            \draw(2,1)--(2,2);
            \draw(6,4)--(6,1)--(5,0)--(4,1)--(4,2);
            \draw(4,3)--(4,4);
            \draw(4,5)--(4,6);
            \draw(6,5)--(6,6);
            \draw(1.5,2) rectangle (4.5,3);
            \draw(3.5,4) rectangle (6.5,5);
            \draw(3.04,2.5) node {\large{$U$}};
            \draw(5.04,4.5) node {\large{$V$}};
            \draw(2.3,1.2) node {\small{$T$}};
            \draw(2,0.6) node {\large{$\rho_{T}$}};
            \draw(3.7,1.2) node {\small{$A$}};
            \draw(3.7,3.5) node {\small{$D$}};
            \draw(3.7,5.7) node {\small{$T$}};
            \draw(4,6.4) node {\large{$\rho_{T}$}};
            \draw(6.3,1.2) node {\small{$B$}};
            \draw(6.3,5.7) node {\small{$E$}};
        \end{tikzpicture}
    \end{minipage}
\vspace{-3mm}
\caption{ {\bf Left}: Hayden-Preskill setup, corresponding to state (\ref{eq:totaState}).
 {\bf Right}: Its decoder}
\label{fig:diagram1}
\end{figure}

The Hayden-Preskill setup is a tractable toy model for studying information flow in evaporating black holes. The setup consists of a  black hole $A$ that has been emitting Hawking radiation $B$. We are particularly interested in the system after the Page time where the black hole has emitted more than half of its original entropy \footnote{We follow the notation of Yoshida-Kitaev \cite{Yoshida:2017non}.}, therefore approximately forming a maximally entangled state $|\epr \ra_{AB}$. Suppose Alice throws  a quantum state $\rho_{T}$ (often called  a diary) into this old black hole. Then as the black hole further evaporates $A \rightarrow C+ D$ by emitting late Hawking radiation $D$, 
information thrown into the black hole  will eventually appear in total Hawking radiation $DB$. Here we denoted by $C$ the remaining black hole after emitting the late radiation $D$, see the left panel of figure \ref{fig:diagram1}.  The analysis of Hayden and Preskill \cite{Hayden:2007cs} showed the diary appears in Hawking radiation almost immediately, namely after the scrambling time. 

To see  this, it is useful to introduce  an additional system called reference $R$ and form a maximally entangled state $|\epr \ra_{RT}$ with the diary $T$. Then in this setup the initial condition of the process is $|\epr \ra_{RT} \otimes |\epr \ra_{AB}$.

 Owing to its chaotic dynamics, information of  the diary thrown into the black hole gets scrambled and spread over the entire degrees of freedom. The resulting  state is given by 
\begin{equation}
	\ket{ \Psi_{HP} } = (I_{R}\otimes U_{T,A \to C,D}\otimes I_{B}) \ket{\epr }_{R,T}\otimes \ket{\epr}_{A,B}, \label{eq:totaState}
\end{equation}
where $U_{T,A \to C,D}$ is a random unitary matrix from $A,T$ to $C,D$, which models the chaotic dynamics of the black hole. By finding the Hilbert space with which $R$ is mostly entangled, one can find where is information of the original diary in the final time slice. See again the left panel of figure \ref{fig:diagram1}.

The surprising results of HP  is summarized into the following inequality.
\begin{equation}
	\overline{ \norm{ \rho_{RC} - \rho_{R}\otimes \rho_{C} }_{1}^{2} } \leq \left( \frac{d_{T}}{ d_{D} } \right)^{2}, \label{eq:decopHP}
\end{equation} 
where $\norm{A}_{1}= \tr \sqrt{A^{\dagger}A} $, $\rho_{RC}, \rho_{R},\rho_{C}$ are the reduced density matrices of \eqref{eq:totaState} on the indicated subsystems, and in the left hand side we take average over random unitaries. This inequality \eqref{eq:decopHP}implies that if one collects a sufficient number of late Hawking quanta so that $d_{D} \gg d_{T}$ the system of the remaining black hole and the reference becomes no longer correlated $\rho_{RC} =\rho_{R} \otimes \rho_{C}$, and there information of the diary has to be encoded in Hawking radiation $DB$.

This result is also natural from the view point  of the framework of quantum error correction\footnote{ We note that the possible maximum number of late Hawking radiation $d_{D}$ is given by the input for the Haar random unitary, implying $d_{D} \leq d_{T}d_{B}$. Due to this bound, the combination $d_{T}/d_{D}$ can not be $0$, but at most $1/d_{B}$. Thus, the exact equality does not hold $\rho_{RC} =\rho_{R} \otimes \rho_{C}$ , as long as $d_{B}$ is finite. This means that strictly speaking  the recovery of the diary from Hawking radiation is at best approximate. 
However, for a sufficiently large dimension of the early radiation, $d_{B} \gg 1$, we can almost ignore the deviation from the exact factorization of $\rho_{RC}$ for late times.}. A quantum error correcting code is a scheme to protect  quantum states (logical states) in code subspace $H_{\rm code}$ against various errors. Such an error is  mathematically modeled by a CPTP map called quantum channel $\mathcal{N}$. The basic idea of quantum error correction is protecting these quantum states in code subspace $H_{\rm code}$  by embedding it to the larger Hilbert space often called physical Hilbert space $H_{phys}$. In  the HP protocol, the Hilbert space of the diary $H_{T}$ corresponds to $H_{\rm code}$ in QEC, and $H_{{\rm phys}}$ is $H_{DB}$. The quantum channel $\mathcal{N}: T \rightarrow DB$, is obtained by tracing out the remaining black hole and the reference system degrees of freedom $C$ and $R$ from $\ket{ \Psi_{HP} }$ in \eqref{eq:totaState} with replacing the reference state $\ket{\epr }_{R,T}$ by $\sqrt{d_{T}\rho_{T}}\ket{\epr }_{R,T}$ ($\rho_{T}$ is an input state),
\begin{equation}
	\begin{aligned}
			\mathcal{N}_{T\to D,B}\left[ \rho_{T} \right] &= \tr_{C} \left[ (U_{T,A \to C,D}\otimes I_{B}) ( \rho_{T} \otimes \ktbra{\epr}{A,B}  )  (U_{T,A \to C,D}^{\dagger} \otimes I_{B}) \right] \\
		&= \frac{1}{d_{B}} \sum_{\tilde{D},\tilde{D}'=1}^{d_{D}}\, \sum_{\tilde{B},\tilde{B}'=1}^{d_{B}}  \ktbrad{ \tilde{D} }{ \tilde{D}' }{D}\otimes \ktbrad{ \tilde{B} }{ \tilde{B}'}{B}\,\sum_{C=1}^{d_{C}}\,\sum_{\tilde{T},\tilde{T}'=1}^{d_{T}} U_{ C,\tilde{D};\tilde{T},\tilde{B} } \, (\rho_{T})_{\tilde{T}\tilde{T}'} \, U_{ C,\tilde{D}';\tilde{T}',\tilde{B}'}^{\dagger}. \label{eq:hpChannel}
	\end{aligned}
\end{equation}
We call this quantum channel the HP channel.

Then a general theorem of QEC\footnote{See e.g., \cite{Schumacher:1996dy,Nielsen_2007} for the theorem.} tells us that the decoupling condition 
 is equivalent to the existence of a recovery map $\mathcal{R}: DB \rightarrow T$  which satisfies
\be 
\mathcal{R} \left[ \mathcal{N} [\rho_{T}] \right] =\rho_{T} \quad \forall \rho_{T} \in H_{T}.
\ee

This again implies the information of the diary is recoverable from Hawking radiation $DB$. See the right panel of figure \ref{fig:diagram1}. Moreover, the concrete expression of the recovery map is known \cite{Barnum2000ReversingQD}, and is called the Petz recovery map
\begin{equation}
\mathcal{R}_{\sigma,\mathcal{N}}^{\text{Petz}}[\tau] = \sigma^{\frac{1}{2}} \mathcal{N}^{\dagger}[ (\mathcal{N}[\sigma])^{-\frac{1}{2}} \tau (\mathcal{N}[\sigma])^{-\frac{1}{2}} ]\sigma^{\frac{1}{2}}. \label{eq:petzMap}
\end{equation}
where $\sigma$ is full rank arbitrary density matrix on the code subspace $H_{{\rm code}}$.  $\mathcal{N}^{-1/2}$ factor of the Petz recovery map is difficult to compute in general. One way for doing this is, as in \cite{Penington:2019kki} first making the replacement $\mathcal{N}^{-1/2} \rightarrow \mathcal{N}^{n}$, where $n$ is a  positive integer, computing it for all $n$, then taking analytic continuation $n \rightarrow - \f{1}{2}$.  Also $\mathcal{N}^{-1/2}$ part is preventing us from having an operational meaning of the map. 

However, in systems exhibiting quantum chaos, we expect that the recovery map gets simplified, because $\mathcal{N}[\sigma]$ has a flat spectrum, therefore the approximation $\mathcal{R} \sim \mathcal{N}^{\dagger}$ appears to be possible\footnote{In appendix \ref{app:Kraus}, we give another equivalent argument supporting our expectation of this simplification in terms of the Kraus representation of the HP channel.}. If this is the case, since $\rho \sim  \mathcal{N}^{\dagger}\left[ \mathcal{N} [\rho]\right]$  for arbitrary density matrix $\rho$ in the code subspace, therefore  the relative entropy between them $S(\rho ||\mathcal{N}^{\dagger} \left[ \mathcal{N} [\rho]\right])$ vanishes. 

For the HP channel, the adjoint HP channel $\mathcal{N}^{\dagger}$ is given by 
\begin{equation}
	 \begin{aligned}
	 	\mathcal{N}_{D,B \to T}^{\dagger} [\mathcal{O}_{DB}] &= \tr_{A,B} \left[ \ktbra{\epr}{A,B}    (U_{T,A \to C,D}^{\dagger} \, \mathcal{O}_{DB} \, U_{T,A \to C,D} ) \right]\\
	 	&= \braTFD{A,B} (U_{T,A \to C,D}^{\dagger} \otimes I_{B}) \, (\mathcal{O}_{DB}\otimes I_{C}) \,( U_{T,A \to C,D}\otimes I_{B} )  \ketTFD{A,B}
	 	. \label{eq:adjoHPChannel}
	 \end{aligned}
\end{equation}
Here, the adjoint channel is defined by the relation\footnote{More generally, for a quantum channel $\mathcal{N}$, its adjoint channel is defined by the similar relation
\begin{equation}
	\tr\left[ \mathcal{N}\left[ \rho \right] O \right] =\tr \left[ \rho \, \mathcal{N}^{\dagger} \left[ O \right] \right] \label{eq:defofAdjoint}
\end{equation}
 }
\begin{equation}
	 \tr_{D,B}\left[ \mathcal{N}_{T\to D,B}\left[ \rho_{T}\right]  \, O_{DB} \right]=\tr_{T}\left[ \rho_{T} \, \mathcal{N}_{D,B \to T}^{\dagger} [\mathcal{O}_{DB}] \right].
\end{equation}

For later convenience, we introduce a correctly normalized recovery map
\begin{equation}
	\mathcal{R}^{\text{Lite}}_{D,B\to T} [\mathcal{O}_{DB}] \coloneqq \frac{1}{N} \cdot \frac{d_{B} d_{D}}{d_{T}} \mathcal{N}_{D,B\to T}^{\dagger}\left[ \mathcal{O}_{DB}\right], \label{eq:petzLite}
\end{equation}
and define it as the \textit{Petz-lite}\footnote{The terminology ``Petz-lite" is introduced in \cite{Penington:2019kki}, and we also use this terminology in this paper.}. Here $N$ is the normalization constant
 \begin{equation}
	N= \left( \frac{d_{D}}{d_{T}} \right)^{2}+1,
\end{equation}
determined by the condition $\overline{\tr_{T}\left[ \mathcal{R}^{\text{Lite}}_{D,B\to T} \left[\, \mathcal{N}_{T\to D,B}[\sigma_{T}] \, \right] \right] } =1$, where $\sigma_{T}$ is some reference state in $T$. In the Haar random case, the choice of the reference state  $\sigma_{T}$ is not important as long as it is normalized.

With this $N$, the Petz-lite can be expressed as
\begin{equation}
	\begin{aligned}
		\mathcal{R}^{\text{Lite}}_{D,B\to T} [\mathcal{O}_{DB}] &=\frac{1}{\left( \dfrac{d_{D}}{d_{T}} \right)^{2}+1}\cdot \frac{d_{B} d_{D}}{d_{T}} \mathcal{N}_{D,B\to T}^{\dagger}\left[ \mathcal{O}_{DB}\right]\\
		&=\frac{1}{ 1+ \left( \dfrac{d_{T}}{d_{D}} \right)^{2}}\cdot  d_{C} \, \mathcal{N}_{D,B\to T}^{\dagger}\left[ \mathcal{O}_{DB}\right] \label{eq:petzLiteFullExpres}
		\end{aligned}
\end{equation}
where in the second line, we used the relation $d_{B} d_{T}=d_{C}d_{D}$ due to the unitarity of the Haar random unitary. For the parameter region $ d_{T}/ d_{D} \ll 1$, the normalization is just given by $d_{C}$, which coincides with an expression obtained from another discussion. In appendix \ref{app:Kraus}, we give the discussion.

\subsection{West-coast notation and replica-wormhole-like objects}
\label{sec:WCP}

In the following, we are interested in the typical properties of the recovery map $\mathcal{R}$ for the HP channel $\mathcal{N}$. To investigate these properties, we will consider replicated quantities, such as ${\rm tr} (\mathcal{N}[\rho_{T}])^{n}$ involving a product of Haar random unitaries and its average. Since such averaging involves Wick type contractions between various pairs of Haar random unitaries  in the product, it is convenient to introduce a graphical notation that manifests which pair of unitaries are contracted. Therefore, here we introduce a notation similar to the one employed in  \cite{Penington:2019kki} for modeling the black hole microstates and their statistical properties, and call this West-coast notation. 

To begin with, let us define  the following black hole microstate on $C$,  involving a Haar random unitary 
\begin{equation}
    \ket{ \psi_{i}^{T} }_{C} \coloneqq \sqrt{d_{C}\, d_{D}}  \sum_{C=1}^{d_{C}} \ket{C}U_{C,T;i} \quad.
\end{equation}
Here $\{ |C\ra\}$ is the set of basis states on the Hilbert space $H_{C}$ and the index $i$ collectively denote the indices for both late radiation $D$ and early radiation $B$, $i:(D,B)$ or more concretely $ \ket{i}=\ket{D}\otimes \ket{B}$, so the label $i$ rums from $1$ to $d_{D}d_{B}\equiv k$.

In the following, we use this type of states 
$ \ket{ \psi_{i}^{T} }_{C}$ to write quantities of our interest, instead of random unitary matrices $U_{C,D;T,B}$. Under this notation, we can write 
\begin{equation}
	 \ipGrapsi{i}{T}{j}{T'} = d_{C}\, d_{D} \sum_{C=1}^{d_{C}} U^{\dagger}_{i;C,T} \, U_{C,T';j} \label{eq:overlapByHaarRandom}
\end{equation}
and therefore the HP channel \eqref{eq:hpChannel} is given by 
\begin{equation}
	\begin{aligned}
		\mathcal{N}_{T\to D,B} [\rho_{T}] &=  \frac{1}{k d_{C}}  \sum_{i,j=1}^{k} \ktbrad{i}{j}{ } \, \cdot  \sum_{\tilde{T},\tilde{T}'=1}^{d_{T}}  \ipGrapsi{j}{\tilde{T}'}{i}{\tilde{T}}\, (\rho_{T})_{\tilde{T}\tilde{T}'} \quad.
	\end{aligned}\label{eq:hpChannelGra}
\end{equation}
 In this notation, we call the subscript index $i$ Hawking radiation index, and the superscript $T$ code index.

West-coast model treats each of these microstate   $| \psi_{i} \ra $ by a single-sided AdS black hole with insertion of “end of the world brane" (or EoW brane in short) labeled by the index  $i$ behind the horizon. This state has a Hartle Hawking type preparation, in terms of an Euclidean path integral 
with the EoW brane which starts from the Euclidean conformal boundary.
In this model the overlap between two such states   $ \la \psi_{i} | \psi_{j} \ra$ is computed by an Euclidean gravitational path integral  on a region of  Euclidean disc enclosed by the part of the asymptotic boundary (an interval) and the brane in the bulk.

With this gravitational path integral picture in mind, here we explain 
the fact that there is a simple diagrammatic prescription to compute a product of such overlaps   $\overline{ \prod_{m=1}^{n} \la \psi_{i_{m}}^{a_{m}}| \psi _{j_{m}}^{b_{m}} \ra}$. \footnote{ In the west-coast paper, this quantity is just called the product of overlap and denoted without the bar, i.e., $\overline{ \prod_{m=1}^{n} \la \psi_{i_{m}}^{a_{m}}| \psi _{j_{m}}^{b_{m}} \ra} \big|_{ours} =\prod_{m=1}^{n} \la \psi_{i_{m}}^{a_{m}}| \psi _{j_{m}}^{b_{m}} \ra_{WC}$. We will use the convention with the bar to keep in mind we do average over random unitaries in the computation. } without directly applying the formulae for the Haar random averages, which becomes quite involved when the number of unitary matrices appearing increases.

Then the prescription is the following:

\begin{enumerate}
\item For each  overlap in the product $\la \psi_{i_{m}}^{a_{m}}| \psi _{j_{m}}^{b_{m}} \ra$  draw an interval with two end points, and associate the labels $(i_{m},a_{m})$ to one end and $(j_{m},b_{m})$ to the other. (In the west-coast model this interval with indices at the end points provides the boundary condition to the gravitational path integral for the product of the overlaps.)

\item The $n$ intervals prepared in this way have 2$n$ endpoints in total. We pick up two of these end points and connect them  by a line, which we  call the end of the world brane. We repeat this until all the endpoints are connected to the other by a brane.  There are many different ways to do this. One possibility is that the end point of $m$-th interval is always connected to the other end point of the same interval. Or the other possibility is that the end point of $m$-th interval is always
connected to the point on the next $(m+1)$-th interval.  

\item Each diagram $D$ constructed in this way contains $n$ end of the world branes. We then associate each brane in the diagram  with a Kronecker delta factor. If the brane is connecting two endpoints with the labels$(i_{l},a_{l})$ and $(j_{m},b_{m})$, then this factor  is given by  $\delta_{i_{l}j_{m}}\delta_{a_{l}b_{m}}$.  We compute this for all branes in the diagram and then multiply these factors. Let us denote this factor for the diagram by $I_{D}$. 

\item  Since each diagram can be regarded as (disjoint union of ) two dimensional surfaces, we can associate a Euler number $\chi_{D}$ to the diagram. We then pick up the factor $(d_{C})^{\chi_{D}}$ which corresponds to the gravitational path integral part in the west-coast model. We then sum the total factor $ I_{D} (d_{C})^{\chi_{D}}$ for all possible diagram $D$.

\item  The average of the overlaps is equal to the sum of these factors over all possible diagrams;

\be
\overline{ \prod_{m=1}^{n} \la \psi_{i_{m}}^{a_{m}}| \psi _{j_{m}}^{b_{m}} \ra} = \sum_{D \in {\rm All \; diagrams}} I_{D} \;(d_{C})^{\chi_{D}}.
\ee
\end{enumerate}

Let us provide a few examples. First, for the  single overlap $\overline{ \braket{ \psi_{i}^{T} }{ \psi_{j}^{T'} } } $. We can easily evaluate it
\begin{equation}
   \begin{aligned}
        \overline{ \braket{ \psi_{i}^{T} }{ \psi_{j}^{T'} }  }  &= d_{C}\, d_{D} \sum_{C=1}^{d_{C}} \overline{ U^{\dagger}_{i;C,T} \, U_{C,T';j} }\\
        &=d_{C}\,  \underbrace{\delta_{D_{i}D_{j}} \,\delta_{B_{i}B_{j}}}_{{\delta_{ij}}} \, \delta_{TT'}\\
    &=d_{C}\, \delta_{ij} \, \delta_{TT'}, \label{eq:overlapNorma}
   \end{aligned}
\end{equation}
where in the second line we used the general result for two Haar random unitaries
\begin{equation}
    \overline{U_{a,b} U_{c,d}^{\dagger}} = \frac{1}{d}\, \delta_{ad} \delta_{bc} \qquad (a,b,c,d=1,\cdots,d).\label{eq:HaarFirstMoment}
\end{equation}
This result is easily reproduced from the west-coast prescription.

\begin{figure}
\vspace{-1cm}
    \begin{minipage}[b]{0.45\linewidth}
      \centering
        \begin{tikzpicture}[x=0.4cm,y=0.4cm]
            \draw[line width=0.8pt](1,0)..controls(0,1)and(0,3)..(1,4);
            \draw[line width=0.8pt](7,0)..controls(8,1)and(8,3)..(7,4);
            \draw[line width=1.2pt,blue](1,0)..controls(3,1)and(3,3)..(1,4);
            \draw[line width=1.2pt,blue](7,0)..controls(5,1)and(5,3)..(7,4);
            \fill(1,0) circle (1.2pt);
            \fill(1,4) circle (1.2pt);
            \fill(7,0) circle (1.2pt);
            \fill(7,4) circle (1.2pt);
            \draw(0.3,0) node {\footnotesize{$\tilde{T}_1$}};
            \draw(1.7,-0.2) node {\footnotesize{$j_1$}};
            \draw(0.3,4) node {\footnotesize{$T_1$}};
            \draw(1.7,4.2) node {\footnotesize{$i_1$}};
            \draw(7.8,0) node {\footnotesize{$\tilde{T}_2$}};
            \draw(6.3,-0.2) node {\footnotesize{$j_2$}};
            \draw(7.8,4) node {\footnotesize{$T_2$}};
            \draw(6.3,4.2) node {\footnotesize{$i_2$}};
        \end{tikzpicture}
    \end{minipage}
    \hspace{2em}
    \begin{minipage}[b]{0.45\linewidth}
      \centering
        \begin{tikzpicture}[x=0.4cm,y=0.4cm]
            \draw[line width=0.8pt](1,0)..controls(0,1)and(0,3)..(1,4);
            \draw[line width=0.8pt](7,0)..controls(8,1)and(8,3)..(7,4);
            \draw[line width=1.2pt,blue](1,4)..controls(3,3)and(5,3)..(7,4);
            \draw[line width=1.2pt,blue](1,0)..controls(3,1)and(5,1)..(7,0);
            \fill(1,0) circle (1.2pt);
            \fill(1,4) circle (1.2pt);
            \fill(7,0) circle (1.2pt);
            \fill(7,4) circle (1.2pt);
            \draw(0.3,0) node {\footnotesize{$\tilde{T}_1$}};
            \draw(1.7,-0.2) node {\footnotesize{$j_1$}};
            \draw(0.3,4) node {\footnotesize{$T_1$}};
            \draw(1.7,4.2) node {\footnotesize{$i_1$}};
            \draw(7.8,0) node {\footnotesize{$\tilde{T}_2$}};
            \draw(6.3,-0.2) node {\footnotesize{$j_2$}};
            \draw(7.8,4) node {\footnotesize{$T_2$}};
            \draw(6.3,4.2) node {\footnotesize{$i_2$}};
        \end{tikzpicture}
    \end{minipage}
\vspace{-3mm}
\caption{ Diagrams for computing the average of overlaps (\ref{eq:OverlapVariance}). A black line connects two points that appear in the same overlap, and the blue lines correspond to the end of the world branes in Haar random averaging. {\bf Left}: The disconnected diagram 
 {\bf Right}: The connected diagram}
\label{fig:Wormholestress}   
\end{figure}
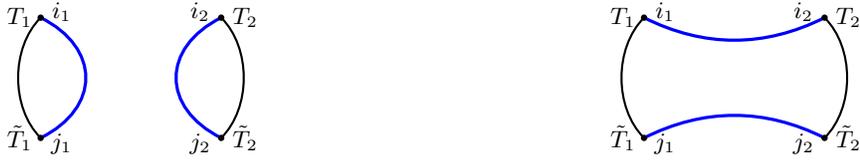

Next, let us evaluate the Haar average of the combination of the overlaps for later convenience,
\begin{equation}
    \braket{ \psi_{i}^{T_{1}} }{ \psi_{j}^{T'_{1}} } \cdot  \braket{ \psi_{j}^{T'_{2}} }{ \psi_{i}^{T_{2}} }.
\end{equation}
Clearly, by setting $T_{1}=T_{2}=T$ and $T_{1}'=T_{2}'=T'$, the above combination reduces to the variance of the overlap $\overline{ \left| \braket{ \psi_{i}^{T} }{ \psi_{j}^{T'} } \right|^{2} } $. We can evaluate the above quantity by the diagrammatic prescription mentioned above (See figure \ref{fig:Wormholestress}),
\begin{equation}
    \begin{aligned}
    	& \overline{ \braket{ \psi_{i}^{T_{1}} }{ \psi_{j}^{T'_{1}} } \cdot  \braket{ \psi_{j}^{T'_{2}} }{ \psi_{i}^{T_{2}} } }\approx  \left(d_{C}\right)^{2}\delta_{ij} \delta_{T_{1}T_{1}'} \cdot \delta_{ji} \delta_{T_{2}'T_{2}} + d_{C}\, \delta_{ii} \delta_{T_{1}T_{2}}\cdot \delta_{jj} \delta_{T_{2}'T_{1}'}. \label{eq:OverlapVariance}
    \end{aligned}
\end{equation}

This coincides with the result obtained by using the Weingarten formula,
\begin{equation}
   \begin{aligned}
   	\overline{U_{a_{1},b_{1}} U_{c_{1},d_{1}}^{\dagger} \cdot U_{a_{2},b_{2}} U_{c_{2},d_{2}}^{\dagger} } &= \frac{1}{d^{2}-1}\left( \delta_{a_{1}d_{1}} \delta_{b_{1}c_{1}} \cdot \delta_{a_{2}d_{2}} \delta_{b_{2}c_{2}} + \delta_{a_{1}d_{2}} \delta_{b_{1}c_{2}} \cdot \delta_{a_{2}d_{1}} \delta_{b_{2}c_{1}} \right)\\
    & \qquad + \frac{1}{ d\left(d^{2}-1 \right)} \left( \delta_{a_{1}d_{1}}  \delta_{a_{2}d_{2}}  \delta_{b_{1}c_{2}} \delta_{b_{2}c_{1}} + \delta_{a_{1}d_{2}} \delta_{a_{2}d_{1}}   \delta_{b_{1}c_{1}}  \delta_{b_{2}c_{2}} \right)\\
   &  \hspace{8cm} (a,b,c,d=1,\cdots,d).
    \end{aligned}\label{eq:HaarSecondMoment}
\end{equation}

 In general, the prescription introduced here correctly computing the average over Haar random unitaries in the product of overlaps, as long as the rank of the  random unitaries $d= d_{C}\;d_{D} =d_{T}\; d_{A}$ is large.

Furthermore the adjoint channel \eqref{eq:adjoHPChannel}, in terms of the west-coast notation is given by 
\begin{equation}
	\begin{aligned}
		\mathcal{N}_{D,B \to T}^{\dagger} [\mathcal{O}_{DB}] &= \frac{1}{k d_{C}} \sum_{T,T'=1}^{d_{T}}  \ktbrad{T'}{T}{ } \, \cdot \sum_{i,j=1}^{k}  \braket{\psi_{j}^{T'}}{\psi_{i}^{T}} \, \bra{j} \mathcal{O}_{DB} \ket{i} \quad . \label{eq:hpAdjChannelGra}
	\end{aligned}
\end{equation}

Below, using this graphical expression, we evaluate several relative entropies to check the validity of the approximation $\mathcal{R}\sim \mathcal{N}^{\dagger}$.

\section{Relative entropy: Sufficiency}\label{sec:sufficiency}

As we have mentioned, the decoupling condition \eqref{eq:decopHP} implies there is a recovery map for the Hayden-Preskill channel \eqref{eq:hpChannel}. Another characterization of the existence of the recovery map $\mathcal{R}$ for given $\mathcal{N}$ is the notion of sufficiency \cite{Petz:1986tvy,Petz:2002eql,ohya2004quantum}.  To state this, let us first recall the fact that relative entropy satisfies
the monotonicity property
\be
 S(\rho ||\sigma) \geq  S(\mn[\rho] || \mn[\sigma])
 \ee
for any CTPT map $\mn$. By repeating this we have
\be
 S(\rho ||\sigma) \geq  S(\mn[\rho] || \mn[\sigma]) \geq S( \mr \left[ \mn[\rho] \right] ||\mr \left[  \mn[\sigma] \right]),
 \ee
therefore if the recovery map exists $\mr \circ \mn  =1_{{\rm code}}$, then  $ S(\rho ||\sigma)=  S(\mn[\rho] || \mn[\sigma])$, for any density matrices on the code subspace. This condition is known as sufficiency, and it was shown that if $\mn$ satisfies this condition, the recovery map is given by \eqref{eq:petzMap}.
Here we would like to check the HP channel \eqref{eq:hpChannel} does satisfy sufficiency, by directory computing the relative entropy $S(\mn[\rho] || \mn[\sigma])$ in the presence of the quantum channel $\mn$ \footnote{See \cite{Vardhan:2021mdy,Kudler-Flam:2022jwd} for related discussions on original Petz map cases.}.

Since our interest is a typical result under the Haar random average, we consider the Haar averaged relative entropy, $\overline{S(\mn[\rho] || \mn[\sigma])}$. To evaluate the relative entropy, we use the replica trick 
\begin{equation}
     \begin{aligned}
         \overline{S( \mathcal{N}[\rho] || \mathcal{N}[\sigma]) } = \lim_{n\to 1} \frac{1}{n-1} \left( \overline{ \log \tr\left[ \mathcal{N}[\rho]^{n} \right] } - \overline{ \log \tr\left[ \mathcal{N}[\rho] \mathcal{N}[\sigma]^{n-1} \right] }  \right).
         \label{eq;replica}
     \end{aligned}
\end{equation}
Generally, since it is difficult to evaluate the Haar average of logarithmic functional, instead of the expression, we consider 
\begin{equation}
   \overline{S( \mathcal{N}[\rho] || \mathcal{N}[\sigma]) } \approx  \lim_{n\to 1} \frac{1}{n-1} \left(  \log \overline{ \tr\left[ \mathcal{N}[\rho]^{n} \right] } -\log  \overline{  \tr\left[ \mathcal{N}[\rho] \mathcal{N}[\sigma]^{n-1} \right] }  \right).
   \label{eq:relaN}
\end{equation}
It is known that in the large Hilbert dimension limit, this quantity is almost equal to the original one. For a moment let us focus on the first term of \eqref{eq:relaN}. Using the west-coast notation \eqref{eq:hpChannelGra}, the trace $\tr\left[ \mathcal{N}[\rho]^{n} \right]$ can be written in terms of overlaps,
\begin{equation}
    \begin{aligned}
        \tr\left[ \mathcal{N}[\rho]^{n} \right]&= 
\frac{1}{\left(k\, d_{C}\right)^{n}} \sum_{\bm{i}=1}^{k}\sum_{\bm{T},\bm{\tilde{T}}=1}^{d_{T}} \,\prod_{m=0}^{n-1}  \left(\la \psi_{i_{m}}^{\tilde{T}_{m}}|\psi_{i_{m+1}}^{T_{m}}\ra \; \rho_{T_{m} \, \tilde{T}_{m}}\right)\\
    \end{aligned}
    \label{eq:renyich}
\end{equation}
where the bold fonts $\bm{i}, \bm{T}$ in the summation symbol means the sum with respect to the set of indices; $ \sum_{\bm{i}=1}^{k}=\sum_{i_{1}=1}^{k}\cdots \sum_{i_{n}=1}^{k}$.

\begin{figure}
\vspace{-1cm}
    \begin{minipage}[b]{0.45\linewidth}
      \centering
        \begin{tikzpicture}[x=0.4cm,y=0.4cm]
            \draw[line width=0.8pt](0,0)..controls(0,1)and(0.5,2.5)..(1,3);
            \draw[line width=0.8pt](2,5)..controls(2.5,6)and(3.5,6.6)..(4.5,7);
            \draw[line width=1.2pt,blue](0,0)..controls(2,0.5)and(2,2.2)..(1,3);
            \draw[line width=1.2pt,blue](2,5)..controls(2.5,3.5)and(4.1,4)..(4.5,7);
            \fill(0,0) circle (1.2pt);
            \fill(1,3) circle (1.2pt);
            \fill(2,5) circle (1.2pt);
            \fill(4.5,7) circle (1.2pt);
            \fill(-0.5,-1.5) circle (0.7pt);
            \fill(-0.68,-2.04) circle (0.7pt);
            \fill(-0.86,-2.58) circle (0.7pt);
            \fill(6,8) circle (0.7pt);
            \fill(6.47,8.32) circle (0.7pt);
            \fill(6.94,8.64) circle (0.7pt);
            \draw(-0.4,-0.1) node {\footnotesize{$\tilde{T}_{m}$}};
            \draw(1.2,-0.35) node {\footnotesize{$i_{m+1}$}};
            \draw(0.4,3.3) node {\footnotesize{$T_{m}$}};
            \draw(1.4,3.5) node {\footnotesize{$i_{m}$}};
            \draw(1,5.5) node {\footnotesize{$\tilde{T}_{m-1}$}};
            \draw(2.6,5) node {\footnotesize{$i_{m}$}};
            \draw(4.5,7.5) node {\footnotesize{$T_{m-1}$}};
            \draw(5.5,6.7) node {\footnotesize{$i_{m-1}$}};
        \end{tikzpicture}
    \end{minipage}
    \hspace{2em}
    \begin{minipage}[b]{0.45\linewidth}
      \centering
        \begin{tikzpicture}[x=0.4cm,y=0.4cm]
            \draw[line width=0.8pt](0,0)..controls(0,1)and(0.5,2.5)..(1,3);
            \draw[line width=0.8pt](2,5)..controls(2.5,6)and(3.5,6.6)..(4.5,7);
            \draw[line width=1.2pt,blue](0,0)..controls(0.2,-0.2)and(0.5,-1)..(0.5,-1.5);
            \draw[line width=1.2pt,blue](1,3)..controls(1.5,2.5)and(2.5,2.5)..(2,5);
            \draw[line width=1.2pt,blue](4.5,7)..controls(4.7,6.2)and(5.2,5.8)..(6,6);
            \fill(0,0) circle (1.2pt);
            \fill(1,3) circle (1.2pt);
            \fill(2,5) circle (1.2pt);
            \fill(4.5,7) circle (1.2pt);
            \fill(-0.5,-1.5) circle (0.7pt);
            \fill(-0.68,-2.04) circle (0.7pt);
            \fill(-0.86,-2.58) circle (0.7pt);
            \fill(6,8) circle (0.7pt);
            \fill(6.47,8.32) circle (0.7pt);
            \fill(6.94,8.64) circle (0.7pt);
            \draw(-0.4,-0.1) node {\footnotesize{$\tilde{T}_{m}$}};
            \draw(1.2,-0.25) node {\footnotesize{$i_{m+1}$}};
            \draw(0.4,3.3) node {\footnotesize{$T_{m}$}};
            \draw(1.4,3.5) node {\footnotesize{$i_{m}$}};
            \draw(1,5.5) node {\footnotesize{$\tilde{T}_{m-1}$}};
            \draw(2.6,5) node {\footnotesize{$i_{m}$}};
            \draw(4.5,7.5) node {\footnotesize{$T_{m-1}$}};
            \draw(5.6,6.7) node {\footnotesize{$i_{m-1}$}};
        \end{tikzpicture}
    \end{minipage}
\vspace{-3mm}
\caption{ {\bf Left}: The dominant diagram for (\ref{eq:renyich}) when  $d_{D} \ll d_{T}$ (disconnected diagram)
 {\bf Right}: The connected  diagram dominating the sum at  $d_{T} \ll d_{D}$. }
\label{fig:diagram2}   
\end{figure}
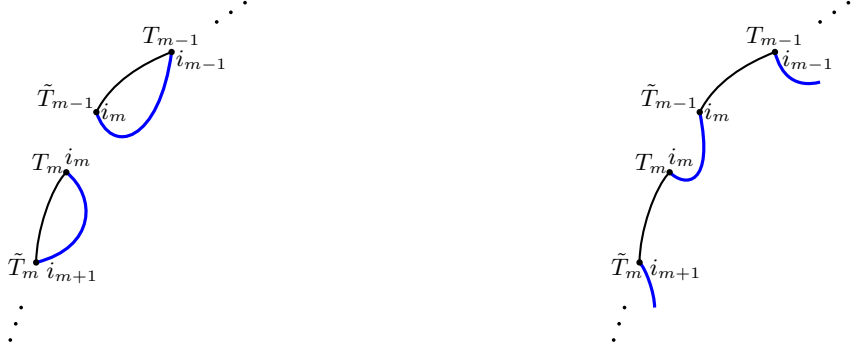

In computing the R\'enyi entropy   \eqref{eq:renyich} we need to evaluate the product of overlaps $\prod_{m=0}^{n-1} \la \psi_{i_{m}}^{\tilde{T}_{m}}|\psi_{i_{m+1}}^{T_{m}}\ra $ with $|\psi_{i_{n}}^{T}\ra \equiv |\psi_{i_{0}}^{T}\ra $ and its Haar random average. We do this using the diagrammatic technique introduced in the previous section. 

Among all possible diagrams, we are particularly interested in the ones dominating the sum, both in 
for early time regime ($d_{D}\ll d_{T}$) and late times ($d_{D}\gg d_{T}$). We now argue that 
fully-disconnected  diagram (the left panel of Figure \ref{fig:diagram2}) where for  all EoW branes starting point and end point are on the same interval  dominates in the early time regime, and the fully connected diagram (the right panel of Figure \ref{fig:diagram2})where the indices form a single loop dominates the late time regime by explicit calculations. The calculation here is very similar to the ones in \cite{Penington:2019kki,Kudler-Flam:2021alo}.

First, let us evaluate the contribution of the fully disconnected diagram.   Since the contribution of this diagram is evaluated as 
\be
\left(\overline{\prod_{m=0}^{n-1} \la \psi_{i_{m}}^{\tilde{T}_{m}}|\psi_{i_{m+1}}^{T_{m}}\ra} \right)_{\text{discon}} = d_{C}^{n}\;  \prod_{m=0}^{n-1} \left(\delta_{i_{m} i_{m+1}} \delta_{\tilde{T}_{m} T_{m}} \right).
\label{eq:disconv}
\ee

The contribution of this diagram to the R\'enyi entropy is 
\begin{equation}
	\begin{aligned}
		\left.\overline{\tr\left[ \mathcal{N}[\rho]^{n} \right]} \right|_{\text{fully discon}} = \frac{1}{\left(k\, d_{C}\right)^{n}}\cdot k \, \left(d_{C}\right)^{n}\, \sum_{\bm{T}=1}^{d_{T}} \rho_{T_{1}T_{1}}\,\rho_{T_{2}T_{2}}\, \cdots \rho_{T_{n} T_{n}} 
	  =\frac{1}{\left( k \right)^{n-1}} \left( \tr\left[ \rho \right] \right)^{n}.
	\end{aligned}
\end{equation}

Similarly, the value of the fully connected diagram is given by 
\be
\left(\overline{\prod_{m=0}^{n-1} \la \psi_{i_{m}}^{\tilde{T}_{m}}|\psi_{i_{m+1}}^{T_{m}}\ra} \right)_{\text{fully conn}}=d_{C}\prod_{m=0}^{n-1} \left(\delta_{\tilde{T}_{m+1}T_{m}} \right) \;\Rightarrow \left.\overline{\tr\left[ \mathcal{N}[\rho]^{n} \right]} \right|_{\text{fully conn}}
 =\frac{1}{\left( d_{C} \right)^{n-1}}  \tr\left[ \rho^{n} \right].
 \label{eq:connv}
   \ee

Combining these two results, $\overline{ \tr\left[ \mathcal{N}[\rho]^{n} \right] }$ is given by
\begin{equation}
    \begin{aligned}
        \overline{\tr\left[ \mathcal{N}[\rho]^{n} \right] } &= \frac{1}{\left( k \right)^{n-1}} \left( \tr\left[ \rho \right] \right)^{n}  + \frac{1}{\left(d_{C} \right)^{n-1}} \tr\left[ \rho^{n} \right] + \cdots,
    \end{aligned}
\end{equation}
where $\cdots$ means contributions coming from partially connected saddles.

Since there are upper and lower bounds on $\tr\left[ \rho^{n} \right]$, that is,   $1/(d_{T})^{n-1} \leq  \tr\left[ \rho^{n} \right] \leq 1$, we can see that 
\begin{equation}
    \begin{aligned}
        \overline{\tr\left[ \mathcal{N}[\rho]^{n} \right] } &= \frac{1}{\left( k \right)^{n-1}} \left( \tr\left[ \rho \right] \right)^{n}  + \frac{1}{\left(d_{C} \right)^{n-1}} \tr\left[ \rho^{n} \right] + \cdots,\\
        &\approx \begin{dcases}
            \frac{1}{\left( k \right)^{n-1}} & k \ll d_{C} \Leftrightarrow d_{T} \ll \left(\frac{d_{T}}{d_{D}}\right)^{2}  \\
            \frac{1}{\left(d_{C} \right)^{n-1}} \tr\left[ \rho^{n} \right] &  d_{C}\, d_{T} \ll k \Leftrightarrow \left(\frac{d_{T}}{d_{D}}\right)^{2} \ll 1
        \end{dcases}.
    \end{aligned}
\end{equation}
Thus, when the necessary condition for the decoupling condition, $d_{T}/ d_{D} \ll 1$, holds, the dominant contribution is given by the fully connected saddle.

We have to carefully evaluate the precise range of $m$ where the value of the connected saddle gets larger than that of the disconnected saddle.
This value of $m$ depends on the density matrix $\rho$ on the code subspace, and gets maximized when it is  the maximally mixed state $\rho =I_{T}/d_{T}$. Therefore after $k> d_{C}d_{T}$ the connected saddle become the dominant one for all density matrices in $H_{{\rm code}}$.

Next, let us evaluate the second term of \eqref{eq:relaN}. This computation is completely parallel to the above computation. In terms of the overlaps, it is given by 
\begin{equation}
    \begin{aligned}
\tr\left[ \mathcal{N}[\rho] \mathcal{N}[\sigma]^{n-1} \right]&= 
\frac{1}{\left(k\, d_{C}\right)^{n}} \sum_{\bm{i}=1}^{k}\sum_{\bm{T},\bm{\tilde{T}}=1}^{d_{T}} \,  \left(\prod_{m=0}^{n-1} \la \psi_{i_{m}}^{\tilde{T}_{m}}|\psi_{i_{m+1}}^{T_{m}}\ra \right) \;\rho_{T_{0} \, \tilde{T}_{0}}\left(\prod_{m=1}^{n-1} \sigma_{T_{m} \, \tilde{T}_{m}}  \right).
    \end{aligned}
\end{equation}

 The contribution of the fully disconnected diagram and the connected diagram to the second term of \eqref{eq:relaN} can be evaluated, again by substituting the result \eqref{eq:disconv} and \eqref{eq:connv}
 \begin{equation}
 	\begin{aligned}
 	\left. \tr\left[ \mathcal{N}[\rho] \mathcal{N}[\sigma]^{n-1} \right]\right|_{\text{discon}} =\frac{1}{\left( k \right)^{n-1}}\, \tr\left[ \rho \right]  \, \left( \tr\left[ \sigma \right] \right)^{n-1}, 
	\end{aligned}
 \quad \begin{aligned}
 	 \overline{\left. \tr\left[ \mathcal{N}[\rho] \mathcal{N}[\sigma]^{n-1} \right]\right|}_{\text{conn}}=\frac{1}{\left( d_{C} \right)^{n-1}}  \tr\left[ \rho \, \sigma^{n -1} \right].
	 \end{aligned}
  \end{equation}

Thus using these results, we obtain 
\begin{equation}
    \begin{aligned}
        \overline{ \tr\left[ \mathcal{N}[\rho] \mathcal{N}[\sigma]^{n-1} \right] } & = \frac{1}{\left( k \right)^{n-1}}  \tr\left[ \rho \right] \left( \tr\left[ \sigma \right] \right)^{n-1}  + \frac{1}{\left(d_{C} \right)^{n-1}} \tr\left[ \rho \sigma^{n-1} \right] + \cdots,\\
         &\approx \begin{dcases}
            \frac{1}{\left( k \right)^{n-1}} & k \ll d_{C}\Leftrightarrow  d_{T} \ll\left(\frac{d_{T}}{d_{D}}\right)^{2}  \\
            \frac{1}{\left(d_{C} \right)^{n-1}} \tr\left[ \rho \sigma^{n-1} \right] & k \gg d_{C} d_{T} \Leftrightarrow \left(\frac{d_{T}}{d_{D}}\right)^{2} \ll 1
        \end{dcases},
    \end{aligned}
\end{equation}
where $\cdots$ again means contributions coming from partially connected saddles, and also, in the second approximate equality we assumed that $1/(d_{T})^{n-1} \lesssim  \tr \left[ \rho \sigma^{n-1} \right] \leq 1$  in order to obtain the conditions\footnote{If the support of the density matrix $\rho$ is not contained in that of $\sigma$, then $ \tr \left[ \rho \sigma^{n-1} \right]=0$, implying the divergent relative entropy $S(\rho||\sigma)=\infty$. In that case, we would need another treatment, so we do not consider such a case in this paper.}.

Now that we have evaluated the two terms appeared in the relative entropy, we can obtain the resulting relative entropy
\begin{equation}
     \begin{aligned}
         \overline{S( \mathcal{N}[\rho] || \mathcal{N}[\sigma]) }&\approx \lim_{n\to 1}  \frac{1}{n-1} \left(  \log \overline{\tr\left[ \mathcal{N}[\rho]^{n} \right] } -  \log \overline{ \tr\left[ \mathcal{N}[\rho] \mathcal{N}[\sigma]^{n-1} \right] }  \right),\\
         & \approx \begin{dcases}
            0 & k \ll d_{C}\Leftrightarrow  d_{T} \ll\left(\frac{d_{T}}{d_{D}}\right)^{2}  \\
            \lim_{n\to 1}  \frac{1}{n-1} \left( \log \tr\left[ \rho^{n} \right] - \log \tr\left[ \rho \sigma^{n-1} \right]    \right) & k \gg d_{C} d_{T} \Leftrightarrow \left(\frac{d_{T}}{d_{D}}\right)^{2} \ll 1
        \end{dcases}\\
        & =\begin{dcases}
            0 & k \ll d_{C}\Leftrightarrow  d_{T} \ll\left(\frac{d_{T}}{d_{D}}\right)^{2}  \\
            S(\rho||\sigma) & k \gg d_{C} d_{T} \Leftrightarrow \left(\frac{d_{T}}{d_{D}}\right)^{2} \ll 1.
        \end{dcases}
     \end{aligned}\label{eq:relativeNrhoNsigma}
\end{equation}
Thus we can conclude that, when the condition $d_{T}/d_{D} \ll 1$ is satisfied, the relative entropies obeys the relation
\begin{equation}
	\overline{S( \mathcal{N}[\rho] || \mathcal{N}[\sigma]) }  \approx S(\rho || \sigma).
\end{equation}

This result implies that the condition of sufficiency holds for the Hayden-Preskill channel when $\left(\frac{d_{T}}{d_{D}}\right)^{2} \ll 1$.

\subsection{Check the recovery map}

We argued that in chaotic systems \eqref{eq:petzMap} the Petz recovery map gets simplified and is reduced to so called Petz-lite map $\mathcal{R}^\text{Lite} $defined in  \eqref{eq:petzLiteFullExpres}.  In this section, we  show this by checking 
\be
\overline{S( \mathcal{R}^\text{Lite} \left[ \mathcal{N}[\rho_{T}]\right]|| \rho_{T})} =0,\quad {\rm when} \; \left(\f{d_{T}}{d_{D}}\right)^{2} \ll 1.
\ee
for any density matrix $\rho_{T}$ on the code subspace.
This means that at  sufficiently late time one can recover $\rho_{T}$ from the state of the Hawking radiation  $\mathcal{N}[\rho_{T}]$ by applying the recovery map  $\mathcal{R}^\text{Lite}$.  

One can show this by computing the relative entropy by the replica trick similar to \eqref{eq;replica}, 
\be
\overline{S( \mathcal{R}^\text{Lite} \left[ \mathcal{N}[\rho_{T}] \right]|| \rho_{T})} =\lim_{n \rightarrow 1} \f{1}{n-1}\left(\log \overline{{\rm tr} (\mathcal{R}^\text{Lite}\left[\mathcal{N}[\rho]\right])^{n}} -\log \overline{ {\rm tr} (\mathcal{R}^\text{Lite}\left[ \mathcal{N}[\rho]\right]\rho^{n-1})} \right).
\label{eq;rep1}
 \ee

In terms of Haar random unitaries,  $\mathcal{R}^\text{Lite}\left[\mathcal{N}[\rho_{T}]\right]$ is given by 
\begin{equation}
	\begin{aligned}
	\mathcal{R}^\text{Lite}\left[\mathcal{N}[\rho_{T}]\right]&= \frac{1}{N} \cdot \frac{d_{B} d_{D}}{d_{T}}\cdot \mathcal{N}^{\dagger}_{D,B\to T} \left[ \mathcal{N}_{T\to D,B}\left[ \rho \right] \right]\\
		&= \frac{1}{N} \sum_{\tilde{T},\tilde{T'}=1 }^{d_{T}} \ktbrad{\tilde{T}}{\tilde{T}'}{T}  \cdot \frac{1}{ k (d_{C})^{2} d_{T} } \sum_{T,T'=1}^{d_{T}} \sum_{i,j=1}^{k} \ipGrapsi{i}{\tilde{T}}{j}{\tilde{T}'} \ipGrapsi{j}{T'}{i}{T} (\rho)_{TT'}.
	\end{aligned}
\end{equation}
Therefore the first term in \eqref{eq;rep1} is given by 
\be
{\rm tr} (\mathcal{R}^\text{Lite}\left[\mathcal{N}[\rho]\right])^{n}=\f{1}{(Nkd_{C}^2 d_{T})^{n}} \sum_{\bm{T},\bm{T}'=1}^{d_{T}} \sum_{\bm{\tilde{T}},\bm{\tilde{T}}'=1}^{d_{T}} \sum_{\bm{i},\bm{j}=1}^{k} \;\prod_{m=1}^{n} \; \left(  \ipGrapsi{i_{m}}{T_{m}}{j_{m}}{T_{m+1}}   \ipGrapsi{j_{m}}{\tilde{T}_{m}}{i_{m}}{\tilde{T}'_{m}}\rho_{\tilde{T}_{m}\tilde{T}'_{m}} \right).
\label{eq:RNn}
\ee

\begin{figure}
\vspace{-1cm}
    \begin{minipage}[b]{0.45\linewidth}
      \centering
        \begin{tikzpicture}[x=0.33cm,y=0.33cm]
            \draw[line width=0.7pt](0,3)..controls(0.2,4)and(1,5.2)..(2,6);
            \draw[line width=0.7pt](3,7)..controls(3.7,8)and(5,8.7)..(6,9);
            \draw[line width=1.05pt,blue](0,3)..controls(2.5,4.5)and(2.5,-1)..(4,0);
            \draw[line width=1.05pt,blue](2,6)..controls(1,3.5)and(6,1.8)..(4,0);
            \draw[line width=1.05pt,blue](3,7)..controls(5,8)and(5,4)..(6,4);
            \draw[line width=1.05pt,blue](6,9)..controls(4.5,7.5)and(7.5,4.5)..(6,4);
            \fill[blue](4,0) circle (0.525pt);
            \fill[blue](6,4) circle (0.525pt);
            \fill(0,3) circle (1.05pt);
            \fill(2,6) circle (1.05pt);
            \fill(3,7) circle (1.05pt);
            \fill(6,9) circle (1.05pt);
            \fill(-0.5,1.5) circle (0.7pt);
            \fill(-0.825,0.85) circle (0.7pt);
            \fill(-1.15,0.2) circle (0.7pt);
            \fill(7,10) circle (0.7pt);
            \fill(7.6,10.4) circle (0.7pt);
            \fill(8.2,10.8) circle (0.7pt);
            \draw(-0.7,3.3) node {\scriptsize{$T_{m}$}};
            \draw(0.6,2.5) node {\scriptsize{$i_{m}$}};
            \draw(1.5,6.6) node {\scriptsize{$T_{m+1}$}};
            \draw(2.6,5.3) node {\scriptsize{$j_{m}$}};
            \draw(2.8,7.9) node {\scriptsize{$\tilde{T}_{m}$}};
            \draw(3.5,6.5) node {\scriptsize{$j_{m}$}};
            \draw(5.3,9.6) node {\scriptsize{$\tilde{T}_{m}^\prime$}};
            \draw(6.3,8.3) node {\scriptsize{$i_{m}$}};
        \end{tikzpicture}
    \end{minipage}
    \hspace{2em}
    \begin{minipage}[b]{0.45\linewidth}
      \centering
        \begin{tikzpicture}[x=0.33cm,y=0.33cm]
            \draw[line width=0.7pt](0,3)..controls(0.2,4)and(1,5.2)..(2,6);
            \draw[line width=0.7pt](3,7)..controls(3.7,8)and(5,8.7)..(6,9);
            \draw[line width=1.05pt,blue](2,6)..controls(1,3.5)and(3,3)..(4.5,4.5);
            \draw[line width=1.05pt,blue](3,7)..controls(5,8)and(5.5,5.5)..(4.5,4.5);
            \draw[line width=1.05pt,blue](0,3)..controls(1,-1.3)and(5,-1.8)..(8,1);
            \draw[line width=1.05pt,blue](6,9)..controls(10,9)and(11.2,4.3)..(8,1);
            \fill[blue](4.5,4.5) circle (0.525pt);
            \fill[blue](8,1) circle (0.525pt);
            \fill(0,3) circle (1.05pt);
            \fill(2,6) circle (1.05pt);
            \fill(3,7) circle (1.05pt);
            \fill(6,9) circle (1.05pt);
            \fill(-0.5,1.5) circle (0.7pt);
            \fill(-0.825,0.85) circle (0.7pt);
            \fill(-1.15,0.2) circle (0.7pt);
            \fill(7,10) circle (0.7pt);
            \fill(7.6,10.4) circle (0.7pt);
            \fill(8.2,10.8) circle (0.7pt);
            \draw(-0.7,3.3) node {\scriptsize{$T_{m}$}};
            \draw(0.7,2.8) node {\scriptsize{$i_{m}$}};
            \draw(1.5,6.6) node {\scriptsize{$T_{m+1}$}};
            \draw(2.5,5.3) node {\scriptsize{$j_{m}$}};
            \draw(2.8,7.9) node {\scriptsize{$\tilde{T}_{m}$}};
            \draw(3.5,6.5) node {\scriptsize{$j_{m}$}};
            \draw(5.3,9.6) node {\scriptsize{$\tilde{T}_{m}^\prime$}};
            \draw(6.3,8.3) node {\scriptsize{$i_{m}$}};
        \end{tikzpicture}
    \end{minipage}
\vspace{-3mm}
\caption{Diagrams for the product of overlaps appearing in the calculation of (\ref{eq:RNn}) . {\bf Left}:  disconnected diagram.
 {\bf Right}: The connected  diagram. }
\label{fig:diagram3}   
\end{figure}
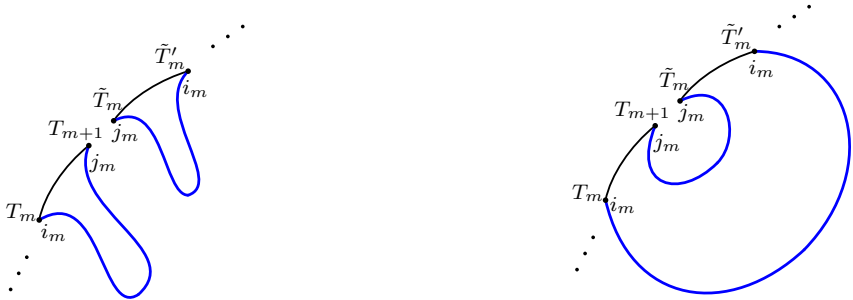

We compute this by following the procedure explained in section \ref{sec:WCP}, namely by preparing an interval for each overlap, and connecting the endpoints of the intervals by EoW branes, then evaluating each diagram generated in this way.  As shown in the figure, $m$ th replica consists of two intervals with indices for Hawking radiation $i_{m},j_{m}$. Therefore it is clear that the dominant diagram when $k=d_{D}d_{B}$ is sufficiently large is the one connecting the endpoint with the index $i_{m}$ in the first interval of to the endpoint of the second replica with the same index in the same replica (the right panel of Figure \ref{fig:diagram3}). Similarly, we connect the endpoints with $j_{m}$ in this replica.   This is because, if there is an EoW brane connecting endpoints with distinct Hawking indices (say $i,j$), then the value of the diagram is significantly reduced in the large $k$ limit  because of the Kronecker delta factor $\delta_{ij}$ coming from the brane.

This means that in the dominant saddle two different replicas are
 not connected by any EoW brane, because they start and end at the same replica. This means that the R\'enyi entropy is a self averaging quantity  
\be
\overline{{\rm tr} (\mathcal{R}^\text{Lite}\left[\mathcal{N}[\rho]\right])^{n}} = {\rm tr} \left( 
\overline{\mathcal{R}^\text{Lite}\left[\mathcal{N}[\rho]\right]}\right)^{n}
\ee

A similar statement holds for  the second term of \eqref{eq;rep1}, therefore we conclude that the relative entropy of our interest is also self averaging, 
\be
\overline{S( \mathcal{R}^\text{Lite} \left[\mathcal{N}[\rho_{T}]\right]\; || \rho_{T})} = S( \overline{\mathcal{R}^\text{Lite} \left[\mathcal{N}[\rho_{T}] \right]} \;|| \rho_{T}).
\label{eq:saing}
\ee
when $k$ is sufficiently large. This implies  that in the relative entropy one can replace  $ \mathcal{R}^\text{Lite} \left[\mathcal{N}[\rho_{T}]\right] $
to its average 
$\overline{\mathcal{R}^\text{Lite} \left[\mathcal{N}[\rho_{T}]\right]}$. The average of the density matrix is given by 
\be
\overline{\mathcal{R}^\text{Lite} \left[\mathcal{N}[\rho_{T}]\right]} 
=\frac{1}{1 + \left(\dfrac{d_{T}}{d_{D}}\right)^{2}}  \left(  \rho + \left(\dfrac{d_{T}}{d_{D}}\right)^{2} \cdot \dfrac{I_{T}}{d_{T}} \right).
\ee

A more precise way to argue this is the following: Let us compute 
\begin{align}
\overline{{\rm tr} \left[\left(\mathcal{R}^\text{Lite} \left[\mathcal{N}[\rho_{T}]\right]  -\overline{\mathcal{R}^\text{Lite} \left[\mathcal{N}[\rho_{T}]\right]} \right)^{2}\right]} = \overline{{\rm tr} \left[\left(\mathcal{R}^\text{Lite} \left[\mathcal{N}[\rho_{T}]\right]  \right)^{2}\right]} -{\rm tr} \left[\left( \overline{\mathcal{R}^\text{Lite} \left[\mathcal{N}[\rho_{T}]\right]  }\right)^{2}\right].
\end{align}
Then, the right hand side of the above equation is given by 
\begin{equation}
	\begin{aligned}
\hspace{-25pt}\frac{(k\, d_{C})^{4}}{ \left(N k (d_{C})^{2} d_{T} \right)^{2} }  \Bigg\{ \frac{1}{k\, d_{C}} \Bigg[ \frac{1}{k^{2}}\left( 2+d_{T} \tr\left[ \rho ^{2}\right] + (d_{T})^{2} \right)&\\
&  \hspace{-3cm}+ \frac{1}{k\, d_{C}} \left(  (d_{T})^{2}  \tr\left[ \rho ^{2}\right]  + 2d_{T}+ 2 \tr\left[ \rho ^{2}\right] \right)+ \frac{1}{(d_{C})^2}   d_{T} \tr\left[ \rho ^{2}\right]  \Bigg] \Bigg\}
\end{aligned}
\end{equation}
which becomes small when $k \gg d_{C} d_{T}$. By plugging this expression, we have
\begin{equation}
	\begin{aligned}
	\overline{S(\mathcal{R}^\text{Lite} \left[\mathcal{N}[\rho_{T}]\right]|| \rho_{T})} 
		& \approx S( \overline{\mathcal{R}^\text{Lite} \left[\mathcal{N}[\rho_{T}]\right]}\;|| \rho_{T}) \\[+10pt]     
  & = \begin{dcases}
             S\bigg(  \rho \, \bigg|\! \bigg| \dfrac{I_{T}}{d_{T}} \bigg) & k \ll d_{C}\Leftrightarrow  d_{T} \ll\left(\frac{d_{T}}{d_{D}}\right)^{2}  \\
            0 & k \gg d_{C} d_{T} \Leftrightarrow \left(\frac{d_{T}}{d_{D}}\right)^{2} \ll 1.
        \end{dcases}
	\end{aligned}
\end{equation}
Thus, for early times $k \ll d_{C}$, the relative entropy is non-vanishing unless $\rho=I_{T}/d_{T}$, but for late times $d_{C}\, d_{T}\ll k$, the relative entropy is vanishing. This result implies that when $k \gg d_{C}d_{T}$,  $\mathcal{R}^\text{Lite}$ indeed works as a recovery map.

\subsection{Relation to the Yoshida-Kitaev protocol}

So far we have shown that when $k \gg d_{C}d_{T}$, the Petz-lite $\mathcal{R}^\text{Lite}\sim \mathcal{N}^{\dagger}$ indeed works as a recovery map. However, we have not discussed the physical interpretation of the Petz-lite. So, in this subsection, we explain the interpretation by showing the equivalence between the Petz-lite and the well-known Yoshida-Kitaev (YK) protocol. The relation between the Yoshida-Kitaev protocol and the Petz map has been suggested by Yoshida \cite{Yoshida:2021xyb,Yoshida:2021haf}.

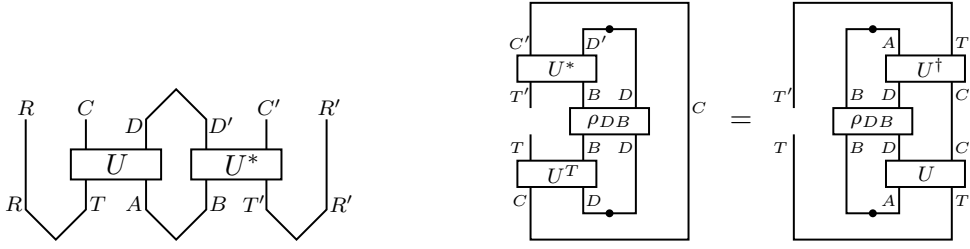
\begin{figure}
\vspace{-1cm}
    \begin{minipage}[b]{0.45\linewidth}
      \centering
        \begin{tikzpicture}[line width=0.75pt,x=0.4cm,y=0.4cm]
            \draw(0,4)--(0,1)--(1,0)--(2,1)--(2,2);
            \draw(2,3)--(2,4);
            \draw(4,3)--(4,4)--(5,5)--(6,4)--(6,3);
            \draw(4,2)--(4,1)--(5,0)--(6,1)--(6,2);
            \draw(8,4)--(8,3);
            \draw(8,2)--(8,1)--(9,0)--(10,1)--(10,4);
            \draw(1.5,2) rectangle (4.5,3);
            \draw(5.5,2) rectangle (8.5,3);
            \draw(3.04,2.5) node {$U$};
            \draw(7.2,2.5) node {$U^{\ast}$};
            \draw(-0.4,1.2) node {\scriptsize{$R$}};
            \draw(0,4.4) node {\scriptsize{$R$}};
            \draw(2.4,1.2) node {\scriptsize{$T$}};
            \draw(2,4.4) node {\scriptsize{$C$}};
            \draw(3.6,1.2) node {\scriptsize{$A$}};
            \draw(3.6,3.8) node {\scriptsize{$D$}};
            \draw(6.4,1.2) node {\scriptsize{$B$}};
            \draw(6.5,3.8) node {\scriptsize{$D^\prime$}};
            \draw(7.6,1.2) node {\scriptsize{$T^\prime$}};
            \draw(8.1,4.4) node {\scriptsize{$C^\prime$}};
            \draw(10.5,1.2) node {\scriptsize{$R^\prime$}};
            \draw(10.1,4.4) node {\scriptsize{$R^\prime$}};
        \end{tikzpicture}
    \end{minipage}
    \hspace{0.5em}
    \begin{minipage}[b]{0.45\linewidth}
      \centering
        \begin{tikzpicture}[line width=0.75pt,x=0.35cm,y=0.35cm]
            \fill(3,1) circle (1.5pt);
            \fill(3,8) circle (1.5pt);
            \fill(13,1) circle (1.5pt);
            \fill(13,8) circle (1.5pt);
            \draw(0,4)--(0,3);
            \draw(0,2)--(0,0)--(6,0)--(6,9)--(0,9)--(0,7);
            \draw(0,6)--(0,5);
            \draw(2,4)--(2,3);
            \draw(2,2)--(2,1)--(4,1)--(4,4);
            \draw(4,5)--(4,8)--(2,8)--(2,7);
            \draw(2,6)--(2,5);
            \draw(-0.5,2) rectangle (2.5,3);
            \draw(-0.5,6) rectangle (2.5,7);
            \draw(1.5,4) rectangle (4.5,5);
            \draw(8,4.5) node {$=$};
            \draw(10,4)--(10,0)--(16,0)--(16,2);
            \draw(16,3)--(16,6);
            \draw(16,7)--(16,9)--(10,9)--(10,5);
            \draw(12,4)--(12,1)--(14,1)--(14,2);
            \draw(14,3)--(14,4);
            \draw(14,5)--(14,6);
            \draw(14,7)--(14,8)--(12,8)--(12,5);
            \draw(11.5,4) rectangle (14.5,5);
            \draw(13.5,2) rectangle (16.5,3);
            \draw(13.5,6) rectangle (16.5,7);
            \draw(1.26,2.5) node {\scriptsize{$U^{T}$}};
            \draw(1.2,6.5) node {\scriptsize{$U^{\ast}$}};
            \draw(15.04,2.5) node {\scriptsize{$U$}};
            \draw(15.2,6.55) node {\scriptsize{$U^{\dagger}$}};
            \draw(3,4.5) node {\scriptsize{$\rho_{DB}$}};
            \draw(13,4.5) node {\scriptsize{$\rho_{DB}$}};
            \draw(-0.5,3.5) node {\tiny{$T$}};
            \draw(-0.4,5.5) node {\tiny{$T^\prime$}};
            \draw(-0.4,1.5) node {\tiny{$C$}};
            \draw(-0.4,7.5) node {\tiny{$C^\prime$}};
            \draw(2.4,1.5) node {\tiny{$D$}};
            \draw(2.4,3.5) node {\tiny{$B$}};
            \draw(2.4,5.5) node {\tiny{$B$}};
            \draw(2.5,7.5) node {\tiny{$D^\prime$}};
            \draw(3.6,3.5) node {\tiny{$D$}};
            \draw(3.6,5.5) node {\tiny{$D$}};
            \draw(6.4,5) node {\tiny{$C$}};
            \draw(9.5,3.5) node {\tiny{$T$}};
            \draw(9.6,5.5) node {\tiny{$T^\prime$}};
            \draw(12.4,3.5) node {\tiny{$B$}};
            \draw(12.4,5.5) node {\tiny{$B$}};
            \draw(13.6,1.5) node {\tiny{$A$}};
            \draw(13.6,3.5) node {\tiny{$D$}};
            \draw(13.6,5.5) node {\tiny{$D$}};
            \draw(13.6,7.5) node {\tiny{$A$}};
            \draw(16.4,1.5) node {\tiny{$T$}};
            \draw(16.4,3.5) node {\tiny{$C$}};
            \draw(16.4,5.5) node {\tiny{$C$}};
            \draw(16.4,7.5) node {\tiny{$T$}};
        \end{tikzpicture}
    \end{minipage}
\vspace{-3mm}
\caption{ {\bf Left}: Yoshida-Kitaev decoding protocol.
 {\bf Right}: operator transpose that providing the key equivalence (\ref{eq:opetransEPR}).}
\label{fig:diagramYK}   
\end{figure}

In \cite{Yoshida:2017non}, Yoshida and Kitaev proposed an interesting recovery protocol for the object thrown into the black hole $T$ from late and early radiation $DB$. Brief summary of their protocol is as follows: 

\begin{enumerate}
\item In addition to the original Hayden-Preskill setup, introduce a copy of the diary and the reference, denoted by $R'T'$. We choose the state on $R'T'$  to be an EPR state. Bob can manipulate Hawking radiation $DB$  and $R'T'$. Before applying the decoding protocol the state of the total system is  
\be
|\Psi_{HP} \ra \otimes  |\epr \ra_{R'T'}
\ee
where $|\Psi_{HP} \ra$ is the state on $RCDB$ given by \eqref{eq:totaState}.

\item  We then use the  early Hawking radiation $B$ and the copy of the diary $T'$ to simulate the black hole dynamic by applying  $U^{*}$  which is the complex conjugate of $U$ for the time evolution of the original system.  After the simulation total system consists of $RCDR'C'D'$, and the state is 
\be
| \Psi_{YK} \ra_{RCDD'C'R'} =\left( I_{RC} \otimes I_{D} \otimes U^{*}_{D'C' \rightarrow BT'} \otimes I_{R} \right) |\Psi_{HP} \ra \otimes  |\epr \ra_{R'T'}
\ee

\item  Post select to the EPR pair on  $DD'$. If it succeeds, the state on $RR'$ is the EPR state with high fidelity, meaning the success of information recovery.

\end{enumerate}
The quantum  circuit for the protocol is shown in the left panel of figure \ref{fig:diagramYK}.  
Combining these steps, the quantum channel $\mathcal{R}_{D,B\to R'}^{KY} $ for the Yoshida-Kitaev (YK) recovery map is given by
\begin{equation}
	\begin{aligned}
		&\mathcal{R}_{D,B\to R'}^{\text{YK}} \left[ \mathcal{O}_{DB} \right] \\
  &=  \frac{1}{\Delta}  \tr_{C'} \left[ \, _{D,D'} \! \bra{\epr} U^{*}_{B,T' \to C',D'} \left( \mathcal{O}_{DB}  \otimes \ktbra{\epr}{T',R'}   \right) U^{T}_{B,T' \to C',D'}  \ket{\epr}_{D,D' } \right],
	\end{aligned} \label{eq:kyRecovery}
\end{equation}
where $\Delta$ is a normalization factor given by
\begin{equation}
    \Delta = \overline{ \left| _{D,D'}\ip{\epr}{ \Psi_{YK} } \right|^{2}} \approx \frac{1}{(d_{T})^{2}}+\frac{1}{(d_{D})^{2}}.\label{eq:normaKYstate}
\end{equation}

For the above YK recovery map, we show the equivalence between the YK recovery map $\mathcal{R}_{D,B\to R'}^{\text{YK}} $ and the Petz-lite \eqref{eq:petzLite}, $\mathcal{R}_{D,B\to T}^{\text{Lite}}$ up to the isomorphism $V_{ T \to R'}$ between systems $T$ and $R'$,
\begin{equation}
	\mathcal{R}_{D,B \to R'}^{KY} \left[ \mathcal{O}_{DB} \right] =    V_{ T \to R'} \, \mathcal{R}_{D,B\to T}^{\text{Lite}}\left[  \mathcal{O}_{DB} \right] \,  V_{T \to R' }^{\dagger} \quad ,  \label{eq:kyPetzRelation}
\end{equation}
where $V_{ T \to R'}$ is explicitly given by 
\begin{equation}
	V_{T\to R' } \coloneqq d_{T} \, _{T,T'} \! \braket{\epr}{\epr}_{T',R'} = \sum_{\tilde{T}=1 }^{d_{T}} \ket{\tilde{T}}_{R'\, T} \!\!\bra{\tilde{T}}. \label{eq:isomorphism}
\end{equation} 

The argument for the equivalence is summarized in the right panel of  figure \ref{fig:diagramYK}.
We start with the YK recovery map \eqref{eq:kyRecovery}. First, we rewrite the trace of subsystem $C'$ in the YK recovery map as 
\begin{equation}
	\tr_{ C' } \left[ \mathcal{O} \right]= d_{C} \, _{C,C'} \! \bra{\epr} \left( I_{C}  \otimes \mathcal{O}\right)\ket{\epr}_{C,C' }, \label{eq:traceToEPR}
\end{equation}
and introduce two EPR states $\ket{\epr}_{D,D' }$ and $\ket{\epr}_{C,C' }$.

Next, by using  \eqref{eq:traceToEPR} and the relation (see appendix \ref{app:opetransEPR} for the derivation)
\begin{equation}
	\begin{aligned}
		U^{T}_{ C',D' \to B,T'} \ket{\epr}_{C,C'} \otimes  \ket{\epr}_{D,D'} &=  U_{A, T \to C,D} \ket{\epr}_{A,B} \otimes \ket{\epr}_{T,T'}, \label{eq:opetransEPR}
	\end{aligned}
\end{equation} 
the KY recovery map \eqref{eq:kyPetzRelation} can be rewritten as 
\begin{equation}
	\begin{aligned}
		&\mathcal{R}_{D,B \to R'}^{KY} \left[ \mathcal{O}_{DB} \right]\\
		& = \frac{d_{C}}{\Delta} \left( \, _{A,B} \! \bra{\epr} \otimes  \, _{T,T'} \! \bra{\epr} \right)  \left[ U_{A, T \to C,D}^{\dagger} \left( \mathcal{O}_{DB}  \otimes  \ktbra{\epr}{T',R'}  \right)U_{A, T \to C,D}  \right]\\
		& \hspace{11cm}  \times \left( \ket{\epr}_{A,B} \otimes \ket{\epr}_{T,T'} \right)\\
		&= \frac{d_{C}}{\Delta} \left(  \, _{T,T'} \! \braket{\epr}{\epr}_{T',R'} \right) \\
  &  \hspace{3cm} \times \, _{A,B} \! \bra{\epr} \left[  U_{T, A \to C,D}^{\dagger}\,  \mathcal{O}_{DB}\,   U_{T,A  \to C,D}  \right] \ket{\epr}_{A,B}\\
  & \hspace{10cm}\times\left(  \, _{T',R'} \! \braket{\epr}{\epr}_{T,T'} \right)\\
		&= \frac{ d_{C} }{ (d_{T})^{2}\,\Delta } \,  V_{T\to R' }  \,\mathcal{N}_{D,B \to T,A} ^{\dagger}\left[ \mathcal{O}_{D,B}  \right] \,  V_{T\to R' }^{\dagger},
	\end{aligned} 
\end{equation}
where, in the final line, we used the definition of the isomorphism \eqref{eq:isomorphism} and the adjoint HP channel \eqref{eq:adjoHPChannel}. Additionally, the above overall constant $\frac{ d_{C} }{ (d_{T})^{2}\,\Delta }$ coincides with that of the Petz-lite \eqref{eq:petzLiteFullExpres}, since 
\begin{equation}
	\frac{ d_{C} }{ (d_{T})^{2}\,\Delta }=\frac{d_{C}}{1+\left( \dfrac{d_{T}}{d_{D}} \right)^{2}},
\end{equation}
where we used the definition of $\Delta$, \eqref{eq:normaKYstate}. Therefore the above expression implies the desired relation \eqref{eq:kyPetzRelation}.

\section{Recovery map for the Hayden-Preskill channel in SYK}\label{sec:SYKHP}

So far we have given the evidence that the Petz-lite works as a recovery map, under the Haar random unitary which is highly chaotic. In this section, we argue that this continues to hold for a more realistic but tractable model of chaotic dynamics; the  Sachdev-Ye-Kitaev (SYK) model \cite{Sachdev:1992fk,KitaevTalks,Sachdev:2015efa}. In this paper, we briefly explain the relevant calculations, leaving details in the upcoming paper \cite{WIP}.

\subsection{Setup of SYK Hayden-Preskill protocol}

In this section, we explain the setup to study the Hayden Preskill like protocol (what we call SYK HP channel) in the SYK model. This was first introduced in \cite{Chandrasekaran:2022qmq,Chandrasekaran:2021tkb}.

The  SYK model is a theory of $N$ Majorana fermions $\psi_{i}$, and its Hamiltonian is given by
\begin{equation}
	H=\left(i\right)^{q/2} \sum_{1 \leq i_{1} < i_{2} < \cdots < i_{q} \leq N} j_{i_{1} i_{2} \cdots i_{q}} \psi_{i_{1}}\psi_{i_{2}} \cdots \psi_{i_{q}},
\end{equation}
where $q \in 2\mathbb{N} \, (q>2)$, $j_{i_{1} i_{2} \cdots i_{q}}$ is a random coefficient drawn from a Gaussian random distribution with zero mean and the variance $\left\langle j_{i_{1} i_{2} \cdots i_{q}}^{2} \right\rangle=J^{2}(q-1)!/N^{q-1}$.

Following \cite{Chandrasekaran:2022qmq}, we consider two copies of the Hilbert space of the SYK model, say left SYK system $L$ and right one $R$. Hereafter we  denote the Majorana fermions  on the left system by $\psi_{i,L}$ and $\psi_{i,R}$ for the right. For notational simplicity, we use the convention 
\begin{equation}
	\left\{ \psi_{i},\psi_{j} \right\} =2\delta_{i,j}, \label{eq:CACR}
\end{equation}
for the anti commutation relation for the fermions on the same side.
In this set up, the right SYK system corresponds to  early radiation degrees of freedom of the original Hayden-Preskill setup, and the left SYK system corresponds to the rest; the union of the diary system  and the initial black hole before the action of the random unitary, or equivalently the  remaining black hole plus  late radiation degrees of freedom  after the unitary evolution. In particular, the left system $L$ is divided into two subsystems, say $\tilde{L}$ and $K$, the former  correspond to the remaining black hole  and the latter to the  late radiation part of the original HP setup.

On the union of  the above SYK systems $L$ and $R$, we consider the following thermo-field double (TFD) state ;
\begin{equation}
	\ketTFD{L,R}=Z^{-1/2}(\beta)\, e^{-\beta (H_{L}+H_{R})/4 } \ket{0}_{L,R}, \label{eq:tfdSYK}
\end{equation}
where $Z(\beta)$ is a normalization factor of the state, and $\ket{0}_{L,R}$ is given by \cite{Qi:2018bje}
\begin{equation}
	\left[ \, \psi_{j,L}(0)+ i \psi_{j,R}(0) \, \right] \ket{0}_{L,R} =0 \quad \text{for } \forall j. \label{eq:relationRL}
\end{equation} 
Note that the thermo-field double state \eqref{eq:tfdSYK} satisfies the relation $(H_{L}-H_{R})\ket{\tfd}=0$. 
This TFD state corresponds to an entangled state between the initial black hole and the early radiation.

 The code subspace (a diary system) of our interest is two dimensional, and let's denote two basis vectors by $\ket{0}$ and $\ket{1}$. This code subspace is embedded into the physical Hilbert space $LR$ by an isometry. The image of the code subspace  is 
  spanned by the TFD state $\ket{\tfd}_{L,R}$ and the excited state $\psi_{i,L}(0)\ket{\tfd}_{L,R}$. Here we assume that the  Majorana fermion $\psi_{i,L}(0)$ acting on the TFD state  lives in the subsystem $\tilde{L}$, $i\in \tilde{L}$. More explicitly,  by the isometry, the  states in the code subspace $\ket{T}\, (T=0,1)$  are mapped to
\begin{equation}
	\left( V_{T,L\to L}  \otimes I_{R}\right) \left( \ket{T}_{T}\otimes \ket{\tfd}_{L,R}  \right) \coloneqq \begin{dcases}
		\ket{\tfd}_{L,R} & \text{ for } T=0\\
		\dfrac{1}{ \left( Z_{\delta} \right)^{\frac{1}{2} } }\psi_{i,L}(i\delta) \ket{\tfd}_{L,R} & \text{ for } T=1, \label{eq:embeddingV}
	\end{dcases}
\end{equation}
 where $\psi_{i,L}( i\delta)$ is the regulated Majorana fermion operator
 \begin{equation}
 	\psi_{i,L}(i\delta)= e^{-\delta H_{L}} \psi_{i,L}(0) e^{ \delta H_{L}},
 \end{equation}
 and  $\delta$ is an infinitesimal cutoff parameter to normalize the  state with the operator insertion even in  the conformal limit, where the SYK model has an effective description  in terms of  the reparametrization modes \cite{Maldacena:2016hyu}. $ Z_{\delta} $ is its normalization factor given by the  two point function
 \begin{equation}
 	\begin{aligned}
 		Z_{\delta}&=\frac{1}{N-K} \sum_{i=1}^{N-K} \frac{1}{Z(\beta)} \tr \left[ e^{-\beta H_{L}} \psi_{i,L}(-i\delta) \psi_{i,L}(i\delta)   \right]\\
 		&= \frac{1}{N-K} \sum_{i=1}^{N-K} \frac{1}{Z(\beta)} \tr \left[ e^{-\beta H_{L}} e^{2\delta H_{L}}  \psi_{i,L}(0) e^{-2\delta H_{L}} \psi_{i,L}(0)   \right]=G_{\beta} (2\delta).
 	\end{aligned}
  \label{eq:average}
 \end{equation}  
 This normalization factor is not for the specific Majorana fermion ``$i$", but averaged over the region $\tilde{L}$ with $N-K$ sites. We expect  the difference between the two only appears  in sub-leading terms with respect to $K/N$ because of typicality. Therefore, we use this normalization factor \eqref{eq:average} for later convenience.

\begin{figure}[ht]
\centering
\includegraphics[scale=0.3]{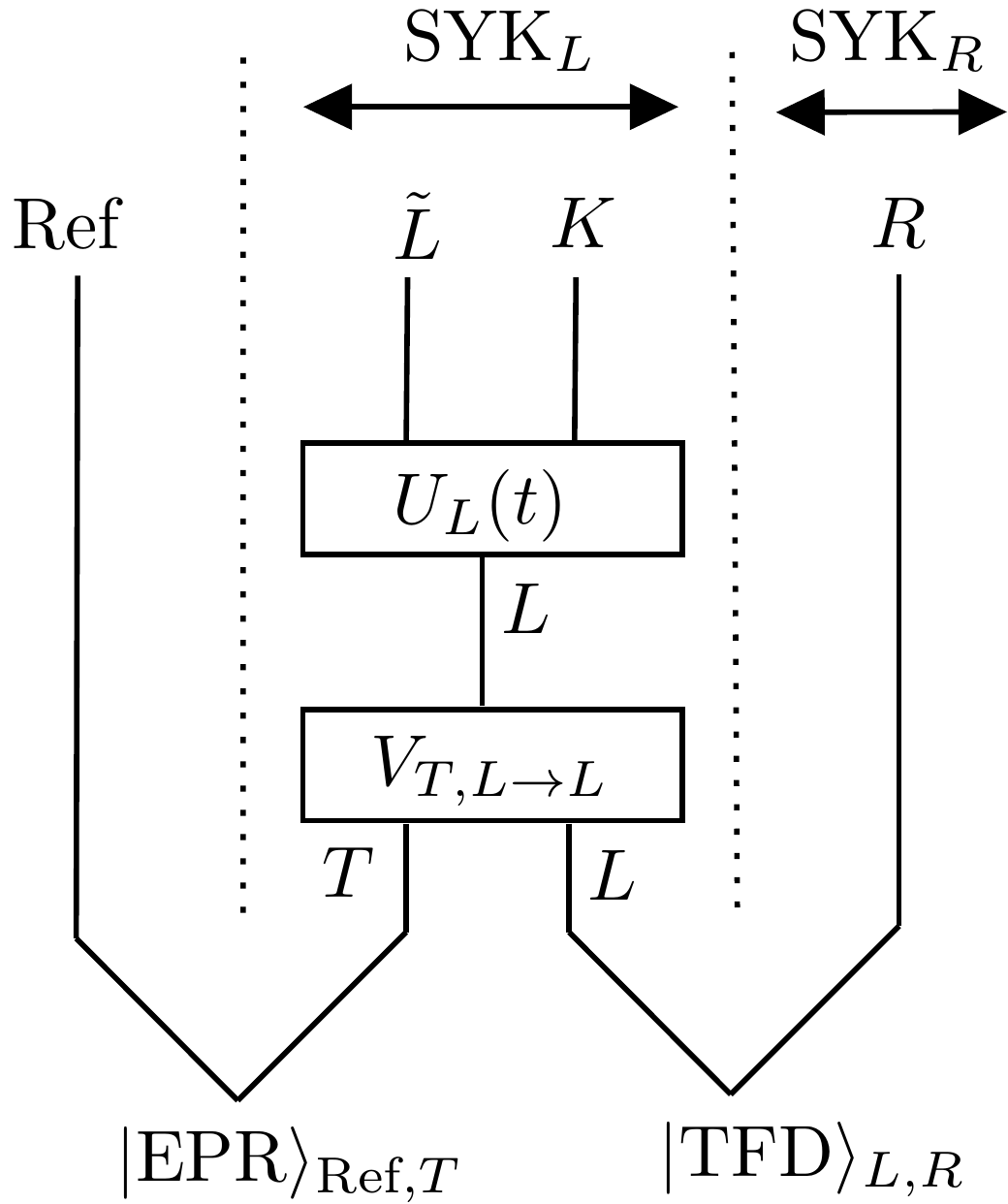}	
\caption{Circuit diagram corresponding to state (\ref{eq:totaStateSYK}).}
\label{fig:SYKHPcircuit}
\end{figure}

Using the above embedding, we can holographically prepare an initial entangle state between the early radiation and an initial black hole containing a diary in the SYK model. For this system, we consider an unitary time evolution on  the left system $L$ by the SYK Hamiltonian $H_{L}$,  
\begin{equation}
	U_{L}(t) = \exp\left(  i t H_{L} \right).
\end{equation}
By this time evolution,  information in the diary gets scrambled and uniformly distributed over the left SYK system after the scrambling time. The resulting state is 
\begin{equation}
	\ket{ \Psi_\text{SYK HP} } = (I_\text{Ref}\otimes U_{L}(t)\otimes I_{R}) \left( I_\text{Ref}\otimes V_{T,L\to L}  \otimes I_{R} \right)\left( \ket{\epr }_{\text{Ref},T}\otimes \ket{\tfd}_{L,R}\right),\label{eq:totaStateSYK}
\end{equation} 
which corresponds to the state \eqref{eq:totaState}. In figure \ref{fig:SYKHPcircuit}, we give the circuit diagram corresponding to the state \eqref{eq:totaStateSYK}.

 We are interested in recovering the diary information from the early and late radiations $R$ and $\tilde{L}$ by using the Petz-lite for the SYK HP protocol. 
As in \eqref{eq:hpChannel}, the SYK HP channel $\mathcal{N}^{ \text{SYK}}_{T\to K,R}$ representing error  is obtained by tracing out the remaining black hole part $\tilde{L}$ in the final state  \eqref{eq:totaStateSYK},
\begin{equation}
	\mathcal{N}^{ \text{SYK}}_{T\to K,R}[\rho_{T}] \coloneqq \tr_{\tilde{L}} \left[ U_{L} V_{T,L\to L} \left( \rho_{T} \otimes \ktbra{\tfd}{L,R}  \right)  V_{T,L\to L}^{\dagger}  U_{L}^{\dagger} \right]. \label{eq:hpSYKchannel}
\end{equation}
This channel maps a density matrix on the diary $T$ to the one on the late and early radiation system $K,R$.
Also, the adjoint $\mathcal{N}^{ \text{SYK} \dagger}_{K,R\to T}$ of the SYK HP channel is given by 
\begin{equation}
	\begin{aligned}
		\mathcal{N}^{ \text{SYK} \dagger}_{K,R\to T}[ \mathcal{O}_{KR}] &\coloneqq \tr_{L,R} \left[ \ktbra{\tfd}{L,R}  \left(  V_{L\to T,L}^{\dagger}   U_{L}^{\dagger} \,   \mathcal{O}_{KR} \,   U_{L} V_{L\to T,L} \right)\right]\\
		& = \, _{L,R} \ev{ \left(  V_{L\to T,L}^{\dagger}   U_{L}^{\dagger} \,   \mathcal{O}_{KR} \,   U_{L} V_{L\to T,L} \right) }{\tfd}_{L,R}
		.\label{eq:hpSYKAdjointchannel}
	\end{aligned}
\end{equation}

The above quantum channels are analogous to the original HP channel and its adjoint for the Haar random unitary. However, we note that the difference that the SYK HP channel and its adjoint includes the embedding map $V$, which induces (fermionic) excitations.

\subsection{Some matrix elements of the Petz-lite and R\'enyi-two correlators}
\label{subsec:ElementPetzlite}

Now that we have prepared the SYK HP channel and its adjoint, we can construct the Petz-lite map for this channel.  As in the Petz-lite for the Haar random case \eqref{eq:petzLite}, we consider the Petz-lite for the SYK case,
\begin{equation}
	\mathcal{R}^{\text{Lite,SYK}}_{K,R\to T} \left[ \mathcal{O}_{KR} \right] = \frac{1}{ N_{\text{SYK}} } \mathcal{N}^{ \text{SYK} \dagger}_{K,R\to T}[ \mathcal{O}_{KR}],\label{eq:petzLiteSYK}
\end{equation}
where $ N_{\text{SYK}}$ is the normalization factor, which is determined by the condition 
\begin{equation}
	\tr_{T} \left[ \mathcal{R}^{\text{Lite,SYK}}_{K,R\to T} \left[ \mathcal{N}^{ \text{SYK}}_{T\to K,R}[\sigma_{T}] \right] \right]=1.
\end{equation}
Here $\sigma_{T}$ is some reference state in $T$ for the normalization. We take it to be $\sigma_{T}=\ktbra{0}{T}$. For this choice, the normalization factor is given by
\begin{equation}
	\begin{aligned}
		N_{\text{SYK}}  &=  \sum_{T=0,1}  \mel{T}{ \mathcal{N}^{\text{SYK} \dagger}_{K,R\to T} \left[ \mathcal{N}^{ \text{SYK}}_{T\to K,R}[\ktbra{0}{T}] ] \right]}{T}.
	\end{aligned} \label{eq:normalizationPetzLiteSYK}
\end{equation}
We note that due to this normalization, we can see that the Petz-lite \eqref{eq:petzLiteSYK} for the SYK HP protocol have a similar overall constant with the Petz-lite for the original HP protocol \eqref{eq:petzLiteFullExpres}. To see the similarity, we first rewrite the Petz-lite \eqref{eq:petzLiteSYK} with the normalization factor \eqref{eq:normalizationPetzLiteSYK} as follows
\begin{equation}
	\begin{aligned}
		\mathcal{R}^{\text{Lite,SYK}}_{K,R\to T} \left[ \mathcal{O}_{KR} \right] =\dfrac{ \ev{ \hat{d}_{\tilde{L}} }_{\beta}  }{ 1 +  \ev{ \hat{d}_{\tilde{L}} }_{\beta} \mel{1}{ \mathcal{N}^{\text{SYK} \dagger}_{K,R\to T} \left[ \mathcal{N}^{ \text{SYK}}_{T\to K,R}[\ktbra{0}{T}] ] \right]}{1} }  \mathcal{N}^{ \text{SYK} \dagger}_{K,R\to T}[ \mathcal{O}_{KR}],
	\end{aligned}\label{eq:petzLiteSYKAnalogous}
\end{equation}
where $\ev{ \hat{d}_{\tilde{L}} }_{\beta}  $ is an effective dimension of subsystem $\tilde{L}$ defined by the purity $\tr_{\tilde{L}} \left[ \left( \rho_{ \tilde{L} }  \right)^{2} \right]$ of the TFD state with respect to the subsystem\footnote{We note that in our setting, subsystem $\tilde{L}$ is smaller than the complement system $KR$.},
\begin{equation}
	\mel{0}{ \mathcal{N}^{\text{SYK} \dagger}_{K,R\to T} \left[ \mathcal{N}^{ \text{SYK}}_{T\to K,R}[\ktbra{0}{T}] ] \right]}{0} =\tr_{KR} \left[ \left( \rho_{ KR }  \right)^{2} \right]=\tr_{\tilde{L}} \left[ \left( \rho_{ \tilde{L} }  \right)^{2} \right] \eqqcolon \frac{1}{  \ev{ \hat{d}_{\tilde{L}} }_{\beta}  }.\label{eq:defOfEffectiveDim}
\end{equation}
The effective dimension is analogous to the dimension of the remaining black hole in the original HP setup. Indeed in the infinite temperature limit $\beta \to 0$, the effective dimension is almost reduced to the actual dimension of subsystem $\tilde{L}$, $d_{\tilde{L}} = 2^{N-K}$. However in general the effective dimension  is smaller than the actual dimension due to the property of the purity and thermal effects;
\begin{equation}
	   1 \, \leq \,  \ev{ \hat{d}_{\tilde{L}} }_{\beta}  \,  \leq \,  d_{\tilde{L}},
\end{equation}
where this effective dimension becomes closed to 1 in $\beta \to \infty$ and  $d_{\tilde{L}}$ in $\beta \to 0$.
With this effective dimension, we can compare the Petz-lite \eqref{eq:petzLiteSYKAnalogous} for the SYK model to that for the original one \eqref{eq:petzLiteFullExpres} in the HP setup
\begin{equation*}
	\begin{aligned}
		\mathcal{R}^{\text{Lite},HP}_{D,B\to T} [\mathcal{O}_{DB}] =\frac{ d_{C} }{ 1+ \left( \dfrac{d_{T}}{d_{D}} \right)^{2}} \, \mathcal{N}_{D,B\to T}^{\dagger}\left[ \mathcal{O}_{DB}\right].
		\end{aligned}
\end{equation*}
 The similarities between the quantities in the HP and the SYK  are summarized in the following identifications;
\begin{equation}
	\begin{aligned}
		 d_{C} \quad &\longleftrightarrow \quad \ev{ \hat{d}_{\tilde{L}} }_{\beta},\\
		\left( \dfrac{d_{T}}{d_{D}} \right)^{2} \quad &\longleftrightarrow \quad \ev{ \hat{d}_{\tilde{L}} }_{\beta} \cdot \mel{1}{ \mathcal{N}^{\text{SYK} \dagger}_{K,R\to T} \left[ \mathcal{N}^{ \text{SYK}}_{T\to K,R}[\ktbra{0}{T}] ] \right]}{1}.
	\end{aligned}
\end{equation}
Also, we have the unitarity constraint on the dimensions of the Hilbert spaces; $d_{T}\, d_{B} = d_{C} \, d_{D}$. By using the relation, we can rewrite the dimension as 
\begin{equation}
	\left( \dfrac{d_{T}}{d_{D}} \right)^{2} = \frac{d_{C} \, d_{T} }{ d_{B} \, d_{D} },
\end{equation}
from which we have  the following identification
\begin{equation}
	\mel{1}{ \mathcal{N}^{\text{SYK} \dagger}_{K,R\to T} \left[ \mathcal{N}^{ \text{SYK}}_{T\to K,R}[\ktbra{0}{T}] ] \right]}{1} \quad   \longleftrightarrow \quad  \frac{1}{d_{C}} \cdot \left( \dfrac{d_{T}}{d_{D}} \right)^{2} = \frac{d_{T} }{ d_{B} \, d_{D} }  = \frac{d_{T}}{k}.
\end{equation}
This might be a good ratio to understand current physics; if we have a sufficiently large amount of Hawking radiation compared with the diary, $d_{T} \ll d_{B}\, d_{D} =k$, the ratio becomes almost 0. As we see soon after, the left quantity also becomes almost 0 around and/or after a critical time.

With this discussion of the normalization factor in mind, we consider a matrix element of $\mathcal{R}^{\text{Lite,SYK}}_{K,R\to T} \left[ \mathcal{N}^{ \text{SYK}}_{T\to K,R}[\rho_{T}] \right]$ for a general density matrix $\rho_{T}$ in the Hilbert space of the diary,
\begin{equation}
	\mel{T}{\mathcal{R}^{\text{Lite,SYK}}_{K,R\to T} \left[ \mathcal{N}^{ \text{SYK}}_{T\to K,R}[\rho_{T}] \right]}{T'}. \label{eq:matElementOfRecove}
\end{equation}
To check whether the Petz-lite works as the recovery map, it is sufficient to see whether the following relation holds (approximately) or not,
\begin{equation}
	\mel{T}{\mathcal{R}^{\text{Lite,SYK}}_{K,R\to T} \left[ \mathcal{N}^{ \text{SYK}}_{T\to K,R}[\rho_{T}] \right]}{T'} \overset{?}{\approx} \mel{T}{\rho_{T}}{T'} \quad \text{for } \forall \rho_{T}.
\end{equation}
Checking the above relation is equivalent to focus on the matrix elements
\begin{equation}
	\mel{T}{\mathcal{R}^{\text{Lite,SYK}}_{K,R\to T} \left[ \mathcal{N}^{ \text{SYK}}_{T\to K,R} \left[ \ktbrad{\tilde{T}}{\tilde{T}'}{T} \right] \right]}{T'} \overset{?}{\approx} \braket{T}{\tilde{T}}\braket{\tilde{T}'}{T'}, \qquad \forall T, T', \tilde{T},\tilde{T}' \label{eq:matElementOfRecoveElemt}
\end{equation}
Generally, we have 16 components of the above matrix, but half of them, including odd Majorana fermions, are trivially vanishing due to the fermionic parity of the SYK model. In other words, matrix elements which satisfies $(T+T'+ \tilde{T}+\tilde{T}') \equiv 1 \mod 2$ are vanishing.

Now, we focus on three of non-zero matrix elements, and briefly explain how we can evaluate them
\footnote{The details of the calculation will be discussed in upcoming paper \cite{WIP}. } .
First, we consider the $T, T', \tilde{T},\tilde{T}'=0$ case. If \eqref{eq:matElementOfRecoveElemt} holds then since its  right-hand side is $1$, and therefore the following identity holds,
\begin{equation}
	\begin{aligned}
		1 \overset{?}{\approx}  \mel{0}{\mathcal{R}^{\text{Lite,SYK}}_{K,R\to T} \left[ \mathcal{N}^{ \text{SYK}}_{T\to K,R} \left[ \ktbrad{0}{0}{T} \right] \right]}{0} 
		&= \left(1+  \ev{ \hat{d}_{\tilde{L}} }_{\beta} \cdot \mel{1}{ \mathcal{N}^{\text{SYK} \dagger}_{K,R\to T} \left[ \mathcal{N}^{ \text{SYK}}_{T\to K,R}[\ktbra{0}{T}] ] \right]}{1}   \right)^{-1}.
	\end{aligned}\label{eq:matrixele0000}
\end{equation}

The second one is for the $T, T'=1$,  $ \tilde{T},\tilde{T}'=0$ case, where the matrix element is expected to become 0. In this case, we can see that this matrix element has the same ratios as above,
\begin{equation}
	\begin{aligned}
		0 \overset{?}{\approx}  \mel{1}{\mathcal{R}^{\text{Lite,SYK}}_{K,R\to T} \left[ \mathcal{N}^{ \text{SYK}}_{T\to K,R} \left[ \ktbrad{0}{0}{T} \right] \right]}{1} = \dfrac{   \ev{ \hat{d}_{\tilde{L}} }_{\beta}  \cdot   \mel{1}{ \mathcal{N}^{\text{SYK} \dagger}_{K,R\to T} \left[ \mathcal{N}^{ \text{SYK}}_{T\to K,R}[\ktbrad{0}{0}{T}] ] \right]}{1}  }{ 1+ \ev{ \hat{d}_{\tilde{L}} }_{\beta} \cdot  \mel{1}{ \mathcal{N}^{\text{SYK} \dagger}_{K,R\to T} \left[ \mathcal{N}^{ \text{SYK}}_{T\to K,R}[\ktbra{0}{T}] ] \right]}{1}   }.
	\end{aligned}\label{eq:matrixele1001}
\end{equation}

The final one is for  $T, \tilde{T}=0$,  $ T',\tilde{T}'=1$, the matrix elements \eqref{eq:matElementOfRecoveElemt}, which is expected to be 1, becomes
\begin{equation}
	\begin{aligned}
	 1 \overset{?}{\approx} 	\mel{0}{\mathcal{R}^{\text{Lite,SYK}}_{K,R\to T} \left[ \mathcal{N}^{ \text{SYK}}_{T\to K,R} \left[ \ktbrad{0}{1}{T} \right] \right]}{1} = \dfrac{ \ev{ \hat{d}_{\tilde{L}} }_{\beta} \cdot  \mel{0}{ \mathcal{N}^{\text{SYK} \dagger}_{K,R\to T} \left[ \mathcal{N}^{ \text{SYK}}_{T\to K,R}[\ktbrad{0}{1}{T}] ] \right]}{1} }{ 1+ \ev{ \hat{d}_{\tilde{L}} }_{\beta} \cdot \mel{1}{ \mathcal{N}^{\text{SYK} \dagger}_{K,R\to T} \left[ \mathcal{N}^{ \text{SYK}}_{T\to K,R}[\ktbra{0}{T}] ] \right]}{1}   }.
	\end{aligned}\label{eq:matrixele0011}
\end{equation}

The rest of  matrix elements
\begin{align*}
    \mel{0}{\mathcal{R}^{\text{Lite,SYK}}_{K,R\to T} \left[ \mathcal{N}^{ \text{SYK}}_{T\to K,R} \left[ \ktbrad{1}{0}{T} \right] \right]}{1}, \quad  \mel{1}{\mathcal{R}^{\text{Lite,SYK}}_{K,R\to T} \left[ \mathcal{N}^{ \text{SYK}}_{T\to K,R} \left[ \ktbrad{1}{1}{T} \right] \right]}{1}.
\end{align*}
are difficult to directly evaluate as we will mention in  footnote \ref{foot:diffeva}. In the next section, we evaluate these matrix elements indirectly from the results of this section.

Thus, to see the recovery \eqref{eq:matElementOfRecoveElemt}, 
 we need to study the behaviours of the matrix elements of $\mathcal{N}^{\dagger}\mathcal{N}$ which appear in the right hand side of \eqref{eq:matrixele0000},  \eqref{eq:matrixele1001}, 
 \eqref{eq:matrixele0011}. In order for the recovery to happen, these have to satisfy
\begin{equation}
	 \ev{ \hat{d}_{\tilde{L}} }_{\beta} \cdot \mel{1}{ \mathcal{N}^{\text{SYK} \dagger}_{K,R\to T} \left[ \mathcal{N}^{ \text{SYK}}_{T\to K,R}[\ktbra{0}{T}] ] \right]}{1}  \overset{?}{\approx}0, \label{eq:ratio1001}
\end{equation}
\begin{equation}
	\ev{ \hat{d}_{\tilde{L}} }_{\beta} \cdot  \mel{0}{ \mathcal{N}^{\text{SYK} \dagger}_{K,R\to T} \left[ \mathcal{N}^{ \text{SYK}}_{T\to K,R}[\ktbrad{0}{1}{T}] ] \right]}{1} \overset{?}{\approx}1, \label{eq:ratio0011}
\end{equation}

We study the behavior of the left hand side of\eqref{eq:ratio1001} \eqref{eq:ratio0011} below.  To this end, it is convenient to rewrite the quantities as correlators. From the definitions of the channels \eqref{eq:hpChannel} and \eqref{eq:hpSYKAdjointchannel}, we obtain the left-left correlators
\begin{equation}
	\begin{aligned}
		&\ev{ \hat{d}_{\tilde{L}} }_{\beta} \cdot \mel{1}{ \mathcal{N}^{\text{SYK} \dagger}_{K,R\to T} \left[ \mathcal{N}^{ \text{SYK}}_{T\to K,R}[\ktbra{0}{T}] ] \right]}{1} =  \frac{1}{Z_{\delta}}\cdot \dfrac{  \mel{\tfd}{ \psi_{i,L}(t-i\delta)  \, \left( I_{\tilde{L}} \otimes \rho_{KR}  \right) \,\psi_{i,L}(t+i\delta)   }{\tfd} }{ \tr_{KR} \left[ \left( \rho_{KR}  \right)^{2} \right] },
	\end{aligned}\label{eq:correlator1001}
\end{equation}
\begin{equation}
	\begin{aligned}
		\ev{ \hat{d}_{\tilde{L}} }_{\beta} \cdot  \mel{0}{ \mathcal{N}^{\text{SYK} \dagger}_{K,R\to T} \left[ \mathcal{N}^{ \text{SYK}}_{T\to K,R}[\ktbrad{0}{1}{T}] ] \right]}{1}&= \frac{1}{Z_{\delta}}\cdot  \dfrac{  \mel{\tfd}{   \psi_{i,L}(t -i\delta)  \, \left( \rho_{\tilde{L}} \otimes I_{KR}  \right) \,\psi_{i,L}(t+i\delta)   }{\tfd} }{ \tr_{KR} \left[ \left( \rho_{ KR }  \right)^{2} \right] },
	\end{aligned}\label{eq:correlator0011}
\end{equation}
where the two fermions are put on the left system, and $\rho_{KR}$ and $\rho_{\tilde{L}}$ are defined by
\begin{equation}
	\rho_{\tilde{L}} = \tr_{KR} \left[ \ktbra{\tfd}{LR} \right], \qquad \rho_{ KR } = \tr_{\tilde{L}} \left[ \ktbra{\tfd}{LR} \right].\label{eq:defOfDensityMatrix}
\end{equation}
We give the derivation of the correlators in appendix \ref{app:derivationChanneltoCorrelator}.

We also note that the numerators in the above correlators can be written as 
\begin{equation}
\begin{aligned}
	&\mel{\tfd}{ \psi_{i,L}(t-i\delta)  \, \left( I_{\tilde{L}} \otimes \rho_{KR}  \right) \,\psi_{i,L}(t+i\delta)   }{\tfd} \\
	&=
	\tr_{KR}\left[  \tr_{\tilde{L}}\left[ \psi_{i,L}(t+i\delta) \ktbra{\tfd}{L,R} \psi_{i,L}(t-i\delta)^{\dagger} \right]  \rho_{KR} \right]
\end{aligned} \label{eq:correlatorKRtrace}
\end{equation}
and
\begin{equation}
\begin{aligned}
	&  \mel{\tfd}{   \psi_{i,L}(t -i\delta)  \, \left( \rho_{\tilde{L}} \otimes I_{KR}  \right) \,\psi_{i,L}(t+i\delta)   }{\tfd}  \\
	&=
	\tr_{\tilde{L}}\left[  \tr_{KR}\left[ \psi_{i,L}(t+i\delta) \ktbra{\tfd}{L,R} \psi_{i,L}(t-i\delta)^{\dagger} \right]  \rho_{\tilde{L}} \right].
\end{aligned}  \label{eq:correlatorTildeLtrace}
\end{equation}
These expressions are also useful to see  that these quantities are related to  ``Renyi-2" quantities as explained below.

Below we would like to evaluate these correlators analytically, but the expressions \eqref{eq:correlator1001} and \eqref{eq:correlator0011} are not suitable for an analytic treatment as they are``specific site" correlators, so we can not apply the large-$N$ techniques to evaluate them. 
However, since we are basically interested in typical behaviors under highly chaotic dynamics  in our setup, the specific choice of the embedding would \textit{not} be essential. Therefore, below we consider the ``typical" embedding of the code information into the whole $\tilde{L}$ system uniformly. Therefore we  replace these correlators with their averages on $\tilde{L}$,
\begin{equation}
	\begin{aligned}
		 &\frac{1}{Z_{\delta}}\cdot \dfrac{  \mel{\tfd}{ \psi_{i,L}(t-i\delta)  \, \left( I_{\tilde{L}} \otimes \rho_{KR}  \right) \,\psi_{i,L}(t+i\delta)   }{\tfd} }{ \tr_{KR} \left[ \left( \rho_{KR}  \right)^{2} \right] }\\
		 &\qquad \to  \frac{1}{N-K} \sum_{i=1}^{N-K} \frac{1}{Z_{\delta}}\cdot \dfrac{  \mel{\tfd}{ \psi_{i,L}(t-i\delta)  \, \left( I_{\tilde{L}} \otimes \rho_{KR}  \right) \,\psi_{i,L}(t+i\delta)   }{\tfd} }{ \tr_{KR} \left[ \left( \rho_{KR}  \right)^{2} \right] }
	\end{aligned}\label{eq:correlator1001Avera}
\end{equation}
and 
\begin{equation}
\begin{aligned}
	&\frac{1}{Z_{\delta}}\cdot  \dfrac{  \mel{\tfd}{   \psi_{i,L}(t -i\delta)  \, \left( \rho_{\tilde{L}} \otimes I_{KR}  \right) \,\psi_{i,L}(t+i\delta)   }{\tfd} }{ \tr_{KR} \left[ \left( \rho_{ KR }  \right)^{2} \right] }\\
	&\qquad \to  \frac{1}{N-K} \sum_{i=1}^{N-K} \frac{1}{Z_{\delta}}\cdot  \dfrac{  \mel{\tfd}{   \psi_{i,L}(t -i\delta)  \, \left( \rho_{\tilde{L}} \otimes I_{KR}  \right) \,\psi_{i,L}(t+i\delta)   }{\tfd} }{ \tr_{KR} \left[ \left( \rho_{ KR }  \right)^{2} \right] }.
\end{aligned}  \label{eq:correlator0011Avera}
\end{equation}
These replacements would change the correlators in sub-leading orders of $N$, but the essential physics would not be changed, because of typicality.

These averaged two point functions are special cases of the (right-left) modular-flowed correlators of the form 
\begin{equation}
	\frac{1}{N-K}\sum_{i=1}^{N-K}\frac{\mel{\tfd}{ \psi_{i,R} (\tau)  \left( \rho_{\tilde{L}}^{n-1-k} \otimes \rho_{KR}^{k} \right)  \psi_{i,L} (\tau') }{\tfd}}{\tr\left[ \rho_{KR}^{n} \right]} \label{eq:replicaModularCorre},
\end{equation}
where one of the fermions is put on the left system and the other one is on the right system. 
In the Euclidean regime, they are computed by using the replica trick  in \cite{Chandrasekaran:2022qmq} when $K \ll N$.

\begin{figure}
  \centering
  \begin{minipage}{0.8\textwidth} 
    \centering
    \includegraphics[width=\textwidth]{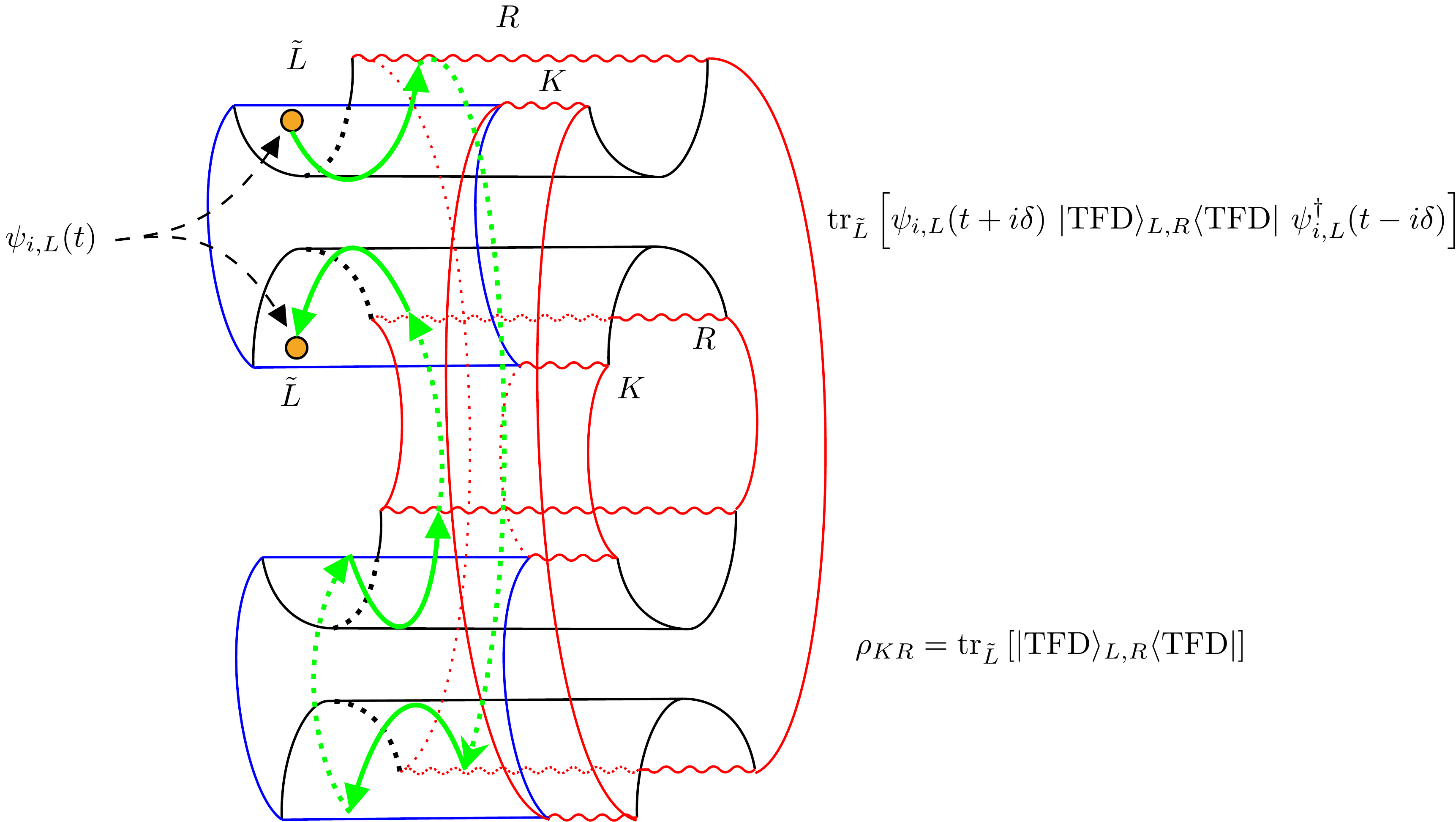}
    
    \vspace{15pt} 
    
    \includegraphics[width=\textwidth]{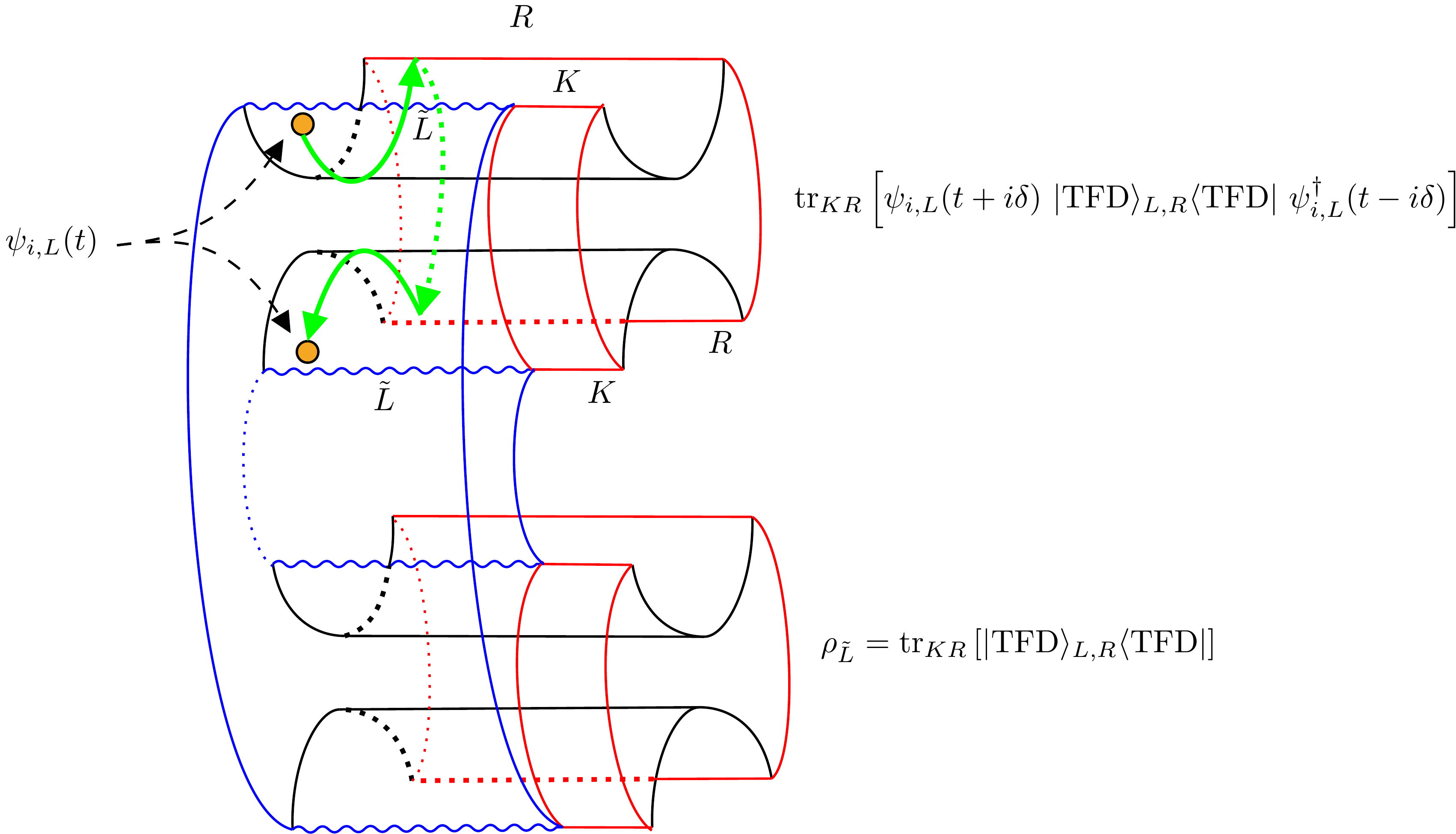}
    
    \caption{Diagrams for the path integral calculation of the correlator (\ref{eq:correlator1001}) with using the relation (\ref{eq:correlatorKRtrace}) (Top), and the other correlator (\ref{eq:correlator0011}) with (\ref{eq:correlatorTildeLtrace}) (Bottom).
     The red regions in the figure correspond to subsystem $RK$, and the blue regions correspond to subsystem $\tilde{L}$. The semicircles correspond to the Euclidean segments that prepares the TFD states. Orange dots represent the insertions of the SYK Majorana fermion with the regularization, $\psi_{i,L}(t+i\delta)$. The combination of the upper two semicircles with the operator insertions corresponds to the density matrix $\tr_{\tilde{L}} [ \psi_{i,L} \ktbra{\tfd}{L,R} \psi_{i,L}^{\dagger} ]$ (and $\tr_{KR} [ \psi_{i,L} \ktbra{\tfd}{L,R} \psi_{i,L}^{\dagger} ]$), and the remaining combination represents the other one $\rho_{KR}$ (and $\rho_{\tilde{L}}$).  Solid green arrows in the figure correspond to $\beta/2$ Euclidean evolutions. The two insertions are separated by  Euclidean time $2\beta$ (Top), and $\beta$ (Bottom). These separations are directly related to $\tau +2\beta$ and $\tau+\beta$ appearing in (\ref{eq:correlatrKRexpanded})and (\ref{eq:correlatorTildeLexpanded}) respectively. }
    \label{fig:correlator_KR_tildeL}
  \end{minipage}
\end{figure}

We use the result to compute ``R\'enyi-2" (left-left) modular-flowed correlators \eqref{eq:correlator1001Avera} and \eqref{eq:correlator0011Avera} from the Euclidean (right-left) correlator \eqref{eq:replicaModularCorre}, by taking the limits $k\to n-1$ (and $ k\to 0$), and  $n\to 2$, then analytically continuing to the Lorentzian regime.
 We note there is a  difference between the above correlator \eqref{eq:replicaModularCorre} computed in \cite{Chandrasekaran:2022qmq} and our correlators \eqref{eq:correlator1001Avera} and \eqref{eq:correlator0011Avera}, namely that in\eqref{eq:replicaModularCorre} two fermions are living on opposite sides but in our correlators they live on the same side. In our setup, one can relate the correlator to the following diagrams (figure \ref{fig:correlator_KR_tildeL}). 

We study the correlators in the large $\beta J$ limit because their analytic expressions of are available in the limit.  One can instead work in the large $q$ limit while keeping the value of $\beta J$ finite. We will not do this here because it is the former limit where the generalization to two-dimensional CFT is straightforward \cite{WIP}.
The right hand side of \eqref{eq:correlator1001Avera} and \eqref{eq:correlator0011Avera} in the Euclidean regime are evaluated in the large  $\beta J$  and $K\ll N$ limit as 
\begin{equation}
	\begin{aligned}
		&\frac{1}{N-K}\sum_{i=1}^{N-K}\dfrac{  \mel{\tfd}{ \psi_{i,L}(\tau)  \, \left( I_{\tilde{L}} \otimes \rho_{KR}  \right) \,\psi_{i,L}(\tau')   }{\tfd} }{ \tr_{KR} \left[ \left( \rho_{KR}  \right)^{2} \right] }\\
		&= G_{2\beta}(\tau+2\beta-\tau') + 2 \frac{K}{N} \left( \mathcal{F}(\tau+2\beta,\tau';\beta,0)-\mathcal{F}_{0}(\tau+2\beta,\tau';\beta,0)  \right)+\mathcal{O}\left( \left(\frac{K}{N}\right)^{2} \right),
	\end{aligned} \label{eq:correlatrKRexpanded}
\end{equation}
\begin{equation}
	\begin{aligned}
		 &\frac{1}{N-K}\sum_{i=1}^{N-K}\dfrac{  \mel{\tfd}{   \psi_{i,L}(\tau)  \, \left( \rho_{\tilde{L}} \otimes I_{KR}  \right) \,\psi_{i,L}(\tau')   }{\tfd} }{ \tr_{KR} \left[ \left( \rho_{ KR }  \right)^{2} \right] } \\
		 &= G_{2\beta}(\tau+\beta-\tau') +2 \frac{K}{N} \left( \mathcal{F}(\tau+\beta,\tau';\beta,0)-\mathcal{F}_{0}(\tau+\beta,\tau';\beta,0)  \right)+\mathcal{O}\left( \left(\frac{K}{N}\right)^{2} \right),
	\end{aligned} \label{eq:correlatorTildeLexpanded}
\end{equation}
Here, $G_{2\beta}(\tau) $ is an Euclidean thermal SYK two point function for subsystem $\tilde{L}$ with periodicity $2\beta$, $ \mathcal{F}_{0}(\tau_{1},\tau_{2};\tau_{3},\tau_{4})$ is the connected SYK four point function, which is related to the bare one $ \mathcal{F}_{0}(\tau_{1},\tau_{2};\tau_{3},\tau_{4})$ by the so-called ladder kernel $K_{c}(\tau_{1},\tau_{2};\tau_{3},\tau_{4})$,
\begin{equation}
	 \begin{aligned}
	 	\mathcal{F}(\tau_{1},\tau_{2};\tau_{3},\tau_{4}) &= \int d\tau \,  \int d\tau' \, \frac{1}{1-K_{c}(\tau_{1},\tau_{2};\tau,\tau')} \mathcal{F}_{0}(\tau,\tau';\tau_{3},\tau_{4}),\\
	 	\mathcal{F}_{0}(\tau_{1},\tau_{2};\tau_{3},\tau_{4}) &= G_{2\beta}(\tau_{13}) G_{2\beta}(\tau_{42}) - G_{2\beta}(\tau_{14}) G_{2\beta}(\tau_{32}), \qquad \tau_{ij} =\tau_{i}-\tau_{j}, \\
	 	  K_{c}(\tau_{1},\tau_{2};\tau_{3},\tau_{4}) &=-J^{2}(q-1) G_{2\beta}(\tau_{13}) G_{2\beta}(\tau_{24}) \left( G_{2\beta}(\tau_{34}) \right)^{q-2} .	  
	 \end{aligned}
\end{equation}

In the SYK model, these two-point and four point functions are well-studied in many papers e.g., \cite{Maldacena:2016hyu,Polchinski:2016xgd,Bagrets:2017pwq,Gross:2017aos,Kitaev:2017awl,Romero-Bermudez:2019vej}. See also \cite{Trunin:2020vwy,Sarosi:2017ykf} for the review and references therein. 

The Euclidean times $\tau, \tau'$ \eqref{eq:correlatrKRexpanded} and \eqref{eq:correlatorTildeLexpanded} are continued to the  Lorentzian time with a regularization parameter $0<\delta \ll 1$;  $\tau \to -i t -\delta, \tau'\to -it+\delta $. In this way, the correlator \eqref{eq:correlatrKRexpanded} is continued to Lorentzian time as an out-of-time ordering correlator (OTOC), $\tau_{1} > \tau_{3} > \tau_{2} >\tau_{4}$, under the condition $ 1\ll \beta J  \ll N/K$.
 This correlator with the ordering is given by \cite{Maldacena:2016hyu,Trunin:2020vwy},
\begin{equation}
	\mathcal{F}(\tau_{1},\tau_{2};\tau_{3},\tau_{4}) = G_{2\beta}(\tau_{12})G_{2\beta}(\tau_{34})\frac{2\beta J}{q^{2} \pi C} \left[ 1- \frac{\pi}{2} \dfrac{ \sin\left( \dfrac{\pi}{\beta} \Delta \tau  \right) }{ \sin\left( \dfrac{\pi}{\beta} \cdot \dfrac{\tau_{12}}{2}  \right) \sin\left( \dfrac{\pi}{\beta} \cdot \dfrac{\tau_{34}}{2}\right)  } \right],
\end{equation}
where $\Delta \tau= (\tau _{1}+\tau_{2})/2-(\tau _{3}+\tau_{4})/2$, and $C$ is a constant related to an overall constant  of the Schwarzian action derived from the Schwinger-Dyson equation of the SYK model \cite{Maldacena:2016hyu,Trunin:2020vwy}.
Thus, we have the following continuation
\begin{equation}
\begin{aligned}
	\mathcal{F}(\tau+2\beta,\tau';\beta,0) \to &\mathcal{F}(-i t -\delta+2\beta,-i t +\delta;\beta,0)\\
	 &= 2G_{2\beta}(2\beta-2\delta) G_{2\beta}(\beta)\cdot \frac{ 2 \beta J}{ q^{2} \pi C} \left[ 1-\frac{\pi}{2} \frac{ \cosh \left( \frac{\pi}{\beta}t \right)}{ \sin \left( \frac{\pi \delta}{\beta} \right) } \right]\\
	&\approx - 2G_{2\beta}(2\beta-2\delta) G_{2\beta}(\beta)\cdot \frac{\beta J}{2  q^{2}  C} \cdot  \frac{ \exp \left( \dfrac{\pi}{\beta}t \right)}{ \sin \left( \dfrac{\pi \delta}{\beta} \right)  }.
\end{aligned}
\end{equation}
In particular, the correlator is exponentially growing in time. 
 On the other hand, the other correlator \eqref{eq:correlatorTildeLexpanded} is continued to Lorentzian time with the ordering $\tau_{3} > \tau_{1} > \tau_{2} >\tau_{4}$ under the condition $ 1\ll \beta J  \ll N/K$, therefore it is not OTOC.
 The correlator with the ordering $\tau_{3} > \tau_{1} > \tau_{2} >\tau_{4}$ is given by
 \begin{equation}
	\begin{aligned}
		&\mathcal{F}(\tau_{1},\tau_{2};\tau_{3},\tau_{4})\\
		& = - G_{2\beta}(\tau_{12})G_{2\beta}(\tau_{34})\frac{2\beta J}{q^{2} \pi C} \left[  \left( \dfrac{\pi\tau_{12} }{2\beta\tan\left( \dfrac{\pi}{\beta} \cdot  \dfrac{\tau_{12}}{2} \right)} + \dfrac{\pi}{ \tan\left( \dfrac{\pi}{\beta} \cdot  \dfrac{\tau_{12}}{2} \right) } -1 \right) \left( \dfrac{\pi\tau_{34} }{2\beta\tan\left( \dfrac{\pi}{\beta}\cdot  \dfrac{\tau_{34}}{2} \right)}-1 \right)   \right],
	\end{aligned}
\end{equation}
and its analytic continuation is 
\begin{equation}
	\begin{aligned}
		 &\mathcal{F}(\tau+\beta,\tau';\beta,0)\\
		 & \to  \mathcal{F}(-it -\delta+\beta,-it +\delta;\beta,0) = - 2G_{2\beta}(\beta-2\delta) G_{2\beta}(\beta)\cdot \frac{ 2 \beta J}{  q^{2}  \pi C} \left[ 1-   \left( \dfrac{\pi}{2} -  \dfrac{\pi\delta}{\beta} \right)\tan\left( \dfrac{\pi\delta}{\beta}  \right)  \right].
	\end{aligned}
\end{equation}
Clearly, this is time-independent unlike the previous case.

We do not evaluate bare four point functions $\mathcal{F}_{0}(\tau_{1},\tau_{2};\tau_{3},\tau_{4}) $ for \eqref{eq:correlatrKRexpanded} and \eqref{eq:correlatorTildeLexpanded}, because they are particular combinations of the thermal SYK two point functions with the power low behavior with respect to time, therefore they do not give dominant contributions to the correlators \eqref{eq:correlatrKRexpanded} and \eqref{eq:correlatorTildeLexpanded}. 

 Combining the above results, we can obtain the analytic expressions of the quantities  \eqref{eq:correlatrKRexpanded} and \eqref{eq:correlatorTildeLexpanded},
\begin{equation}
	\begin{aligned}
		& \ev{ \hat{d}_{\tilde{L}} }_{\beta} \cdot \mel{1}{ \mathcal{N}^{\text{SYK} \dagger}_{K,R\to T} \left[ \mathcal{N}^{ \text{SYK}}_{T\to K,R}[\ktbra{0}{T}] ] \right]}{1}\\
		&\approx \frac{1}{Z_{\delta} } \left[ G_{2\beta} (2\beta-2\delta ) -  G_{2\beta}(2\beta-2\delta) G_{2\beta}(\beta)\cdot \frac{2 \beta J}{  q^{2}  C} 
 \cdot  \frac{K}{N} \frac{ \exp \left( \dfrac{\pi}{\beta}t \right)}{ \sin \left( \dfrac{\pi \delta}{\beta} \right) }  +\mathcal{O}\left( \left(\frac{K}{N}\right)^{2} \right) \right] \\
 & \approx  \frac{G_{2\beta} (2\beta-2\delta )}{ G_{\beta}(2\delta) } \left[ 1  -  \frac{G_{2\beta}(\beta)}{\sin \left( \dfrac{\pi \delta}{\beta} \right) }\cdot \frac{2 \beta J}{ q^{2}  C} 
 \cdot  \frac{K}{N} \exp \left( \dfrac{\pi}{\beta}t \right)  \right]  ,
			\end{aligned}\label{eq:correlator1001Result}
\end{equation}
and 
\begin{equation}
	\begin{aligned}
		&\ev{ \hat{d}_{\tilde{L}} }_{\beta} \cdot  \mel{0}{ \mathcal{N}^{\text{SYK} \dagger}_{K,R\to T} \left[ \mathcal{N}^{ \text{SYK}}_{T\to K,R}[\ktbrad{0}{1}{T}] ] \right]}{1}\\
		&\approx \frac{1}{Z_{\delta} } \left[ G_{2\beta} (\beta-2\delta ) -  G_{2\beta}(\beta-2\delta) G_{2\beta}(\beta)\cdot \frac{ 8\beta J}{  q^{2}  \pi  C} 
 \cdot  \frac{K}{N} \left[ 1-   \left( \dfrac{\pi}{2} -  \dfrac{\pi\delta}{\beta} \right)\tan\left( \dfrac{\pi\delta}{\beta}  \right) \right] +\mathcal{O}\left( \left(\frac{K}{N}\right)^{2} \right) \right]\\
 & \approx  \frac{G_{2\beta} (\beta-2\delta )}{ G_{\beta}(2\delta) }  \left[ 1  -  G_{2\beta}(\beta)\cdot \frac{ 8 \beta J}{  q^{2}  \pi  C} 
 \cdot  \frac{K}{N} \left[ 1-   \left( \dfrac{\pi}{2} -  \dfrac{\pi\delta}{\beta} \right)\tan\left( \dfrac{\pi\delta}{\beta}  \right)  \right] \right],
			\end{aligned}\label{eq:correlator0011Result}
\end{equation}
where we ignored would-be sub-leading terms, coming from the replacements \eqref{eq:correlator1001Avera} and \eqref{eq:correlator0011Avera} in \eqref{eq:correlator1001} and \eqref{eq:correlator0011}, and the sub-sub-leading terms of the averaged correlators.

\begin{figure}[ht]
  \centering
  \begin{minipage}{0.8\textwidth} 
    \centering
    \includegraphics[width=\textwidth]{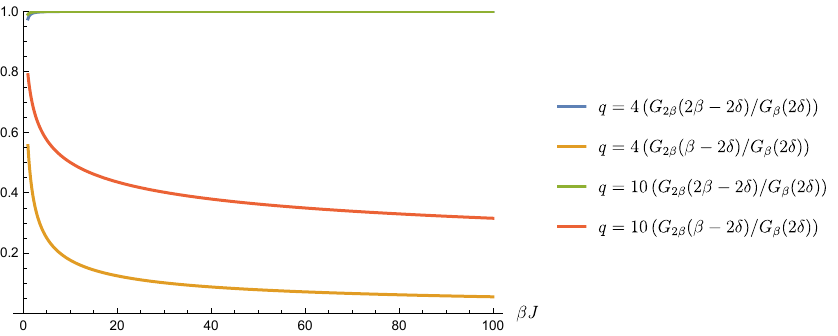}
    
    \vspace{15pt} 
    
    \includegraphics[width=\textwidth]{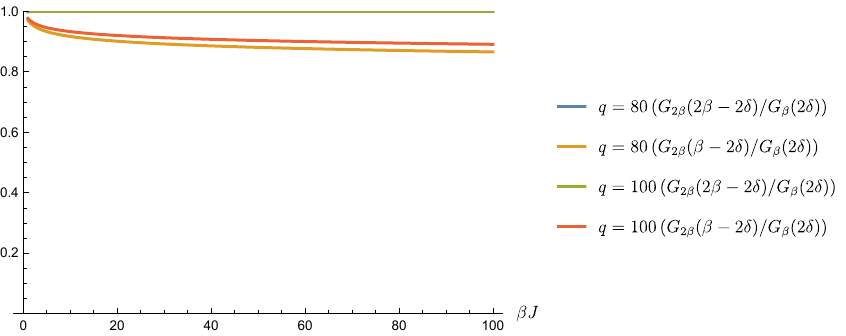}

    \caption{Plots of ratios (\ref{eq:ratio2beta}) and (\ref{eq:ratiobeta}) as a function of $\beta J$ for smaller $q$ (Top), and for larger $q$ (Bottom). Here, we set $\delta J=0.1$. For large $q$ regions, all the ratios become close to $1$. }
    \label{fig:Normaratio}
  \end{minipage}
\end{figure}

Let us consider the consequences of the above results. First, we focus on the ratios $G_{2\beta} (2\beta-2\delta ) / G_{\beta}(2\delta)  $ and $G_{2\beta} (\beta-2\delta ) / G_{\beta}(2\delta)  $ appearing in the above results. Since the SYK two point function under the conformal limit $\beta J \gg 1$ is given by  \cite{Maldacena:2016hyu},
\begin{equation}
	G_{\beta}(\tau)= b \left[ \frac{\pi}{\beta \sin \frac{\pi \tau }{\beta}} \right]^{2\Delta}, \qquad \Delta=\frac{1}{q} \quad J^{2}b^{q}\pi =\left( \frac{1}{2} -\Delta \right)\tan \pi \Delta,
\end{equation}
we can evaluate the ratios as follows
\begin{equation}
	\begin{aligned}
		\frac{G_{2\beta} (2\beta-2\delta )}{ G_{\beta}(2\delta) }= \cos^{2\Delta} \left( \frac{\pi \delta}{\beta} \right),
	\end{aligned}\label{eq:ratio2beta}
\end{equation}
and 
\begin{equation}
	\begin{aligned}
		\frac{G_{2\beta} (\beta-2\delta )}{ G_{\beta}(2\delta) }= \sin^{2\Delta} \left( \frac{\pi \delta}{\beta} \right).
	\end{aligned}\label{eq:ratiobeta}
\end{equation}
Thus, these ratios can not be $1$ simultaneously for general $\delta$ and $\beta$. However since $\Delta=1/q$  when $q$ is large these ratios are close to  $1$. We give plots of the above two functions for several $q$ in figure \ref{fig:Normaratio}.
As we can see from plots  \ref{fig:Normaratio} or directly from \eqref{eq:ratio2beta} and \eqref{eq:ratiobeta}, we need to consider a (relatively) large-$q$ regime, which implies that the SYK Majorana fermion has a small conformal dimension, $\Delta=1/q \ll 1$, in order to achieve recovery. 

One may wonder why here we take the large $q$ limit,  because the (SYK$)_{q}$ is chaotic for all $q \geq 4$ so the identities \eqref{eq:ratio1001}, \eqref{eq:ratio0011} are expected to hold for any value of $q$ in this range. Nevertheless here we have to take the large $q$ limit because  we define the code subspace using the SYK Majorana fermion operator $\psi_{i, L}$ and  the calculations of the relevant correlation functions can be possible only in the large $\beta J$ limit where the entanglement between $L$ and $R$ is weak. Because of the weakness of the entanglement, the recovery is only possible when the dimension of the operator that defines the code subspace is small, implying the necessity of taking the large $q$ limit. 

\begin{figure}[ht]
\centering
\includegraphics{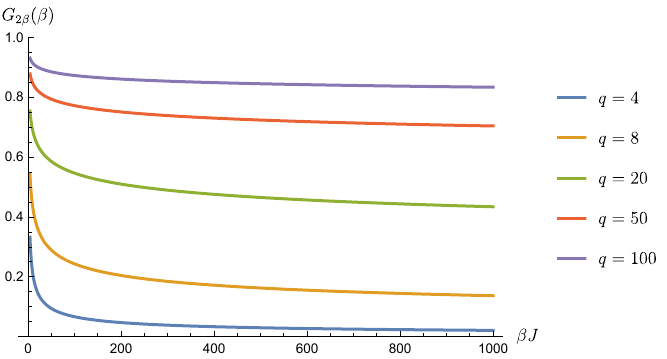}	
\caption{ Plots of the SYK two point function $ G_{2\beta} (\beta ) $, (\ref{eq:GbetaBe}) as a function of $\beta J$ for several $q=\Delta^{-1}$.}
\label{fig:g2betabeta}
\end{figure}

\begin{figure}[ht]
\centering
\includegraphics{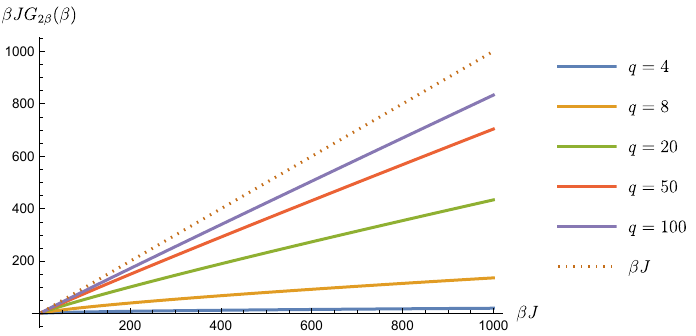}
\caption{ Plots of $ \beta JG_{2\beta} (\beta ) $ as a function of $\beta J$ for several $q=\Delta^{-1}$. The dotted line is just $\beta J$, which is equivalent to $\beta JG_{2\beta} (\beta ) $ under the $q\to \infty$ limit.}
\label{fig:betaJ_g2betabeta}
\end{figure}

Next, we consider the two point function  $ G_{2\beta} (\beta )$ appearing in the sub-leading terms. The two point function $ G_{2\beta} (\beta )$ can be written as
\begin{equation}
 G_{2\beta} (\beta ) = b \left[ \frac{\pi}{2\beta \sin \frac{\pi }{2}} \right]^{2\Delta} = \left[ \left( \frac{1}{2} -\Delta \right) \frac{ \pi \tan \pi \Delta}{(2\beta J)^{2} } \right]^{\Delta}.
		\label{eq:GbetaBe}
\end{equation}
The above expression includes $(1/\beta J)^{\Delta}$, so in $\beta J\to \infty$ limit, the SYK two point function $ G_{2\beta} (\beta )$ vanishes.
We also note the $q$-dependence of the SYK two point function.
Plots of the above function and $ \beta JG_{2\beta} (\beta ) $ for several $q=\Delta^{-1}$ are given in figure \ref{fig:g2betabeta} and \ref{fig:betaJ_g2betabeta} respectively. The plots show that as $q$ increases, the two point function $G_{2\beta} (\beta ) $ and $ \beta JG_{2\beta} (\beta ) $ take larger values.


Thus, from the above discussion, in the strict  $\beta J\to \infty$ limit\footnote{We note that to consider the perturbative expansion, we have assumed $\beta J \ll N/K$, and also implicitly assumed  $q \ll N/K$ for large $q$.
  Thus, we can not take the $\beta J\to \infty$ or $q\to \infty$ limits, unless we take the $N/K \to \infty $ limit. However, the limit $N/K \to \infty $ implies that there is almost no Hawking radiation compared to the entire Hawking radiation  $K/N \to 0$. Intuitively, in such a situation, we would not be able to recover the diary information from the Hawking radiation. }, we have $G_{2\beta}(\beta)\to 0 $, hence the second terms including $G_{2\beta}(\beta)$ in \eqref{eq:correlator1001Result} and \eqref{eq:correlator0011Result} vanish if we keep the exponential factor $\exp\left(\pi t/\beta \right)$ in \eqref{eq:correlator0011Result} fixed. Therefore, in this strict $\beta J\to \infty$ limit,
  we can not have contributions from the second terms including $G_{2\beta}(\beta)$ in \eqref{eq:correlator1001Result} and \eqref{eq:correlator0011Result}. These terms are of order $K/N$ and crucial for the following discussion.

Finally, let us focus on the time dependence of the results \eqref{eq:correlator1001Result} and \eqref{eq:correlator0011Result}.
 First, we focus on the second case \eqref{eq:correlator0011Result}. This result is time-independent at least up to the $K/N$-order, and the second term is always suppressed by the time-independent factor at the $K/N$-order, so the second terms is very small compared with the first term. This implies that the quantity \eqref{eq:correlator0011Result} is almost given by the ratio $G_{2\beta} (\beta-2\delta ) / G_{\beta}(2\delta)  $, which becomes close to $1$ when $q$ is large.

Next, we focus on  \eqref{eq:correlator1001Result}. Because of the exponential time dependent factor, this correlator has crucially different behavior as a function of time from \eqref{eq:correlator0011Result}. For early times $t  \ll 1$, the exponential in the second term can be approximated by $1$, they are similar. However, because of the exponentially growing factor, the perturbative expansion with respect to $K/N$ breaks down, similar to the fact that the perturbative calculations of   OTOCs in $1/N$ become invalid. The time scale of this break down can be estimated by equating the second term with the first term in \eqref{eq:correlator1001Result}. From the condition, we can find a critical time $t_\text{*}$\footnote{
In defining the critical time, we might have the ambiguity that which factors should be included into the critical time (or correspondingly the scrambling time), e.g., $\beta J$ and also $G_{2\beta}(\beta)$. However, as we saw before, the two point function is typically order one $G_{2\beta}(\beta)=\mathcal{O}(1)$, so we might need not to include the factor to the scrambling time. Another factor $1/\sin \left( \frac{\pi \delta}{\beta} \right)$ can be set to be $\mathcal{O}(1)$ by setting the cutoff $\delta$ suitably. For the other factor $\beta J$, since we have the condition $\beta J \ll N/K$, the factor can not give a significant contribution compared to the leading factor $N/K$, thus including the factor would be redundant. Therefore, the critical time here would be the simplest choice.},
\begin{equation}
	\frac{K}{N}\exp \left( \frac{\pi}{\beta}t_{*} \right) \sim 1 \qquad \Longrightarrow \quad t_{*} = \frac{\beta}{\pi} \log\left(\frac{N}{K} \right) = 2 t_\text{Scram},
\end{equation}
where we introduce the usual scrambling time $t_\text{Scram}$, \cite{Sekino:2008he} given by 
\begin{equation}
	t_\text{Scram}  = \frac{\beta}{2\pi }\log\left(\frac{N}{K} \right).
\end{equation}
Using this time scale, we can rewrite the correlator \eqref{eq:correlator1001Result} as 
\begin{equation}
	\begin{aligned}
		&\ev{ \hat{d}_{\tilde{L}} }_{\beta} \cdot \mel{1}{ \mathcal{N}^{\text{SYK} \dagger}_{K,R\to T} \left[ \mathcal{N}^{ \text{SYK}}_{T\to K,R}[\ktbra{0}{T}] ] \right]}{1}\\
 & \approx \frac{G_{2\beta} (2\beta-2\delta )}{ G_{\beta}(2\delta) } \left[ 1  -  \frac{G_{2\beta}(\beta)}{\sin \left( \dfrac{\pi \delta}{\beta} \right) }\cdot \frac{2\beta J}{q^{2} C} 
 \cdot  \exp \left( \frac{\lambda_{L}}{2} \left( t -2t_\text{Scram} \right) \right)  \right]  ,
			\end{aligned}\label{eq:correlator1001Lyapu}
\end{equation}
where we introduce the Lyapunov exponent $\lambda_{L}$ for a black hole with temperature $\beta$,
\begin{equation}
	\lambda_{L} = \frac{2\pi}{\beta}.
\end{equation}
Thus around the critical time, which is twice the scrambling time, we can see that the overall coefficient of $ G_{2\beta} (2\beta -2\delta )/G_{2\beta} (2\delta )$ becomes very small as usual OTOC correlators. This reproduces the expected result \eqref{eq:ratio1001} under the condition $\beta J \gg1$.

From the discussion so far, we have confirmed that the matrix elements \eqref{eq:matrixele0000}, \eqref{eq:matrixele1001} and \eqref{eq:matrixele0011} do behave as we expect them to under the condition $ 1\ll \beta J  \ll N/K$.

\section{Expected properties of the Petz-lite under the SYK dynamics} \label{sec:SYKconjecture}

So far, we have confirmed that the matrix elements we computed \eqref{eq:ratio1001}and \eqref{eq:ratio0011} reproduce our expected results under the conditions of relatively large-$q$ interaction, after the critical time $t_{*}=2 t_{{\rm Scram}}$. Additionally, of course, the following trivial matrix element is equal to $1$ by the definition,
\begin{equation}
	 \ev{ \hat{d}_{\tilde{L}} }_{\beta} \cdot \mel{0}{ \mathcal{N}^{\text{SYK} \dagger}_{K,R\to T} \left[ \mathcal{N}^{ \text{SYK}}_{T\to K,R}[\ktbra{0}{T}] ] \right]}{0}=1. \label{eq:ratio0000}
\end{equation}
Also, we can obtain the same consequences for two related matrix elements. Let us explain them. First, the matrix element \eqref{eq:ratio1001}, which becomes close to $0$, is directly related to 
\begin{equation}
	 \ev{ \hat{d}_{\tilde{L}} }_{\beta} \cdot \mel{1}{ \mathcal{N}^{\text{SYK} \dagger}_{K,R\to T} \left[ \mathcal{N}^{ \text{SYK}}_{T\to K,R}[\ktbra{0}{T}] ] \right]}{1}=\ev{ \hat{d}_{\tilde{L}} }_{\beta} \cdot \mel{0}{ \mathcal{N}^{\text{SYK} \dagger}_{K,R\to T} \left[ \mathcal{N}^{ \text{SYK}}_{T\to K,R}[\ktbra{1}{T}] ] \right]}{0}  \label{eq:ratio0110}
\end{equation}
via the definition of the adjoint channel \eqref{eq:defofAdjoint}. Thus, this matrix element also becomes close to $0$ after the critical time, and the behavior is consistent with our expectation. 

Next, for the matrix element \eqref{eq:ratio0011}, being almost equal to $1$, we have the following relation through the definition of the adjoint channel \eqref{eq:defofAdjoint} again,
\begin{equation}
	\ev{ \hat{d}_{\tilde{L}} }_{\beta} \cdot  \mel{0}{ \mathcal{N}^{\text{SYK} \dagger}_{K,R\to T} \left[ \mathcal{N}^{ \text{SYK}}_{T\to K,R}[\ktbrad{0}{1}{T}] ] \right]}{1} =\ev{ \hat{d}_{\tilde{L}} }_{\beta} \cdot  \mel{1}{ \mathcal{N}^{\text{SYK} \dagger}_{K,R\to T} \left[ \mathcal{N}^{ \text{SYK}}_{T\to K,R}[\ktbrad{1}{0}{T}] ] \right]}{0}.  \label{eq:ratio1100}
\end{equation}

Thus, although we have eight non-trivial matrix elements \eqref{eq:matElementOfRecoveElemt} that should be checked, we already know the behavior of the above five matrix elements, and there are still three matrix elements. However, since two of them are related by the complex conjugation, essentially
 we need to investigate following two matrix elements
\begin{equation}
	\begin{aligned}
		 \ev{ \hat{d}_{\tilde{L}} }_{\beta} \cdot \mel{0}{ \mathcal{N}^{\text{SYK} \dagger}_{K,R\to T} \left[ \mathcal{N}^{ \text{SYK}}_{T\to K,R}[\ktbrad{1}{0}{T}] ] \right]}{1},
	\end{aligned} \label{eq:ratio0101}
\end{equation}
and
\begin{equation}
	\begin{aligned}
		 \ev{ \hat{d}_{\tilde{L}} }_{\beta} \cdot \mel{1}{ \mathcal{N}^{\text{SYK} \dagger}_{K,R\to T} \left[ \mathcal{N}^{ \text{SYK}}_{T\to K,R}[\ktbrad{1}{1}{T}] ] \right]}{1}.
	\end{aligned} \label{eq:ratio1111}
\end{equation}
Here, the first matrix element is related to the following one
\begin{equation}
	\begin{aligned}
		 \left( \ev{ \hat{d}_{\tilde{L}} }_{\beta} \cdot \mel{0}{ \mathcal{N}^{\text{SYK} \dagger}_{K,R\to T} \left[ \mathcal{N}^{ \text{SYK}}_{T\to K,R}[\ktbrad{1}{0}{T}] ] \right]}{1} \right)^{*}= \ev{ \hat{d}_{\tilde{L}} }_{\beta} \cdot \mel{1}{ \mathcal{N}^{\text{SYK} \dagger}_{K,R\to T} \left[ \mathcal{N}^{ \text{SYK}}_{T\to K,R}[\ktbrad{0}{1}{T}] ] \right]}{0}.
	\end{aligned} \label{eq:ratio1010} 
\end{equation}

In evaluating these matrix elements, we can not directly use the technique of \cite{Chandrasekaran:2022qmq} unlike the cases for the matrix elements \eqref{eq:correlator1001} and \eqref{eq:correlator0011}\footnote{We briefly explain the reason why the evaluations of the matrix elements \eqref{eq:ratio1111} and \eqref{eq:ratio1010} are difficult. The reason is that they do not have simple expressions like \eqref{eq:correlatorKRtrace} and \eqref{eq:correlatorTildeLtrace} naively. Of course, for matrix element \eqref{eq:ratio1111}, we can consider the similar expression like \eqref{eq:correlatorKRtrace} with replacing the TFD state with the excited state $\psi_{i,L}\ketTFD{L,R}$, but in that case, we can no longer use the techniques in \cite{Chandrasekaran:2022qmq}, and we need to consider the modular operator for the excited state. For the other matrix element \eqref{eq:ratio1010}, we naively need to introduce transition matrices, not density matrices, to write it in terms of a correlator.\label{foot:diffeva}}. In the upcoming paper \cite{WIP}, we will report their results, but here we explain their expected behaviors from our obtained results. To this end, it would be useful to introduce the Kraus representation of the quantum channel \eqref{eq:hpSYKchannel},
\begin{equation}
	\mathcal{N}^{ \text{SYK}}_{T\to K,R}[\rho_{T}] = \sum_{m=1}^{ d_{\tilde{L}} } E_{m}^{\text{SYK}} \rho_{T} E_{m}^{\text{SYK}\dagger} \label{eq:krausHpSYK}
\end{equation}
given by
\begin{equation}
	E_{m}^{\text{SYK}} = \, _{\tilde{L}}\! \bra{m}U_{L} V_{T,L\to L}  \ketTFD{LR}.
\end{equation}
We can obtain this Kraus representation by introducing an orthonormal basis of the subsystem $\tilde{L}$ as $\left\{ \ket{m}_{\tilde{L}} \right\}_{m=1}^{d_{\tilde{L}} }$.
We also note that the adjoint channel \eqref{eq:hpSYKAdjointchannel} can be  written as
\begin{equation}
	\mathcal{N}^{ \text{SYK} \dagger}_{K,R \to T}[\mathcal{O}_{KR}] = \sum_{m=1}^{d_{\tilde{L}}} E_{m}^{\text{SYK}\dagger} \mathcal{O}_{KR} E_{m}^{\text{SYK}}. \label{eq:krausHpAdjointSYK}
\end{equation}

Using this Kraus representation, it is possible to extract the very important ``typical" relation from our results. Here, the ``typical" means that the relation almost does not depend on the detail of a specific state $ \ket{m}_{\tilde{L}}$ in the subsystem $\tilde{L}$, corresponding to a black hole microstate.
 First, the matrix elements \eqref{eq:ratio0000} is equal to $1$ and can be expressed as 
\begin{equation}
	\begin{aligned}
		\ev{ \hat{d}_{\tilde{L}} }_{\beta} \cdot \mel{0}{ \mathcal{N}^{\text{SYK} \dagger}_{K,R\to T} \left[ \mathcal{N}^{ \text{SYK}}_{T\to K,R}[\ktbra{0}{T}] ] \right]}{0} &= \ev{ \hat{d}_{\tilde{L}} }_{\beta}  \sum_{m,n=1}^{ d_{\tilde{L}} }  \, _{T}\mel{0}{ E_{m}^{\text{SYK} \dagger} E_{n}^{\text{SYK}} \ktbra{0}{T}  E_{n}^{\text{SYK} } E_{m}^{\text{SYK}\dagger}   }{0}_{T}\\
		&= \ev{ \hat{d}_{\tilde{L}} }_{\beta}  \sum_{m,n=1}^{ d_{\tilde{L}} }  \left|  \, _{T} \mel{0}{  E_{m}^{\text{SYK} \dagger} E_{n}^{\text{SYK}}  }{0}_{T} \right| ^{2},
	\end{aligned}
\end{equation}
and we expect the typical relation
\begin{equation}
	 \, _{T} \mel{0}{  E_{m}^{\text{SYK} \dagger} E_{n}^{\text{SYK}}  }{0}_{T}  \sim \frac{1}{ \sqrt{ d_{\tilde{L}} \cdot \ev{ \hat{d}_{\tilde{L}} }_{\beta}  }} \delta_{mn}. \label{eq:expect0EE0}
\end{equation}

Next, we focus on the matrix element \eqref{eq:ratio0011}. This matrix element is also equal to $1$, and we can express the matrix element in terms of the Kraus operators,
\begin{equation}
	\begin{aligned}
		\ev{ \hat{d}_{\tilde{L}} }_{\beta} \cdot \mel{0}{ \mathcal{N}^{\text{SYK} \dagger}_{K,R\to T} \left[ \mathcal{N}^{ \text{SYK}}_{T\to K,R}[\ktbrad{0}{1}{T}] ] \right]}{1} &= \ev{ \hat{d}_{\tilde{L}} }_{\beta}  \sum_{m,n=1}^{ d_{\tilde{L}} }  \, _{T}\mel{0}{ E_{m}^{\text{SYK} \dagger} E_{n}^{\text{SYK}} \ktbrad{0}{1}{T}  E_{n}^{\text{SYK} } E_{m}^{\text{SYK}\dagger}   }{1}_{T}.
	\end{aligned}
\end{equation}
By using the relation \eqref{eq:expect0EE0}, we extract  a similar relation,
\begin{equation}
	 \, _{T} \mel{1}{  E_{m}^{\text{SYK} \dagger} E_{n}^{\text{SYK}}  }{1}_{T}  \sim \frac{1}{ \sqrt{ d_{\tilde{L}} \cdot \ev{ \hat{d}_{\tilde{L}} }_{\beta}  }} \delta_{mn}. \label{eq:expect1EE1}
\end{equation}

Finally, the time-dependent matrix element \eqref{eq:ratio1001}, which almost vanishes around the critical time $t_{*}$, can be written as
\begin{equation}
	\begin{aligned}
		\ev{ \hat{d}_{\tilde{L}} }_{\beta} \cdot \mel{1}{ \mathcal{N}^{\text{SYK} \dagger}_{K,R\to T} \left[ \mathcal{N}^{ \text{SYK}}_{T\to K,R}[\ktbra{0}{T}] ] \right]}{1} &= \ev{ \hat{d}_{\tilde{L}} }_{\beta}  \sum_{m,n=1}^{ d_{\tilde{L}} }  \, _{T}\mel{1}{ E_{m}^{\text{SYK} \dagger} E_{n}^{\text{SYK}} \ktbrad{0}{0}{T}  E_{n}^{\text{SYK} } E_{m}^{\text{SYK}\dagger}   }{1}_{T}\\
		&= \ev{ \hat{d}_{\tilde{L}} }_{\beta}  \sum_{m,n=1}^{ d_{\tilde{L}} }  \left|  \, _{T} \mel{1}{  E_{m}^{\text{SYK} \dagger} E_{n}^{\text{SYK}}  }{0}_{T} \right| ^{2}.
	\end{aligned}
\end{equation}
From this expression, we expect the following relation and its complex conjugation,
\begin{equation}
	 \, _{T} \mel{1}{  E_{m}^{\text{SYK} \dagger} E_{n}^{\text{SYK}}  }{0}_{T}  \sim 0\label{eq:expect1EE0},
\end{equation}
around and/or after the critical time. 

Combing the above expectations, we obtain the typically expected relation\footnote{Here, we check the Knill-Laflamme condition from our obtained results. However, in principle, it would be possible to investigate the Knill-Laflamme condition directly by introducing a basis \cite{Qi:2018bje}. It would be interesting to investigate this topic.}
\begin{equation}
	 \, _{T} \mel{T}{  E_{m}^{\text{SYK} \dagger} E_{n}^{\text{SYK}}  }{T'}_{T}  \sim \frac{1}{ \sqrt{ d_{\tilde{L}} \cdot \ev{ \hat{d}_{\tilde{L}} }_{\beta}  }} \delta_{mn}\delta_{TT'} \qquad \text{ for } t \gtrsim t_{*},\label{eq:expectTEETprime}
\end{equation}
which corresponds to the Knill-Laflamme condition \cite{Knill:1996ny}.

Using this relation, the remaining matrix elements \eqref{eq:ratio0101}, \eqref{eq:ratio1111} are expected to behave as follows
\begin{equation}
	\begin{aligned}
		 \ev{ \hat{d}_{\tilde{L}} }_{\beta} \cdot \mel{0}{ \mathcal{N}^{\text{SYK} \dagger}_{K,R\to T} \left[ \mathcal{N}^{ \text{SYK}}_{T\to K,R}[\ktbrad{1}{0}{T}] ] \right]}{1} &= \ev{ \hat{d}_{\tilde{L}} }_{\beta}  \sum_{m,n=1}^{ d_{\tilde{L}} }  \, _{T}\mel{0}{ E_{m}^{\text{SYK} \dagger} E_{n}^{\text{SYK}} \ktbrad{1}{0}{T}  E_{n}^{\text{SYK} } E_{m}^{\text{SYK}\dagger}   }{1}_{T}\\
		 & \sim 0 \qquad \text{ for } t \gtrsim t_{*},
	\end{aligned} 
\end{equation}
and
\begin{equation}
	\begin{aligned}
		 \ev{ \hat{d}_{\tilde{L}} }_{\beta} \cdot \mel{1}{ \mathcal{N}^{\text{SYK} \dagger}_{K,R\to T} \left[ \mathcal{N}^{ \text{SYK}}_{T\to K,R}[\ktbrad{1}{1}{T}] ] \right]}{1}&= \ev{ \hat{d}_{\tilde{L}} }_{\beta}  \sum_{m,n=1}^{ d_{\tilde{L}} }  \, _{T}\mel{1}{ E_{m}^{\text{SYK} \dagger} E_{n}^{\text{SYK}} \ktbra{1}{T}  E_{n}^{\text{SYK} } E_{m}^{\text{SYK}\dagger}   }{1}_{T}\\
		 &\sim 1.
	\end{aligned} 
\end{equation}
These results are, of course, consistent with our original expectation \eqref{eq:matElementOfRecoveElemt}, but the discussion so far using the typical relation is indirect \eqref{eq:expectTEETprime}. Nevertheless, since this typicality is strong enough for a highly chaotic theory, we expect that nearly identical results can be obtained by direct calculations of the matrix elements \eqref{eq:ratio0101} and \eqref{eq:ratio1111}.

\section{Discussion}\label{sec:discussion}

In this paper, we studied a recovery map for the Hayden-Preskill type scrambling channel $\mathcal{N}$. We showed that 
 one can use a simplified recovery map, called Petz-lite, consisting of the adjoint channel $\mathcal{N}^{\dagger}$ with a suitable normalization factor for this purpose. We considered two examples, the Hayden-Preskill setup and the SYK model, and show that in both cases the Petz-lite indeed works as a recovery map. Also, we find that if the Petz-lite for the SYK case is used to recover information of given code subspace, it takes twice the scrambling time for the recovery. However, the SYK model case includes we did not evaluate all of the matrix elements necessary to show the recovery because of technical difficulties. Instead, we evaluate them in an indirect way in section \ref{sec:SYKconjecture}. In the upcoming paper \cite{WIP}, we will explain their results, and also some generalizations of our results. 
 
Let us discuss our results. First, we focus on the physical interpretation of the critical time given by twice the scrambling time, $t_{*}=2t_\text{Scram}$, when the matrix elements gives the input information, $\mr[\mn[\rho]]\sim \rho$. It was argued in \cite{Sekino:2008he} that information of a diary thrown into a black hole appears after the scrambling time. This means that, after the scrambling time, the HP scrambling  channel $\mathcal{N}$  maps the diary information to Hawking radiation completely. However, even if the diary information appears in the Hawking radiation, it is difficult to get it directly since the information is uniformly embedded into the Hawking radiation. To extract the information, we need a recovery operation given by the Petz-lite $\mathcal{R}\sim \mathcal{N}^{\dagger}$. Since it is the adjoint of the HP channel
$\mathcal{N}$, it again takes the scrambling time to apply the recovery map. Thus, in total, we need to wait for twice the scrambling time for the identity \eqref{eq:matElementOfRecoveElemt} to get satisfied.

Next, let us explain the bulk interpretation of our results\footnote{We note that since  currently there is no clear understanding of a dual gravitational theory  for a subset of the SYK Majorana fermions (or Majorana spin chain), we can not check the interpretation using the gravity side explicitly at least in the context of NAdS$_{2}$/NCFT$_{1}$ context. However, there are several proposals for such a gravitational treatment, e.g., in \cite{Chandrasekaran:2022qmq}. One would be able to use them to check the bulk interpretation.}. The bulk interpretation comes from the island prescription \cite{Penington:2019npb,Almheiri:2019psf}. 
First, the Hayden-Preskill setup concerns post-Page time regimes. In these regimes, there is an island, which is a non-trivial entanglement wedge of Hawking radiation in the black hole interior. Thus, if one throws a diary into a black hole and waits for the scrambling time, then the diary enters the island region, implying that the diary is encoded into the Hawking radiation in a very complicated way. The mechanism that the thrown diary is encoded into the Hawking radiation corresponds to our quantum channel $\mathcal{N}$. To recover the diary information from the Hawking radiation, we need to consider the recovery operation corresponding to the map $\mathcal{R}\sim \mathcal{N}^{\dagger}$. The recovery map is given by the adjoint channel of the quantum channel $\mathcal{N}$. In the bulk side, the action of the adjoint channel $\mathcal{N}^{\dagger}$ means that the ``reverse" process of the original quantum channel $\mathcal{N}$\footnote{Here, we note that in these two processes, we need to use two different (remaining) black holes since, in defining the quantum channel, (remaining) black holes are treated as internal degrees of freedom of the quantum channel.}.   More precisely, the ``reverse" process is given as follows: First, we start from the output state provided by the action of the quantum channel $\mathcal{N}$, implying the diary is located on the island at some time slice $\Sigma$. The application of the adjoint channel $\mathcal{N}^{\dagger}$ then is interpreted as replacing the future of this time slice $\Sigma $  by a white hole.  Because of the replacement, the diary on the island region of the original black hole is coming out from the horizon of the white hole. 
Here, the reason why the white hole appears is that the adjoint channel includes the Hermit conjugation of unitaries $U$ (and $U^{\dagger}$) compared to the quantum channel $\mathcal{N}$.  Thus, the diary thrown into the black hole reappears from the white hole induced by $\mathcal{N}^{\dagger}$. This bulk interpretation is consistent with the critical time. This is because, after throwing the diary, it takes the scrambling time for the diary to enter the island region, and in the ``reverse" process, it would also take the scrambling time for the diary to go outside the island region and the horizon.

Finally, we end with discussing some of our in-progress works and future directions:

\paragraph{Analysis in high temperature regime, \texorpdfstring{$\beta J \ll 1$}{beta J }}

In this paper, we have focused on the large $\beta J$ limit (low-temperature limit) in the SYK model to make the calculation analytic and for the purpose of the generalization to the CFT $_{2}$ case.  In the limit, we can use emergent conformal symmetry of the SYK model and also we would be able to use semi-classical intuition of the dual Jackiw-Teitelboim gravity, but we have a relatively weak initial entangle state $\ketTFD{L,R}$ between the left and right SYK systems. Due to this weak entangle state, we would require some conditions to consider successful recovery protocol, e.g., large-$q$ regime. Thus, analysis without taking the large $\beta J$ limit would be interesting. In that case, we would need to consider numerical approaches.

\paragraph{Direct bulk analysis and relation to other protocols}

In this paper, we studied the recovery protocol from the boundary CFT perspective. One would be able to consider corresponding bulk computations. Also, it would be interesting to figure out the relation between other proposed protocols e.g., \cite{Gao:2019nyj,Brown:2019hmk,Schuster:2021uvg,Nezami:2021yaq} and ours\footnote{For  such protocols, one can characterize protocol by computing ``price", ``distance", etc. \cite{Pastawski:2016qrs,Chandrasekaran:2022qmq,Bentsen:2023xlu}. One would be able to find the relation between our results and such quantities.}.

\paragraph{Generalization to (Holographic) CFT$_{2}$ and other systems} While this paper focuses on the SYK model, which is a $0+1$-dimensional quantum system, it can also be interpreted as a spin chain with $q$-body SYK interactions. Thus, we can interpret that the SYK model has a spatial direction effectively.  As a result, we expect that a similar analysis can be applied to a two-dimensional CFT exhibiting chaos, e.g., holographic CFT$_{2}$. Indeed, one of the Hayden-Preskill setups in a two-dimensional holographic CFT is introduced in \cite{Chandrasekaran:2021tkb}.

Also, there are other possibilities for generalizations to other systems exhibiting chaos. For example, studying the Petz-lite in a chaotic spin chain would be interesting.

\paragraph{Chaotic-Integrable transition} In this paper, the chaotic nature is important for the simplification of the Petz map to the Petz-lite. Thus, if a system do not exhibit the chaotic nature, in other words, the system is integrable, then the Petz-lite (also the original Petz map) is not expected to works correctly. This is because, in an integrable system, the decoupling condition is not expected to hold.  In the framework of the SYK model, we can prepare integrable and non-integrable (chaotic) situation by adding two-body interaction \cite{Garcia-Garcia:2017bkg}. Using the setup, we would be able to study Petz-lite.

\paragraph{Higher dimensional code sub-space?} The SYK version of the HP setup studied in this paper treats the two-dimensional code sub-space spanned by the vacuum and the excited state. However, in a more realistic situation, one needs to deal with code sub-spaces with dimensions greater than two. For example, the interior of a black hole, when it is viewed as a code subspace embedded into the Hawking radiation, the dimension of its Hilbert space has to be large enough to accommodate a part of the semi-classical QFT degrees of freedom to have a geometric interpretation of the black hole interior\footnote{Of course, the interior degrees of freedom may appear to be infinite, but almost all of them can not contribute due to post-selection \cite{Akers:2022qdl}. Even in that case, there can be degrees of freedom with Bekenstein-Hawking entropy.}. To this end, one would need  to consider a more complicated embedding involving for example states like, $\psi_{i,L}\psi_{j\neq i,L}\ket{\tfd}_{L,R}$. In that case, we can evaluate corresponding matrix elements in principle, but it would be difficult to them analytically since we encounter higher-point functions.
 
 Another possibility for higher dimensional code sub-space is to consider a random embedding and the double-scaling limit. For example, we might be able to use the state $\kappa_{ij}\psi_{i,L}\psi_{j,L}\ket{\tfd}_{L,R}$, where  $\kappa_{ij}$ is random like observables in the double-scaled SYK model \cite{Berkooz:2018jqr}. In this case, by taking the double-scaling limit and using chord diagram techniques, we might be able to evaluate the resulting matrix element analytically.  Also, this might open up an interesting connection between QEC in the SYK model and recent discussions of the von Neumann algebra of quantum gravity, in particular, \cite{Lin:2023trc}.

\section*{Acknowledgments}
We thank Yoshifumi Nakata for discussions. AM thanks Norihiro Iizuka, Tomoki Nosaka, Masahiro Nozaki and Jia Tian for comments. AM also thanks Chen Bai for related discussions.
AM thanks the workshop ``Beijing Osaka String/Gravity/Black Hole Workshop” at KITS, where this work was presented. AM also thanks  the long-term work shop ``Quantum Information, Quantum Matter and Quantum Gravity” YITP-T-23-01 at YITP, where this work was also presented. YN was supported by JST, the establishment of university fellowships towards the creation of science technology innovation, Grant Number JPMJFS2123. TU was supported in part by JSPS Grant-in-Aid for Young Scientists 19K14716 and in part by MEXT KAKENHI Grant-in-Aid for Transformative Research Areas~A ``Extreme Universe'' No.21H05184.

\appendix

\section{Derivation of the Petz lite using Kraus representation}
\label{app:Kraus}

In this appendix, we derive the Petz-lite with a different normalization factor based on paper \cite{barnum2000reversing}. See e.g., \S 10.3 of \cite{nielsen_chuang_2010} for related reviews.

We start with the Kraus representation of the HP channel \eqref{eq:hpChannel}. The Kraus representation can be introduced by expressing the trace as 
\begin{equation}
	\begin{aligned}
		\mathcal{N}_{T\to D,B}\left[ \rho_{T} \right] &= \tr_{C} \left[ (U_{T,A \to C,D}\otimes I_{B}) ( \rho_{T} \otimes \ktbra{\epr}{A,B}  )  (U_{T,A \to C,D}^{\dagger} \otimes I_{B}) \right] \\
		&=\sum_{m=1}^{d_{C}} \,_{C}\mel{m}{(U_{T,A \to C,D}\otimes I_{B}) \ket{\epr}_{A,B} \, \rho_{T}   \, _{A,B}\bra{\epr} (U_{T,A \to C,D}^{\dagger} \otimes I_{B}) }{m}_{C}\\
		&=\sum_{m=1}^{d_{C}} E_{m} \rho_{T} E_{m}^{\dagger},
	\end{aligned}
\end{equation}
where $\ket{m}_{C}$ is an orthonormal basis of subsystem $C$, and $E_{m}$ is the Kraus operator defined by
\begin{equation}
	E_{m}= \,_{C}\bra{m}(U_{T,A \to C,D}\otimes I_{B}) \ket{\epr}_{A,B}.\label{eq:hpKraus}
\end{equation}
Here, we note that since the state $\ket{m}_{C}$ is a basis state of the remaining black hole $C$. We also note that the adjoint HP channel is expressed in terms of the Kraus operators,
\begin{equation}
	\mathcal{N} \left[\mathcal{O}  \right] = \sum_{m=1}^{d_{C}} E_{m}^{\dagger} \mathcal{O}  E_{m}.
 \end{equation}

Using this Kraus operator, let us investigate the Knill-Laflamme condition \cite{Knill:1996ny},
\begin{equation}
	P_{code}E_{m}^{\dagger} E_{n} P_{code}=\alpha_{mn} P_{code} \quad \left(\alpha_{mn} = \alpha_{nm}^{*} \in \mathbb{C} \, \text{with } \sum_{m=1}^{d_{C}}\alpha_{mm}=1 \right), \,  \text{for } \forall m,n=1,\cdots, d_{C}.
\end{equation}
where $P_{code}$ is a projection operator onto a code subspace in general, but in our setup, $P_{code}$ is assumed to be just given by the identity operator $P_{code}=I_{T}$, since all input states should be recoverable under the Hayden-Preskill setup.
 If this condition holds, we can construct a recovery map\footnote{See e.g.,  \S 10.3, in particular, theorem 10.1, of \cite{nielsen_chuang_2010} for the review.}.
 
 Under Haar random averaging, we can easily evaluate the Knill-Laflamme condition from the expression \eqref{eq:hpKraus} and Haar average \eqref{eq:HaarFirstMoment},
 \begin{equation}
 	\begin{aligned}
 		\overline{ E_{m}^{\dagger} E_{n} }= \frac{1}{d_{C}} \delta_{mn} I_{T}
 	\end{aligned}\label{eq:HaarAveKLcondi}
 \end{equation}
This result appears to imply that the Knill-Laflamme condition holds \textit{always} under the averaging, but this is not correct. This is because, even if the Knill-Laflamme condition is satisfied, higher moments of the Knill-Laflamme condition, e.g., $\overline{\left|P_{code}E_{m}^{\dagger} E_{n} P_{code} \right|^{2}} $,  might not hold due to contributions coming from Weingarten calculus.  We can see their contributions by directly evaluating the second moment\footnote{See also \cite{Liu:2020sqb} for related discussions.}, 
\begin{equation}
	 \begin{aligned}
	 	\overline{ \left|P_{code}E_{m}^{\dagger} E_{n} P_{code} \right|^{2}}\approx \frac{1}{(d_{C})^{2}}\cdot I_{T}\left[ \delta_{mn} +\frac{d_{T}}{ d_{D}d_{B} }  \right]
	 \end{aligned}\label{eq:HaarAveSecondKL}
\end{equation}
where we used the know result \eqref{eq:HaarSecondMoment} with large-$d$  approximation. 
Thus, when we do not have enough Hawking radiation $D,B$ compared to the diary $T$, that is, $d_{D}d_{B} \gtrsim d_{T}$, we can not ignore the second term, implying the break down of the Knill-Laflamme condition. On the other hand, in the opposite limit $d_{D}d_{B} \gg d_{T}$, where we have enough Hawking radiation, we can ignore the second term, and we get the Knill-Laflamme condition. We note that this is consistent with the decoupling condition \eqref{eq:decopHP}, since the unitarity means the relation
\begin{equation}
	\frac{d_{T}}{ d_{D}d_{B} } =\frac{1}{d_{C}}\cdot \left(\frac{d_{T}}{d_{D}}\right)^2,
\end{equation}
and the factor $(d_{T}/d_{D})^{2}$ gives an upper bound of the decoupling condition \eqref{eq:decopHP}. 

Next, we construct a recovery map for the HP quantum channel. With the Knill-Laflamme condition in mind, we consider the following map, which is equal to the adjoint HP channel up to the overall factor $d_{C}$,
\begin{equation}
	\mathcal{R}[\mathcal{O}]\coloneqq d_{C} \sum_{m=1}^{d_{C}} E^{\dagger}_{m} \mathcal{O} E_{m} =  d_{C}\mathcal{N}^{\dagger}\left[\mathcal{O}\right]. \label{eq:RecoverConstuct}
\end{equation}
Under the Haar random average, this map gives
\begin{equation}
	\begin{aligned}
		\overline{\mathcal{R}[\mathcal{N}\left[ \rho_{T} \right]]}&= d_{C} \sum_{m,n=1}^{d_{C}}  \overline{ E^{\dagger}_{m} E_{n}  \rho_{T}   E^{\dagger}_{n} E_{m}}\\
		&\approx d_{C} \sum_{m,n=1}^{d_{C}} \left[   \overline{ E^{\dagger}_{m} E_{n}}  \rho_{T} \overline{  E^{\dagger}_{n} E_{m}} + \overline{ E^{\dagger}_{m}  \overline{  E_{n} \rho_{T}  E^{\dagger}_{n} } E_{m} }    \right]\\
		&= d_{C} \sum_{m,n=1}^{d_{C}}  \frac{1}{(d_{C})^{2}} \left[ \delta_{mn}\, \rho_{T} +  \frac{\tr\left[ \rho_{T}\right] }{ d_{D}d_{B} }  I_{T}  \right]\\
		&= \rho_{T} + \left( \frac{d_{T}}{d_{D}} \right)^{2}\cdot \frac{1}{d_{T}} I_{T}, \label{eq:recoveredResult}
	\end{aligned}
\end{equation}
where in the second line we used the fact that in the large-Hilbert space dimension limit, Weingarten calculus reduces to Wick calculus, and in the final line, we used $\tr\rho_{T}=1$ and the relation $d_{T}d_{B}=d_{C}d_{D}$. In the third line, we encountered the Knill-Laflamme condition for the first term \eqref{eq:HaarAveKLcondi}, and the second terms disturb the Knill-Laflamme condition. These two terms in the third line correspond to the first and second terms in \eqref{eq:HaarAveSecondKL}. 
 Thus, under the situation $d_{B}d_{D} \gg d_{T}$ where the Knill-Laflamme condition holds (approximately), we can ignore the second term of the above result, implying that the map\eqref{eq:RecoverConstuct} works as a recovery map. This is a quantum information theoretic derivation of the Petz-lite. However, we note that the recovery map here is little bit different from the one \eqref{eq:petzLiteFullExpres} up to the overall factor, but the difference almost vanishes when the condition $d_{B}d_{D} \gg d_{T}$ is satisfied. 

Finally, we end this appendix by giving the connection between the Petz map and the Petz-lite in terms of the Kraus operator and the Knill-Laflamme condition. Generally, since the coefficients $(\alpha_{mn})$ is Hermitian, we can diagonalize the Knill-Laflamme condition by some unitary $(U_{mn})$ as follows \cite{nielsen_chuang_2010},
\begin{equation}
	P_{code}F_{m}^{\dagger} F_{n} P_{code}=\lambda_{m}\delta_{mn} P_{code} \quad \left(\lambda_{m}  \in \mathbb{R}, \text{with } \sum_{m=1}^{d_{C}}\lambda_{m}=1,\, \lambda_{m}> 0
	\right),\, \text{for } \forall m,n=1,\cdots, d_{C},
\end{equation}
where $F_{m}=\sum_{n}U_{mn}E_{n}$ is the newly defined Kraus operator. Using this Kraus operator, one can define the following map
\begin{equation}
	\mathcal{R}[\mathcal{O}]\coloneqq  \sum_{m=1}^{d_{C}} \frac{1}{\lambda_{m}}  P_{code} F^{\dagger}_{m} \mathcal{O} F_{m}  P_{code}.\label{eq:PetzKraus}
\end{equation}
This map can be also expressed in terms of the original quantum channel with introducing some full rank reference state $\sigma$ as follows \cite{barnum2000reversing}
\begin{equation}
	\mathcal{R}[\mathcal{O}] = \sigma^{1/2} \mathcal{N}^{\dagger} \left[ \left(\mathcal{N}[\sigma]\right)^{-1/2} \mathcal{O}   \left(\mathcal{N}[\sigma]\right)^{-1/2}  \right]\sigma^{1/2},
\end{equation}
and this is exactly the Petz map. In the recovery map \eqref{eq:PetzKraus}, the factor $\lambda_{m}$ prevents us from directly giving the adjoint channel $\mathcal{N}^{\dagger}$, and we need to introduce the curios factors $\left(\mathcal{N}[\sigma]\right)^{-1/2}$ and $\sigma^{1/2}$. However, for the case where $\lambda_{m}=1/d_{C}\, (m=1,\cdots,d_{C})$, one can consider the map \eqref{eq:RecoverConstuct} instead of the above map. As we have seen, the Haar random case with the Knill-Laflamme condition \eqref{eq:HaarAveKLcondi} is certainly this case.

\section{Operator Transpose for the EPR state} \label{app:opetransEPR}

In this appendix, we derive the relation \eqref{eq:opetransEPR} algebraically. We can show the relation directly as follows;
\begin{equation}
	\begin{aligned}
		&U^{T}_{ C',D' \to B,T'} \ket{\epr}_{C,C'} \otimes  \ket{\epr}_{D,D'} \\
		&=  \left( I_{C} \otimes I_{D} \otimes U^{T}_{ C',D' \to B,T'}  \right) \ket{\epr}_{C,C'} \otimes  \ket{\epr}_{D,D'} \\
		&= \frac{1}{ \sqrt{ d_{C} d_{D} } } \sum_{\tilde{C}=1}^{d_{C}} \sum_{\tilde{D}=1}^{d_{D}} \ket{\tilde{C},\, \tilde{D}}_{C,D} \otimes \left(U^{T}_{ C',D' \to B,T'}  \ket{\tilde{C},\tilde{D}}_{C',D'}  \right) \\
		&= \frac{1}{ \sqrt{ d_{C} d_{D} } } \sum_{\tilde{C} =1}^{d_{C}} \sum_{\tilde{D}=1}^{d_{D}} \sum_{\tilde{B}=1}^{d_{B}} \sum_{\tilde{T}=1}^{d_{T}} \ket{\tilde{C},\, \tilde{D}}_{C,D} \otimes \ket{\tilde{B},\tilde{T}}_{B,T'}  \cdot \Big. _{B,T'} \! \bra{\tilde{B},\tilde{T}} U^{T}_{ C',D' \to B,T'}  \ket{\tilde{C},\tilde{D}}_{C',D'} \\
		&= \frac{1}{ \sqrt{ d_{C} d_{D} } } \sum_{\tilde{C} =1}^{d_{C}} \sum_{\tilde{D}=1}^{d_{D}} \sum_{\tilde{B}=1}^{d_{B}} \sum_{\tilde{T}=1}^{d_{T}} \ket{\tilde{C},\, \tilde{D}}_{C,D} \otimes \ket{\tilde{B},\tilde{T}}_{B,T'}  \cdot \Big. _{C,D} \! \bra{\tilde{C},\tilde{D}} U_{ A,T \to C,D}  \ket{\tilde{B},\tilde{T}}_{A,T} \\
		&= \frac{1}{ \sqrt{ d_{B} d_{T} } } \sum_{\tilde{B}=1}^{d_{B}} \sum_{\tilde{T}=1}^{d_{T}} \left( U_{ A,T \to C,D}  \ket{\tilde{B},\tilde{T}} _{A,T} \right)  \otimes \ket{\tilde{B},\tilde{T}}_{B,T'} \\
		& = \left( U_{A,T \to C,D} \otimes I_{B} \otimes I_{T'} \right) \ket{\epr}_{A,B} \otimes \ket{\epr}_{T,T'} \\
		& = U_{A,T \to C,D}  \ket{\epr}_{A,B} \otimes \ket{\epr}_{T,T'},
	\end{aligned}
\end{equation}
where in the fifth equality, we used the unitarity condition of the Hilbert space dimensions $d_{T}\,d_{B}=d_{C}\,d_{D}$.
\qed

\begin{figure}
\vspace{-1cm}
    \begin{minipage}[b]{0.45\linewidth}
      \centering
        \includegraphics[scale=0.3]{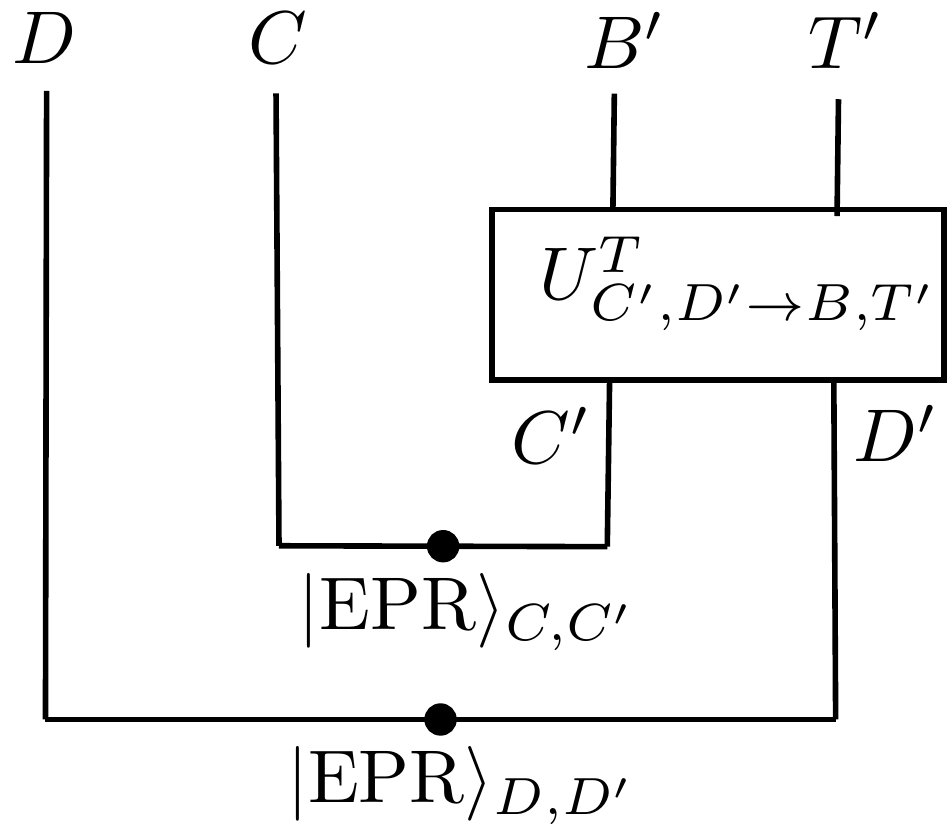}
    \end{minipage}
    \hspace{1em}
    \begin{minipage}[b]{0.45\linewidth}
      \centering
        \includegraphics[scale=0.3]{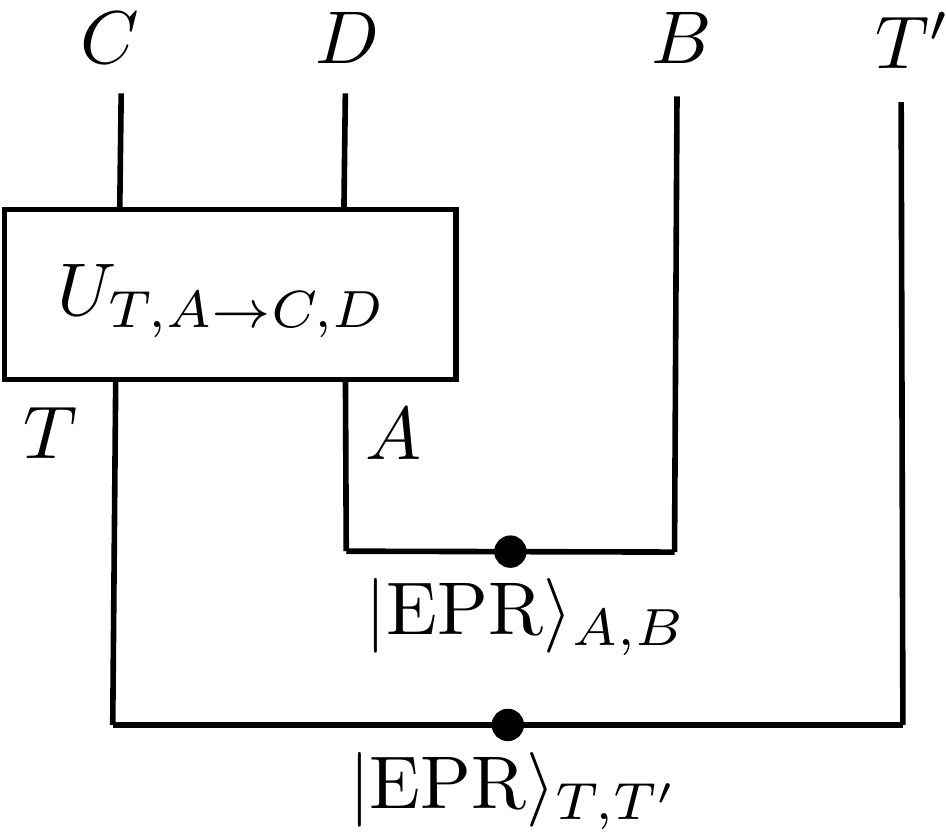}
    \end{minipage}
\vspace{-3mm}
\caption{Diagrams representing left and right hand sides of the relation (\ref{eq:opetransEPR}). {\bf Left}: The left hand side of the relation.
 {\bf Right}: The right hand side of the relation. The left and right diagrams are equivalent. }
\label{fig:diagramEquivEPR}   
\end{figure}

The above relation implies that the left and right diagrams in figure \ref{fig:diagramEquivEPR} are equivalent.

\section{Convention in the SYK Hayden-Preskill protocol}\label{app:conventionSYKHP}

In this appendix, we gather some important definitions and conventions which we use in section \ref{sec:SYKHP}.

\paragraph{Majorana SYK fermions}
\begin{itemize}
	\item Anti-commutation relation
		\begin{equation*}
			\left\{ \psi_{i},\psi_{j}  \right\} = 2\delta_{ij}
		\end{equation*}
	\item The unitary time evolution operator
		\begin{equation*}
			U_{\alpha} = U_{\alpha}(t) = \exp\left( i t H_{\alpha} \right) \quad \left( \alpha =L,R \right)
		\end{equation*}
	\item Positive direction of time evolutions in left and right SYK systems (in Lorentzian signature)
		\begin{equation*}
			 \psi_{i,L}(t) \equiv  U_{L}  \psi_{i,L} (0) U_{L}^{\dagger} = e^{itH_{L}} \psi_{i,L} (0) e^{-itH_{L}},
		\end{equation*}
		\begin{equation*}
			 \psi_{i,R}(t) \equiv  U_{R}^{\dagger}  \psi_{i,R} (0) U_{R} = e^{-itH_{R}} \psi_{i,R} (0) e^{itH_{R}},
		\end{equation*}
		which can be written as 
		\begin{equation}
			\psi_{i,\alpha} (t) =\Delta_{L}^{-i\frac{t}{\beta}} \psi_{i,\alpha}(0)\Delta_{L}^{i\frac{t}{\beta}} = \Delta_{R}^{i\frac{t}{\beta}} \psi_{i,\alpha}(0)\Delta_{R}^{-i\frac{t}{\beta}}  \quad (\alpha= L,R),
		\end{equation}
		where $\Delta_{L}=\Delta_{R}^{-1}$ is the modular operator defined by
		\begin{equation}
			\Delta_{L} = \rho_{L} \otimes \rho_{R}^{-1} = e^{-K_{L}} \otimes e^{K_{R}} = e^{-(K_{L}-K_{R})}, \qquad K_{\alpha} \equiv \beta H_{\alpha} \quad (\alpha= L,R).
		\end{equation}
		Here $\rho_{\alpha} \, (\alpha= L,R)$ is defined by
		\begin{equation}
			\rho_{L} = \tr_{R} \left[ \ktbra{\tfd}{L,R} \right], \quad \rho_{R} = \tr_{L} \left[ \ktbra{\tfd}{L,R} \right]
		\end{equation}
		In the Euclidean signature, one can rewrite the above formal formula as
			\begin{equation}
			\psi_{i,\alpha} (\tau) =\Delta_{L}^{\frac{\tau}{\beta}} \psi_{i,\alpha}(0)\Delta_{L}^{-\frac{\tau}{\beta}}\quad (\alpha= L,R), \label{eq:eucliSYKmajo}
		\end{equation}
		and recover the Lorentzian operator by the analytic continuation $\tau \to -it$.
		
		\item Euclidean regularization parametrized by the cutoff $\delta$
			 \begin{equation}
 				\psi_{i,L}(t+i\delta) \equiv e^{i(t+i\delta)H_{L}} \psi_{i,L}(0) e^{-i(t+i\delta)H_{L}} =e^{(-\delta +it) H_{L}}  \psi_{i,L}(0) e^{(\delta-it) H_{L}}
		 \end{equation}
		 This regularized operator is related to the Euclidean evolved operator \eqref{eq:eucliSYKmajo} by continuation $\tau \to -it +\delta$.
		 	
\end{itemize}

\paragraph{SYK Hayden-Preskill channel}
\begin{itemize}
	\item SYK Hayden-Preskill channel \eqref{eq:hpSYKchannel}
		\begin{equation*}
	\mathcal{N}^{ \text{SYK}}_{T\to K,R}[\rho_{T}] \coloneqq \tr_{\tilde{L}} \left[ U_{L} V_{T,L\to L} \left( \rho_{T} \otimes \ktbra{\tfd}{L,R}  \right)  V_{T,L\to L}^{\dagger}  U_{L}^{\dagger} \right]. \label{}
	\end{equation*}
	
	\item Adjoint SYK Hayden-Preskill channel \eqref{eq:hpSYKAdjointchannel}
		\begin{equation*}
	\begin{aligned}
		\mathcal{N}^{ \text{SYK} \dagger}_{K,R\to T}[ \mathcal{O}_{KR}] &\coloneqq \tr_{L,R} \left[ \ktbra{\tfd}{L,R}  \left(  V_{L\to T,L}^{\dagger}   U_{L}^{\dagger} \,   \mathcal{O}_{KR} \,   U_{L} V_{L\to T,L} \right)\right]\\
		& = \, _{L,R} \ev{ \left(  V_{L\to T,L}^{\dagger}   U_{L}^{\dagger} \,   \mathcal{O}_{KR} \,   U_{L} V_{L\to T,L} \right) }{\tfd}_{L,R}
	\end{aligned}
	\end{equation*}
\end{itemize}

\section{Derivation of correlator from quantum channels}\label{app:derivationChanneltoCorrelator}

In this appendix, we give the derivation of the relation \eqref{eq:correlator1001} and \eqref{eq:correlator0011}. We can derive the relation graphically, but below we give an algebraic derivation of the relation.

We start with the derivation of the relation  \eqref{eq:correlator1001}, which can be obtained straightforwardly from the definition of the quantum channels \eqref{eq:hpSYKchannel} and \eqref{eq:hpSYKAdjointchannel}. We first note that, from the definition of the quantum channel \eqref{eq:hpSYKchannel}, the state $\ktbra{0}{T}$ is mapped to 
\begin{equation}
	\begin{aligned}
		\mathcal{N}^{ \text{SYK}}_{T\to K,R}[\ktbra{0}{T}] ]&= \tr_{\tilde{L}} \left[ U_{L} \ktbra{\tfd}{L,R} U_{L}^{\dagger}   \right]\\
		& = U_{R}\, \rho_{KR}\,U_{R}^{\dagger},
	\end{aligned}
\end{equation}
where we used the fact that $(H_{L}-H_{R})\ketTFD{L,R}$ leading to $U_{L}\ketTFD{L,R}=U_{R}\ketTFD{L,R}$, and $\rho_{KR}$ is defined by \eqref{eq:defOfDensityMatrix}.
For this density matrix, we consider the action of the adjoint channel \eqref{eq:hpSYKAdjointchannel}, and take the following matrix element;
\begin{equation}
	 \ev{ \hat{d}_{\tilde{L}} }_{\beta} \cdot \mel{1}{ \mathcal{N}^{\text{SYK} \dagger}_{K,R\to T} \left[ \mathcal{N}^{ \text{SYK}}_{T\to K,R}[\ktbra{0}{T}] ] \right]}{1} = \frac{ \mel{1}{ \mathcal{N}^{\text{SYK} \dagger}_{K,R\to T} \left[ \, U_{R}\, \rho_{KR}\,U_{R}^{\dagger} \,  \right]}{1}  }{ \tr \left[ \left(\rho_{KR}\right)^{2} \right] },
\end{equation}
where we used the definition \eqref{eq:defOfEffectiveDim}. Using the definition \eqref{eq:hpSYKAdjointchannel}, we can evaluate the denominator as
\begin{equation}
	\begin{aligned}
		 &\mel{1}{ \mathcal{N}^{\text{SYK} \dagger}_{K,R\to T} \left[  \,  U_{R}\, \rho_{KR}\,U_{R}^{\dagger} \,  \right]}{1}\\
		  &=  \big( \, _{L,R} \bra{\tfd}  \otimes \, _{T}\bra{1} \big){ \left(  V_{L\to T,L}^{\dagger}   U_{L}^{\dagger} \,    U_{R}\, \rho_{KR}\,U_{R}^{\dagger} \,   U_{L} V_{L\to T,L} \right) } \big(\ket{\tfd}_{L,R}\otimes \ket{1}_{T}\big)\\
		 &=\frac{1}{Z_{\delta}}  \cdot  \, _{L,R} \bra{\tfd}{ \left(  \psi_{i,L}^{\dagger}(-i\delta) \,   U_{L}^{\dagger} \,    U_{R}\, \rho_{KR}\,U_{R}^{\dagger} \,   U_{L}\,  \psi_{i,L}(i\delta)  \right) } \ket{\tfd}_{L,R}\\
		 & = \frac{1}{Z_{\delta}}  \cdot \, _{L,R} \bra{\tfd}{ \left( U_{R}\psi_{i,L}^{\dagger}(-i\delta) \,   U_{L}^{\dagger} \,     \rho_{KR} \,   U_{L}\,  \psi_{i,L}(i\delta)\,  U_{R}^{\dagger} \right) } \ket{\tfd}_{L,R}\\
		 &= \frac{1}{Z_{\delta}}  \cdot   \, _{L,R} \bra{\tfd}{ \left( U_{L} \psi_{i,L}^{\dagger}(-i\delta) \,   U_{L}^{\dagger} \,     \rho_{KR} \,   U_{L}\,  \psi_{i,L}(i\delta)\,  U_{R}^{\dagger} \right) } \ket{\tfd}_{L,R}\\
		 &= \frac{1}{Z_{\delta}}  \cdot   \, _{L,R} \bra{\tfd}{ \left(  \psi_{i,L}^{\dagger}(t-i\delta) \,       \rho_{KR} \,  \psi_{i,L}(t+i\delta) \right) } \ket{\tfd}_{L,R},
	\end{aligned}
\end{equation}
where in the 4-th equality, we used the relation $U_{L}\ketTFD{L,R}=U_{R}\ketTFD{L,R}$. 
Thus, by combining the above expressions, we obtain the relation  \eqref{eq:correlator1001},
\begin{equation*}
	\begin{aligned}
		&\ev{ \hat{d}_{\tilde{L}} }_{\beta} \cdot \mel{1}{ \mathcal{N}^{\text{SYK} \dagger}_{K,R\to T} \left[ \mathcal{N}^{ \text{SYK}}_{T\to K,R}[\ktbra{0}{T}] ] \right]}{1} =  \frac{1}{Z_{\delta}}\cdot \dfrac{  \mel{\tfd}{ \psi_{i,L}(t-i\delta)  \, \left( I_{\tilde{L}} \otimes \rho_{KR}  \right) \,\psi_{i,L}(t+i\delta)   }{\tfd} }{ \tr_{KR} \left[ \left( \rho_{KR}  \right)^{2} \right] }.
	\end{aligned}
\end{equation*}

Next, we derive the relation \eqref{eq:correlator0011}. Since $\ev{ \hat{d}_{\tilde{L}} }_{\beta}^{-1}=\tr_{KR} \left[ \left( \rho_{ KR }  \right)^{2} \right] $ by the definition \eqref{eq:defOfEffectiveDim}, we focus on the remaining factor $\mel{0}{ \mathcal{N}^{\text{SYK} \dagger}_{K,R\to T} \left[ \mathcal{N}^{ \text{SYK}}_{T\to K,R}[\ktbrad{0}{1}{T}] ] \right]}{1}$. To evaluate the factor, we use the definition of the adjoint channel \eqref{eq:defofAdjoint},
\begin{equation}
	\begin{aligned}
		&\mel{0}{ \mathcal{N}^{\text{SYK} \dagger}_{K,R\to T} \left[ \mathcal{N}^{ \text{SYK}}_{T\to K,R}[\ktbrad{0}{1}{T}] ] \right]}{1}\\
		&= \tr _{K,R}\left[\,   \mathcal{N}^{ \text{SYK}}_{T\to K,R}[\ktbrad{0}{1}{T}]\,  \mathcal{N}^{ \text{SYK}}_{T\to K,R}[\ktbrad{1}{0}{T}]  \, \right]\\
		&= \tr _{K,R}\left[\,   \tr_{\tilde{L}}\left[ U_{L} \ktbra{\tfd}{L,R} \psi_{i,L}^{\dagger}(-i\delta) U_{L}^{\dagger} \right]\, \tr_{\tilde{L}}\left[ U_{L} \psi_{i,L}(i\delta) \ktbra{\tfd}{L,R} U_{L}^{\dagger} \right] \, \right]\\
		&= \tr _{K,R}\left[\,   \tr_{\tilde{L}}\left[ U_{R} \ktbra{\tfd}{L,R} \psi_{i,L}^{\dagger}(-i\delta) U_{L}^{\dagger} \right]\, U_{R}U_{R}^{\dagger} \, \tr_{\tilde{L}}\left[ U_{L} \psi_{i,L}(i\delta) \ktbra{\tfd}{L,R} U_{R}^{\dagger} \right] \, \right]\\
		&= \tr _{K,R}\left[\, U_{R}   \tr_{\tilde{L}}\left[ \ktbra{\tfd}{L,R}U_{R}\, \psi_{i,L}^{\dagger}(-i\delta) U_{L}^{\dagger} \right]\,  \, \tr_{\tilde{L}}\left[ U_{L} \psi_{i,L}(i\delta) U_{R}^{\dagger} \ktbra{\tfd}{L,R} \right] \, U_{R}^{\dagger} \right]\\
		&= \tr _{K,R}\left[\,  \tr_{\tilde{L}}\left[ \ktbra{\tfd}{L,R}U_{L}\, \psi_{i,L}^{\dagger}(-i\delta) U_{L}^{\dagger} \right]\,  \, \tr_{\tilde{L}}\left[ U_{L} \psi_{i,L}(i\delta) U_{L}^{\dagger} \ktbra{\tfd}{L,R} \right] \, \right]\\
		&= \tr _{K,R}\left[\,  \tr_{\tilde{L}}\left[ \ktbra{\tfd}{L,R}\, \psi_{i,L}^{\dagger}(t-i\delta)  \right]\,  \, \tr_{\tilde{L}}\left[  \psi_{i,L}(t+i\delta)  \ktbra{\tfd}{L,R} \right] \, \right].
	\end{aligned}
\end{equation}
By explicitly introducing bases for the traces, we can rewrite the last expression as follows,
\begin{equation}
	\begin{aligned}
		&\tr _{K,R}\left[\,  \tr_{\tilde{L}}\left[ \ktbra{\tfd}{L,R}\, \psi_{i,L}^{\dagger}(t-i\delta)  \right]\,  \, \tr_{\tilde{L}}\left[  \psi_{i,L}(t+i\delta)  \ktbra{\tfd}{L,R} \right] \, \right]\\
		&=\sum_{\alpha,\alpha'=1}^{d_{K}d_{R}}\sum_{a,a'=1}^{d_{\tilde{L}}} \left( \,  _{KR} \bra{\alpha}\otimes \, _{\tilde{L}} \bra{a}   \right) \left(  \ktbra{\tfd}{L,R}\, \psi_{i,L}^{\dagger}(t-i\delta) \right) \left( \ket{\alpha'}_{KR} \otimes \ket{a}_{\tilde{L}} \right)\\
		& \hspace{4cm} \times \left( \,  _{KR} \bra{\alpha'}\otimes \, _{\tilde{L}} \bra{a'}   \right) \left(  \psi_{i,L}(t+i\delta)  \ktbra{\tfd}{L,R}\right) \left( \ket{\alpha}_{KR} \otimes \ket{a'}_{\tilde{L}} \right)\\
		&= \sum_{\alpha,\alpha'=1}^{d_{K}d_{R}}\sum_{a,a'=1}^{d_{\tilde{L}}}  \, _{L,R} \bra{\tfd}  \psi_{i,L}^{\dagger}(t-i\delta) \left( \ket{\alpha'}_{KR} \otimes \ket{a}_{\tilde{L}} \right)\\
		&  \hspace{4cm} \times \left( \,  _{KR} \bra{\alpha}\otimes \, _{\tilde{L}} \bra{a}   \right) \ktbra{\tfd}{L,R}\left( \ket{\alpha}_{KR} \otimes \ket{a'}_{\tilde{L}} \right)\\
		& \hspace{8cm}  \times \left( \,  _{KR} \bra{\alpha'}\otimes \, _{\tilde{L}} \bra{a'}   \right)  \psi_{i,L}(t+i\delta) \ket{\tfd}_{L,R}\\
		&=   \, _{L,R} \bra{\tfd}  \psi_{i,L}^{\dagger}(t-i\delta) \left(\tr_{KR} \left[\ktbra{\tfd}{L,R} \right] \otimes I_{KR}  \right)  \psi_{i,L}(t+i\delta) \ket{\tfd}_{L,R}\\
		&= \, _{L,R} \bra{\tfd}  \psi_{i,L}^{\dagger}(t-i\delta) \left( \rho_{\tilde{L}} \otimes I_{KR}  \right)  \psi_{i,L}(t+i\delta) \ket{\tfd}_{L,R}.
	\end{aligned}
\end{equation}

Therefore, we get the relation \eqref{eq:correlator0011},
\begin{equation*}
	\begin{aligned}
		\ev{ \hat{d}_{\tilde{L}} }_{\beta} \cdot  \mel{0}{ \mathcal{N}^{\text{SYK} \dagger}_{K,R\to T} \left[ \mathcal{N}^{ \text{SYK}}_{T\to K,R}[\ktbrad{0}{1}{T}] ] \right]}{1}&= \frac{1}{Z_{\delta}}\cdot  \dfrac{  \mel{\tfd}{   \psi_{i,L}(t -i\delta)  \, \left( \rho_{\tilde{L}} \otimes I_{KR}  \right) \,\psi_{i,L}(t+i\delta)   }{\tfd} }{ \tr_{KR} \left[ \left( \rho_{ KR }  \right)^{2} \right] }.
	\end{aligned}
\end{equation*}

\bibliographystyle{JHEP}
\bibliography{island.bib}

\providecommand{\href}[2]{#2}\begingroup\raggedright\begin{thebibliography}{10}

\bibitem{Penington:2019npb}
G.~Penington, \emph{{Entanglement Wedge Reconstruction and the Information Paradox}}, \href{https://doi.org/10.1007/JHEP09(2020)002}{\emph{JHEP} {\bfseries 09} (2020) 002} [\href{https://arxiv.org/abs/1905.08255}{{\ttfamily 1905.08255}}].

\bibitem{Almheiri:2019psf}
A.~Almheiri, N.~Engelhardt, D.~Marolf and H.~Maxfield, \emph{{The entropy of bulk quantum fields and the entanglement wedge of an evaporating black hole}}, \href{https://doi.org/10.1007/JHEP12(2019)063}{\emph{JHEP} {\bfseries 12} (2019) 063} [\href{https://arxiv.org/abs/1905.08762}{{\ttfamily 1905.08762}}].

\bibitem{Almheiri:2019hni}
A.~Almheiri, R.~Mahajan, J.~Maldacena and Y.~Zhao, \emph{{The Page curve of Hawking radiation from semiclassical geometry}}, \href{https://doi.org/10.1007/JHEP03(2020)149}{\emph{JHEP} {\bfseries 03} (2020) 149} [\href{https://arxiv.org/abs/1908.10996}{{\ttfamily 1908.10996}}].

\bibitem{Penington:2019kki}
G.~Penington, S.~H. Shenker, D.~Stanford and Z.~Yang, \emph{{Replica wormholes and the black hole interior}}, \href{https://doi.org/10.1007/JHEP03(2022)205}{\emph{JHEP} {\bfseries 03} (2022) 205} [\href{https://arxiv.org/abs/1911.11977}{{\ttfamily 1911.11977}}].

\bibitem{Almheiri:2019qdq}
A.~Almheiri, T.~Hartman, J.~Maldacena, E.~Shaghoulian and A.~Tajdini, \emph{{Replica Wormholes and the Entropy of Hawking Radiation}}, \href{https://doi.org/10.1007/JHEP05(2020)013}{\emph{JHEP} {\bfseries 05} (2020) 013} [\href{https://arxiv.org/abs/1911.12333}{{\ttfamily 1911.12333}}].

\bibitem{Hayden:2007cs}
P.~Hayden and J.~Preskill, \emph{{Black holes as mirrors: Quantum information in random subsystems}}, \href{https://doi.org/10.1088/1126-6708/2007/09/120}{\emph{JHEP} {\bfseries 09} (2007) 120} [\href{https://arxiv.org/abs/0708.4025}{{\ttfamily 0708.4025}}].

\bibitem{Verlinde:2012cy}
E.~Verlinde and H.~Verlinde, \emph{{Black Hole Entanglement and Quantum Error Correction}}, \href{https://doi.org/10.1007/JHEP10(2013)107}{\emph{JHEP} {\bfseries 10} (2013) 107} [\href{https://arxiv.org/abs/1211.6913}{{\ttfamily 1211.6913}}].

\bibitem{Papadodimas:2012aq}
K.~Papadodimas and S.~Raju, \emph{{An Infalling Observer in AdS/CFT}}, \href{https://doi.org/10.1007/JHEP10(2013)212}{\emph{JHEP} {\bfseries 10} (2013) 212} [\href{https://arxiv.org/abs/1211.6767}{{\ttfamily 1211.6767}}].

\bibitem{Barnum2000ReversingQD}
H.~Barnum and E.~Knill, \emph{Reversing quantum dynamics with near-optimal quantum and classical fidelity}, {\emph{Journal of Mathematical Physics} {\bfseries 43} (2000) 2097}.

\bibitem{Petz:1986tvy}
D.~Petz, \emph{{Sufficient subalgebras and the relative entropy of states of a von Neumann algebra}}, \href{https://doi.org/10.1007/BF01212345}{\emph{Commun. Math. Phys.} {\bfseries 105} (1986) 123}.

\bibitem{ohya2004quantum}
M.~Ohya and D.~Petz, \emph{Quantum Entropy and Its Use}, Theoretical and Mathematical Physics. Springer Berlin Heidelberg, 2004.

\bibitem{Yoshida:2021xyb}
B.~Yoshida, \emph{{Recovery algorithms for Clifford Hayden-Preskill problem}},  \href{https://arxiv.org/abs/2106.15628}{{\ttfamily 2106.15628}}.

\bibitem{Yoshida:2021haf}
B.~Yoshida, \emph{{Decoding the Entanglement Structure of Monitored Quantum Circuits}},  \href{https://arxiv.org/abs/2109.08691}{{\ttfamily 2109.08691}}.

\bibitem{Chandrasekaran:2022qmq}
V.~Chandrasekaran and A.~Levine, \emph{{Quantum error correction in SYK and bulk emergence}}, \href{https://doi.org/10.1007/JHEP06(2022)039}{\emph{JHEP} {\bfseries 06} (2022) 039} [\href{https://arxiv.org/abs/2203.05058}{{\ttfamily 2203.05058}}].

\bibitem{Nakata:2023hwg}
Y.~Nakata and M.~Tezuka, \emph{{Hayden-Preskill Recovery in Hamiltonian Systems}},  \href{https://arxiv.org/abs/2303.02010}{{\ttfamily 2303.02010}}.

\bibitem{WIP}
Y.~Nakayama, A.~Miyata and T.~Ugajin, \emph{{Work in progress}}, .

\bibitem{Yoshida:2017non}
B.~Yoshida and A.~Kitaev, \emph{{Efficient decoding for the Hayden-Preskill protocol}},  \href{https://arxiv.org/abs/1710.03363}{{\ttfamily 1710.03363}}.

\bibitem{Schumacher:1996dy}
B.~Schumacher and M.~A. Nielsen, \emph{{Quantum data processing and error correction}}, \href{https://doi.org/10.1103/PhysRevA.54.2629}{\emph{Phys. Rev. A} {\bfseries 54} (1996) 2629} [\href{https://arxiv.org/abs/quant-ph/9604022}{{\ttfamily quant-ph/9604022}}].

\bibitem{Nielsen_2007}
M.~A. Nielsen and D.~Poulin, \emph{Algebraic and information-theoretic conditions for operator quantum error correction}, \href{https://doi.org/10.1103/physreva.75.064304}{\emph{Physical Review A} {\bfseries 75} (2007) }.

\bibitem{Petz:2002eql}
D.~Petz, \emph{{MONOTONICITY OF QUANTUM RELATIVE ENTROPY REVISITED}}, \href{https://doi.org/10.1142/S0129055X03001576}{\emph{Rev. Math. Phys.} {\bfseries 15} (2003) 79} [\href{https://arxiv.org/abs/quant-ph/0209053}{{\ttfamily quant-ph/0209053}}].

\bibitem{Vardhan:2021mdy}
S.~Vardhan, J.~Kudler-Flam, H.~Shapourian and H.~Liu, \emph{{Mixed-state entanglement and information recovery in thermalized states and evaporating black holes}}, \href{https://doi.org/10.1007/JHEP01(2023)064}{\emph{JHEP} {\bfseries 01} (2023) 064} [\href{https://arxiv.org/abs/2112.00020}{{\ttfamily 2112.00020}}].

\bibitem{Kudler-Flam:2022jwd}
J.~Kudler-Flam and P.~Rath, \emph{{Large and small corrections to the JLMS Formula from replica wormholes}}, \href{https://doi.org/10.1007/JHEP08(2022)189}{\emph{JHEP} {\bfseries 08} (2022) 189} [\href{https://arxiv.org/abs/2203.11954}{{\ttfamily 2203.11954}}].

\bibitem{Kudler-Flam:2021alo}
J.~Kudler-Flam, V.~Narovlansky and S.~Ryu, \emph{{Distinguishing Random and Black Hole Microstates}}, \href{https://doi.org/10.1103/PRXQuantum.2.040340}{\emph{PRX Quantum} {\bfseries 2} (2021) 040340} [\href{https://arxiv.org/abs/2108.00011}{{\ttfamily 2108.00011}}].

\bibitem{Sachdev:1992fk}
S.~Sachdev and J.~Ye, \emph{{Gapless spin fluid ground state in a random, quantum Heisenberg magnet}}, \href{https://doi.org/10.1103/PhysRevLett.70.3339}{\emph{Phys. Rev. Lett.} {\bfseries 70} (1993) 3339} [\href{https://arxiv.org/abs/cond-mat/9212030}{{\ttfamily cond-mat/9212030}}].

\bibitem{KitaevTalks}
A.~Kitaev, ``A simple model of quantum holography..'' \url{http://online.kitp.ucsb.edu/online/entangled15/kitaev/},\url{http://online.kitp.ucsb.edu/online/entangled15/kitaev2/}.

\bibitem{Sachdev:2015efa}
S.~Sachdev, \emph{{Bekenstein-Hawking Entropy and Strange Metals}}, \href{https://doi.org/10.1103/PhysRevX.5.041025}{\emph{Phys. Rev. X} {\bfseries 5} (2015) 041025} [\href{https://arxiv.org/abs/1506.05111}{{\ttfamily 1506.05111}}].

\bibitem{Chandrasekaran:2021tkb}
V.~Chandrasekaran, T.~Faulkner and A.~Levine, \emph{{Scattering strings off quantum extremal surfaces}}, \href{https://doi.org/10.1007/JHEP08(2022)143}{\emph{JHEP} {\bfseries 08} (2022) 143} [\href{https://arxiv.org/abs/2108.01093}{{\ttfamily 2108.01093}}].

\bibitem{Qi:2018bje}
X.-L. Qi and A.~Streicher, \emph{{Quantum Epidemiology: Operator Growth, Thermal Effects, and SYK}}, \href{https://doi.org/10.1007/JHEP08(2019)012}{\emph{JHEP} {\bfseries 08} (2019) 012} [\href{https://arxiv.org/abs/1810.11958}{{\ttfamily 1810.11958}}].

\bibitem{Maldacena:2016hyu}
J.~Maldacena and D.~Stanford, \emph{{Remarks on the Sachdev-Ye-Kitaev model}}, \href{https://doi.org/10.1103/PhysRevD.94.106002}{\emph{Phys. Rev. D} {\bfseries 94} (2016) 106002} [\href{https://arxiv.org/abs/1604.07818}{{\ttfamily 1604.07818}}].

\bibitem{Polchinski:2016xgd}
J.~Polchinski and V.~Rosenhaus, \emph{{The Spectrum in the Sachdev-Ye-Kitaev Model}}, \href{https://doi.org/10.1007/JHEP04(2016)001}{\emph{JHEP} {\bfseries 04} (2016) 001} [\href{https://arxiv.org/abs/1601.06768}{{\ttfamily 1601.06768}}].

\bibitem{Bagrets:2017pwq}
D.~Bagrets, A.~Altland and A.~Kamenev, \emph{{Power-law out of time order correlation functions in the SYK model}}, \href{https://doi.org/10.1016/j.nuclphysb.2017.06.012}{\emph{Nucl. Phys. B} {\bfseries 921} (2017) 727} [\href{https://arxiv.org/abs/1702.08902}{{\ttfamily 1702.08902}}].

\bibitem{Gross:2017aos}
D.~J. Gross and V.~Rosenhaus, \emph{{All point correlation functions in SYK}}, \href{https://doi.org/10.1007/JHEP12(2017)148}{\emph{JHEP} {\bfseries 12} (2017) 148} [\href{https://arxiv.org/abs/1710.08113}{{\ttfamily 1710.08113}}].

\bibitem{Kitaev:2017awl}
A.~Kitaev and S.~J. Suh, \emph{{The soft mode in the Sachdev-Ye-Kitaev model and its gravity dual}}, \href{https://doi.org/10.1007/JHEP05(2018)183}{\emph{JHEP} {\bfseries 05} (2018) 183} [\href{https://arxiv.org/abs/1711.08467}{{\ttfamily 1711.08467}}].

\bibitem{Romero-Bermudez:2019vej}
A.~Romero-Berm\'udez, K.~Schalm and V.~Scopelliti, \emph{{Regularization dependence of the OTOC. Which Lyapunov spectrum is the physical one?}}, \href{https://doi.org/10.1007/JHEP07(2019)107}{\emph{JHEP} {\bfseries 07} (2019) 107} [\href{https://arxiv.org/abs/1903.09595}{{\ttfamily 1903.09595}}].

\bibitem{Trunin:2020vwy}
D.~A. Trunin, \emph{{Pedagogical introduction to the Sachdev\textendash{}Ye\textendash{}Kitaev model and two-dimensional dilaton gravity}}, \href{https://doi.org/10.3367/UFNe.2020.06.038805}{\emph{Usp. Fiz. Nauk} {\bfseries 191} (2021) 225} [\href{https://arxiv.org/abs/2002.12187}{{\ttfamily 2002.12187}}].

\bibitem{Sarosi:2017ykf}
G.~S\'arosi, \emph{{AdS$_{2}$ holography and the SYK model}}, \href{https://doi.org/10.22323/1.323.0001}{\emph{PoS} {\bfseries Modave2017} (2018) 001} [\href{https://arxiv.org/abs/1711.08482}{{\ttfamily 1711.08482}}].

\bibitem{Sekino:2008he}
Y.~Sekino and L.~Susskind, \emph{{Fast Scramblers}}, \href{https://doi.org/10.1088/1126-6708/2008/10/065}{\emph{JHEP} {\bfseries 10} (2008) 065} [\href{https://arxiv.org/abs/0808.2096}{{\ttfamily 0808.2096}}].

\bibitem{Knill:1996ny}
E.~Knill and R.~Laflamme, \emph{{A Theory of quantum error correcting codes}}, \href{https://doi.org/10.1103/PhysRevLett.84.2525}{\emph{Phys. Rev. Lett.} {\bfseries 84} (2000) 2525} [\href{https://arxiv.org/abs/quant-ph/9604034}{{\ttfamily quant-ph/9604034}}].

\bibitem{Gao:2019nyj}
P.~Gao and D.~L. Jafferis, \emph{{A traversable wormhole teleportation protocol in the SYK model}}, \href{https://doi.org/10.1007/JHEP07(2021)097}{\emph{JHEP} {\bfseries 07} (2021) 097} [\href{https://arxiv.org/abs/1911.07416}{{\ttfamily 1911.07416}}].

\bibitem{Brown:2019hmk}
A.~R. Brown, H.~Gharibyan, S.~Leichenauer, H.~W. Lin, S.~Nezami, G.~Salton et~al., \emph{{Quantum Gravity in the Lab. I. Teleportation by Size and Traversable Wormholes}}, \href{https://doi.org/10.1103/PRXQuantum.4.010320}{\emph{PRX Quantum} {\bfseries 4} (2023) 010320} [\href{https://arxiv.org/abs/1911.06314}{{\ttfamily 1911.06314}}].

\bibitem{Schuster:2021uvg}
T.~Schuster, B.~Kobrin, P.~Gao, I.~Cong, E.~T. Khabiboulline, N.~M. Linke et~al., \emph{{Many-Body Quantum Teleportation via Operator Spreading in the Traversable Wormhole Protocol}}, \href{https://doi.org/10.1103/PhysRevX.12.031013}{\emph{Phys. Rev. X} {\bfseries 12} (2022) 031013} [\href{https://arxiv.org/abs/2102.00010}{{\ttfamily 2102.00010}}].

\bibitem{Nezami:2021yaq}
S.~Nezami, H.~W. Lin, A.~R. Brown, H.~Gharibyan, S.~Leichenauer, G.~Salton et~al., \emph{{Quantum Gravity in the Lab. II. Teleportation by Size and Traversable Wormholes}}, \href{https://doi.org/10.1103/PRXQuantum.4.010321}{\emph{PRX Quantum} {\bfseries 4} (2023) 010321} [\href{https://arxiv.org/abs/2102.01064}{{\ttfamily 2102.01064}}].

\bibitem{Pastawski:2016qrs}
F.~Pastawski and J.~Preskill, \emph{{Code properties from holographic geometries}}, \href{https://doi.org/10.1103/PhysRevX.7.021022}{\emph{Phys. Rev. X} {\bfseries 7} (2017) 021022} [\href{https://arxiv.org/abs/1612.00017}{{\ttfamily 1612.00017}}].

\bibitem{Bentsen:2023xlu}
G.~Bentsen, P.~Nguyen and B.~Swingle, \emph{{Approximate Quantum Codes From Long Wormholes}},  \href{https://arxiv.org/abs/2310.07770}{{\ttfamily 2310.07770}}.

\bibitem{Garcia-Garcia:2017bkg}
A.~M. Garc\'\i{}a-Garc\'\i{}a, B.~Loureiro, A.~Romero-Berm\'udez and M.~Tezuka, \emph{{Chaotic-Integrable Transition in the Sachdev-Ye-Kitaev Model}}, \href{https://doi.org/10.1103/PhysRevLett.120.241603}{\emph{Phys. Rev. Lett.} {\bfseries 120} (2018) 241603} [\href{https://arxiv.org/abs/1707.02197}{{\ttfamily 1707.02197}}].

\bibitem{Akers:2022qdl}
C.~Akers, N.~Engelhardt, D.~Harlow, G.~Penington and S.~Vardhan, \emph{{The black hole interior from non-isometric codes and complexity}},  \href{https://arxiv.org/abs/2207.06536}{{\ttfamily 2207.06536}}.

\bibitem{Berkooz:2018jqr}
M.~Berkooz, M.~Isachenkov, V.~Narovlansky and G.~Torrents, \emph{{Towards a full solution of the large N double-scaled SYK model}}, \href{https://doi.org/10.1007/JHEP03(2019)079}{\emph{JHEP} {\bfseries 03} (2019) 079} [\href{https://arxiv.org/abs/1811.02584}{{\ttfamily 1811.02584}}].

\bibitem{Lin:2023trc}
H.~W. Lin and D.~Stanford, \emph{{A symmetry algebra in double-scaled SYK}},  \href{https://arxiv.org/abs/2307.15725}{{\ttfamily 2307.15725}}.

\bibitem{barnum2000reversing}
H.~Barnum and E.~Knill, \emph{Reversing quantum dynamics with near-optimal quantum and classical fidelity},  2000.

\bibitem{nielsen_chuang_2010}
M.~A. Nielsen and I.~L. Chuang, \emph{Quantum Computation and Quantum Information: 10th Anniversary Edition}. Cambridge University Press, 2010, \href{https://doi.org/10.1017/CBO9780511976667}{10.1017/CBO9780511976667}.

\bibitem{Liu:2020sqb}
J.~Liu, \emph{{Scrambling and decoding the charged quantum information}}, \href{https://doi.org/10.1103/PhysRevResearch.2.043164}{\emph{Phys. Rev. Res.} {\bfseries 2} (2020) 043164} [\href{https://arxiv.org/abs/2003.11425}{{\ttfamily 2003.11425}}].

\end{thebibliography}\endgroup

\end{document}